\documentclass{JHEP3}

\usepackage{amsmath}
\usepackage{graphicx}
\usepackage{epstopdf}
\usepackage{subfigure}
\usepackage{multirow}
\usepackage{paralist}

\newcommand{\vsh}{\vspace{-.5cm}}

\newcommand{\GeV}{\text{GeV}}

\DeclareRobustCommand{\Sec}[1]{Sec.~\ref{#1}}

\DeclareRobustCommand{\App}[1]{App.~\ref{#1}}
\DeclareRobustCommand{\Tab}[1]{Table~\ref{#1}}

\DeclareRobustCommand{\Fig}[1]{Fig.~\ref{#1}}

\DeclareRobustCommand{\Eq}[1]{Eq.~(\ref{#1})}

\DeclareRobustCommand{\Ref}[1]{Ref.~\cite{#1}}
\DeclareRobustCommand{\Refs}[1]{Refs.~\cite{#1}}

\newcommand{\be}{\begin{equation}}
\newcommand{\ee}{\end{equation}}

\title{Identifying Boosted Objects with N-subjettiness}

\author{Jesse Thaler and Ken Van Tilburg \\
Center for Theoretical Physics, Massachusetts Institute of Technology,
Cambridge, MA 02139, USA \vspace{0.05in} \\
E-mail: \email{jthaler@jthaler.net}, \email{kvt@mit.edu}  
}

\preprint{MIT-CTP 4191}

\abstract{We introduce a new jet shape---$N$-subjettiness---designed to identify boosted hadronically-decaying objects like electroweak bosons and top quarks.  Combined with a jet invariant mass cut, $N$-subjettiness is an effective discriminating variable for tagging boosted objects and rejecting the background of QCD jets with large invariant mass.  In efficiency studies of boosted $W$ bosons and top quarks, we find tagging efficiencies of 30\% are achievable with fake rates of 1\%.  We also consider the discovery potential for new heavy resonances that decay to pairs of boosted objects, and find significant improvements are possible using $N$-subjettiness.  In this way, $N$-subjettiness combines the advantages of jet shapes with the discriminating power seen in previous jet substructure algorithms.}

\begin{document}

\section{Introduction}

The Large Hadron Collider (LHC) will search for new physics by probing a previously unexplored kinematic regime.  Most new physics scenarios that provide a solution to the hierarchy problem predict that the LHC will produce new heavy particles with decay channels involving top quarks, $W$/$Z$ bosons, and Higgs bosons.  In addition, many extensions of the standard model  including technicolor and Higgs compositeness invoke new heavy resonances within the LHC reach with large branching fractions to pairs of gauge bosons and top quarks.  Therefore, a key task in the search for physics beyond the standard model is to efficiently identify final state electroweak gauge bosons and top quarks in a variety of kinematic configurations.

With its current center-of-mass energy of $\sqrt{s} = 7$ TeV, the LHC is already able to produce new TeV-scale resonances which can decay to highly boosted electroweak bosons and/or top quarks.  For a large enough boost factor, the decay and fragmentation of such a boosted object yields a collimated spray of hadrons which a standard jet algorithm would reconstruct as a single jet.  Thus, standard reconstruction methods for electroweak bosons and top quarks become ineffective due to the immensely large background of ordinary QCD jets.  One possibility is to focus on channels where the boosted object decays leptonically, though such methods discard much of the original signal and may therefore not be optimal for detecting new heavy resonances. 

Recently, there has been considerable progress in identifying boosted hadronically-decaying objects using jet substructure techniques.  Algorithmic methods use information from the jet clustering procedure to extract the internal structure of jets \cite{Seymour:1993mx,Butterworth:2002tt,YSplitter,Brooijmans:2008,Thaler:2008ju,Kaplan:2008ie}, and are able to successfully distinguish between jets originating from boosted electroweak boson and top quarks (denoted here as ``$W$ jets'', ``top jets'', etc.) and those originating from light quarks or gluons (``QCD jets'').   Jet shape methods efficiently tag boosted objects with jet-based observables that take advantage of the different energy flow in the decay pattern of signal jets and background jets. \cite{Almeida:2008yp}  In addition, there are ``jet grooming'' techniques such as filtering \cite{Butterworth:2008iy, Butterworth:2008tr}, pruning \cite{Ellis:2009su,Ellis:2009me}, trimming \cite{Krohn:2009th}, and their combinations \cite{Soper:2010xk} which aid in the identification of boosted objects by reducing the smearing effects of jet contamination from initial state radiation, underlying event activity, and pileup.  Taken together, these jet substructure methods show much promise for enhancing searches for new physics in all-hadronic decay channels.\footnote{Additional recent related work appears in \Refs{Butterworth:2009qa,Plehn:2009rk,Kribs:2009yh,Gallicchio:2010sw,Chen:2010wk,Falkowski:2010hi,Kribs:2010hp,Almeida:2010pa,Plehn:2010st,Bhattacherjee:2010za,Rehermann:2010vq,Hackstein:2010wk,Englert:2010ud,Katz:2010mr}.}

In this paper, we introduce a new tagging method for boosted objects based on a novel jet shape dubbed ``$N$-subjettiness'' and denoted by $\tau_N$.  This variable is adapted from the event shape ``$N$-jettiness'' advocated in \Ref{Stewart:2010tn} as a way to veto additional jet emissions and define an exclusive jet cross section.  Here, we take advantage of the multi-body kinematics in the decay pattern of boosted hadronic objects, and use $N$-subjettiness to effectively ``count'' the number of subjets in a given jet.  We find that  $\tau_2/\tau_1$ is an effective discriminating variable to identify two-prong objects like boosted $W$, $Z$, and Higgs bosons, and $\tau_3/\tau_2$ is effective for three-prong objects like boosted top quarks.  

Compared to previous jet substructure techniques, $N$-subjettiness has a number of advantages.  First, to identify boosted objects, one wants to find jets that contain two or more lobes of energy.   While previous jet shape measures do capture the deviation of a jet from a one-lobe QCD-like configuration, $N$-subjettiness is a more direct measure of the desired energy flow properties.  Second, it is convenient to be able to adjust the relative degree of signal efficiency and background rejection without having to perform computationally intensive algorithmic adjustments.  Like for other jet shape methods, $\tau_N$ can be calculated for every jet, and a flexible one-dimensional cut on a function $f(\tau_1, \dots, \tau_N)$ can determine the efficiency/rejection curve.  Similarly, the set of $\tau_N$ values can be used as input to a multivariate discriminant method for further optimization.  Third, $N$-subjettiness is an inclusive jet shape and can be defined and calculated without reference to a jet substructure algorithm.\footnote{As we will discuss below, for computational purposes, we use a definition of $N$-subjettiness that still has residual dependence on a clustering procedure.  See \Sec{subsec:candidate_subjets} for further discussion.}  This will likely make $N$-subjettiness more amenable to higher-order perturbative calculations and resummation techniques  (see, e.g.~recent work in~\Ref{Ellis:2010rw,Banfi:2010pa}) compared to algorithmic methods for studying substructure.  Finally, $N$-subjettiness gives favorable efficiency/rejection curves compared to other jet substructure methods.  While a detailed comparison to other methods is beyond the scope of this work, we are encouraged by these preliminary results.

The remainder of this paper is organized as follows.  In \Sec{sec:nsubjettiness}, we define $N$-subjettiness and discuss some of its properties.  We present tagging efficiency studies in \Sec{sec:efficiency}, where we use $N$-subjettiness to identify individual hadronic $W$ bosons and top quarks, and compare our method against the YSplitter technique \cite{Butterworth:2002tt,YSplitter,Brooijmans:2008} and the Johns Hopkins Top Tagger \cite{Kaplan:2008ie}.  We then apply $N$-subjettiness in \Sec{sec:case} to reconstruct hypothetical heavy resonances decaying to pairs of boosted objects.  Our conclusions follow in \Sec{sec:conclusions}, and further information appears in the appendices.

\section{Boosted Objects and $N$-subjettiness} 
\label{sec:nsubjettiness}

\begin{figure}[tp]
  \begin{center}
    \subfigure[]{\label{fig:wwbar}\includegraphics[trim = 0mm 0mm 0mm 10mm, clip, scale=0.25]{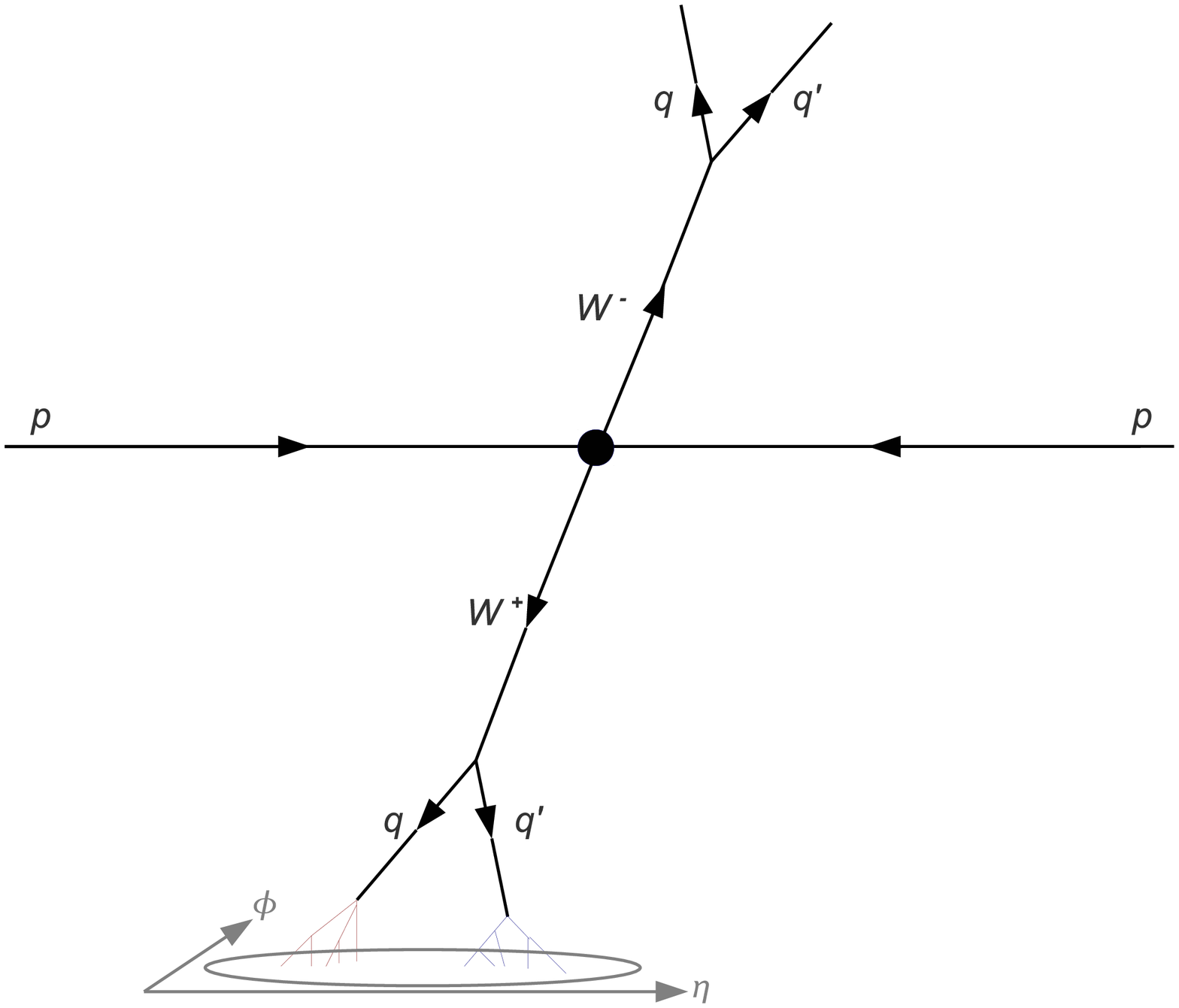}}
    \subfigure[]{\label{fig:eventDisplayW}\includegraphics[trim = 0mm 0mm 0mm 0mm, clip, scale=0.4]{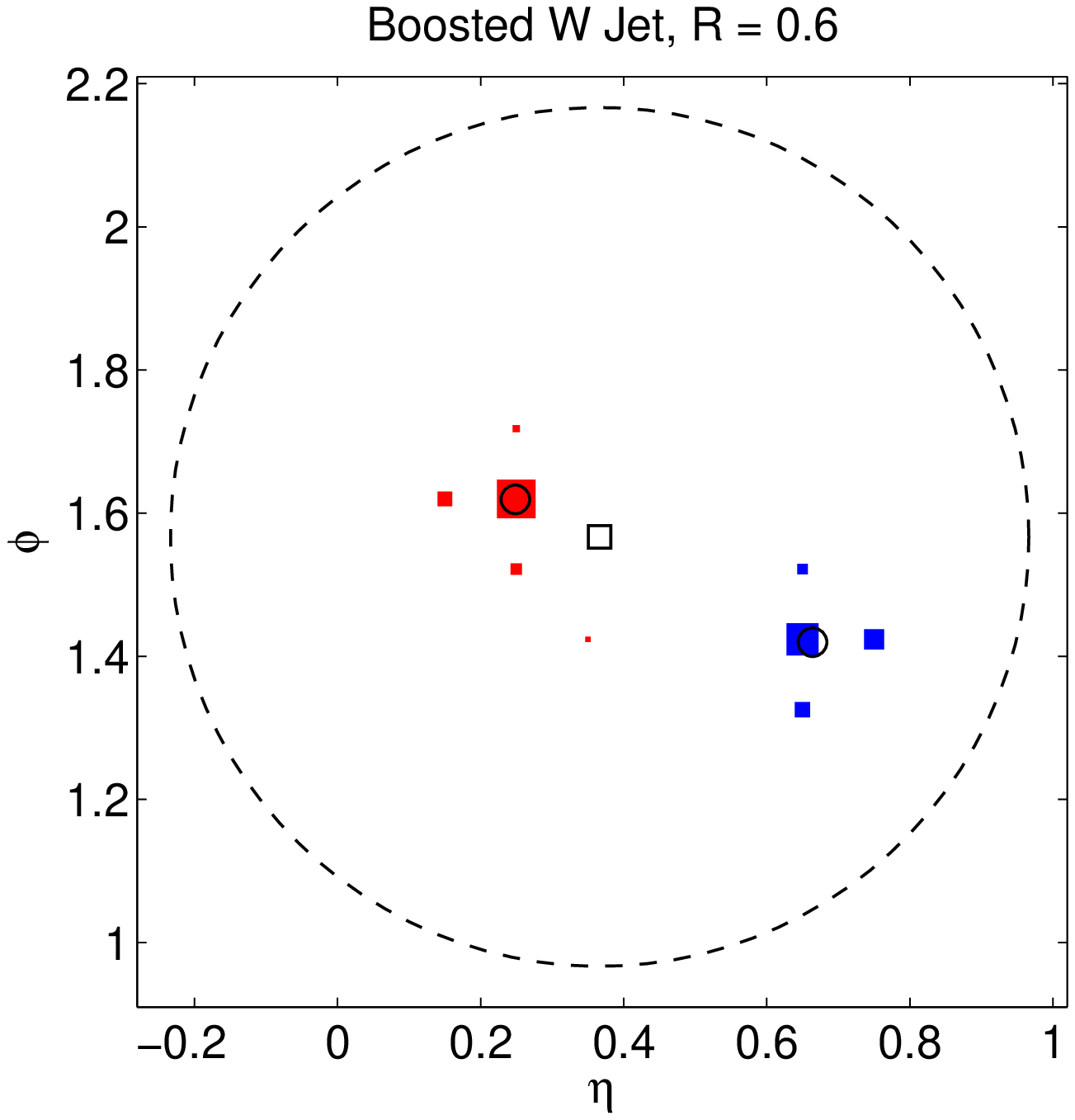}}
    \subfigure[]{\label{fig:qcdW}\includegraphics[trim = 0mm 10mm 0mm 0mm, clip, scale=0.25]{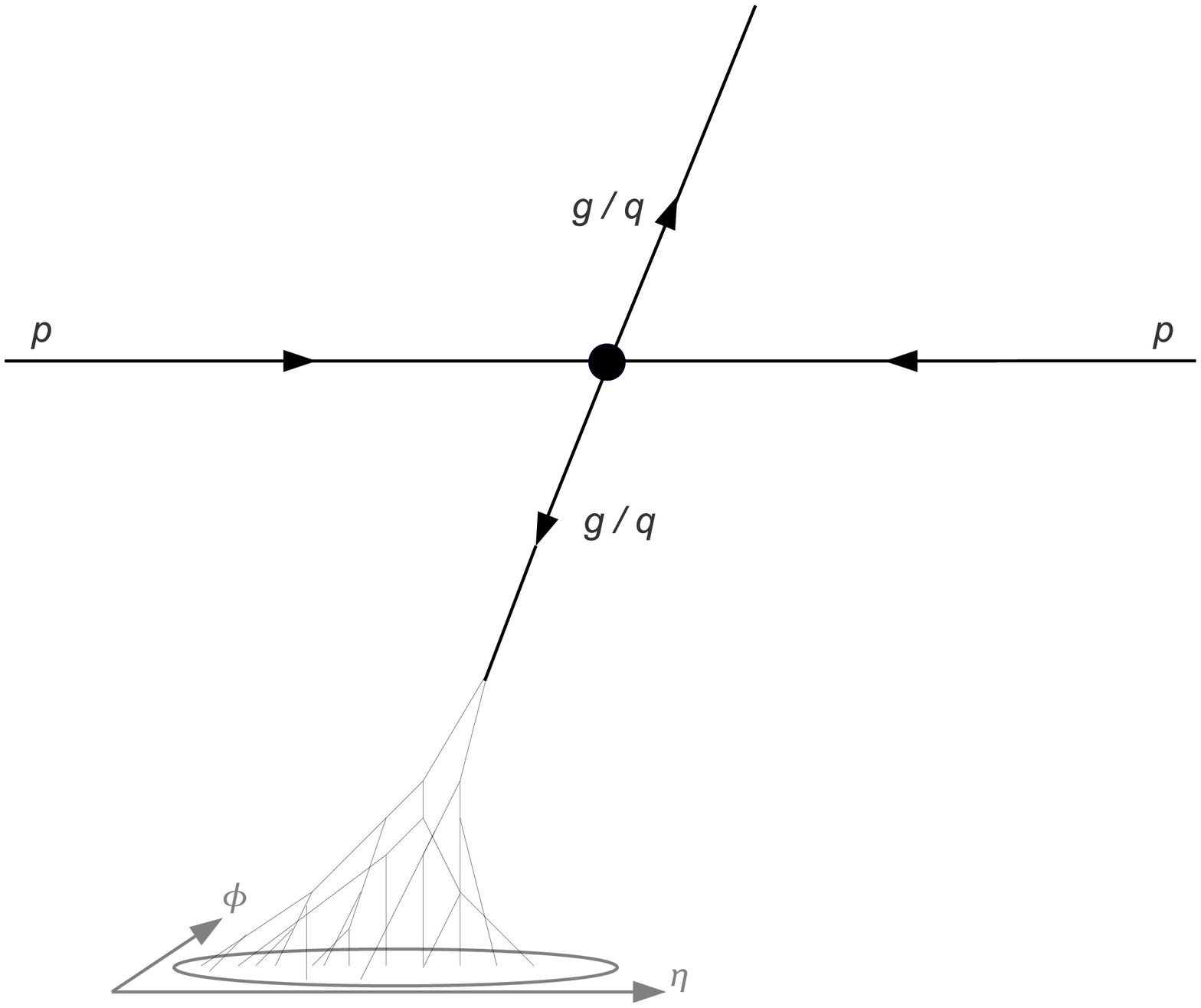}}
    \subfigure[]{\label{fig:eventDisplayQCDW}\includegraphics[trim = 0mm 0mm 0mm 0mm, clip, scale=0.4]{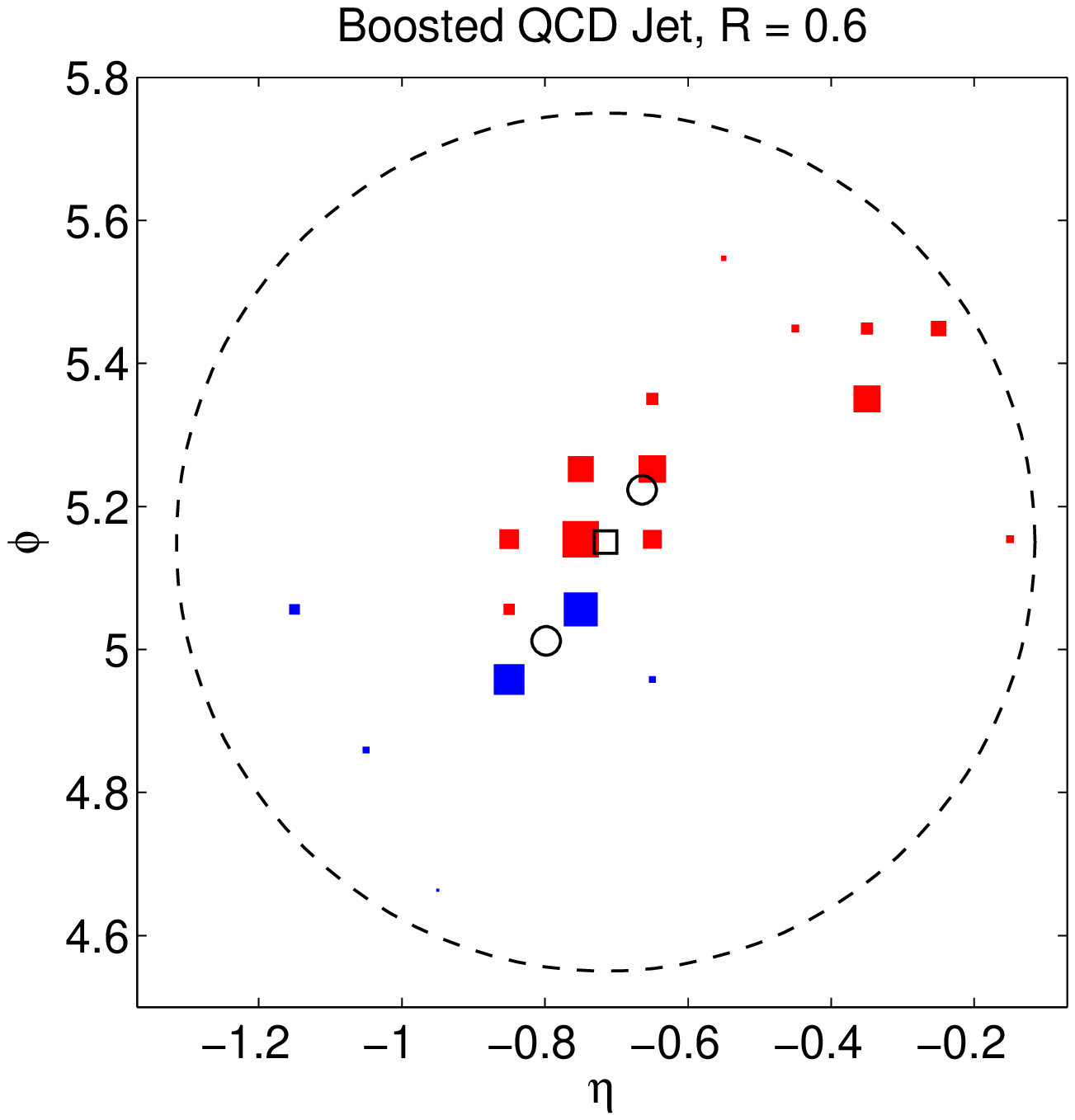}}
  \end{center}
  \caption{Left:  Schematic of the fully hadronic decay sequences in (a) $W^+ W^-$ and (c) dijet QCD events.  Whereas a $W$ jet is typically composed of two distinct lobes of energy, a QCD jet acquires invariant mass through multiple splittings.  Right:  Typical event displays for (b) $W$ jets and (d) QCD jets with invariant mass near $m_W$.  The jets are clustered with the anti-$k_T$ jet algorithm \cite{Cacciari:2008gp} using $R= 0.6$, with the dashed line giving the approximate boundary of the jet.   The marker size for each calorimeter cell is proportional to the logarithm of the particle energies in the cell.  The cells are colored according to how the exclusive $k_T$ algorithm divides the cells into two candidate subjets.  The open square indicates the total jet direction and the open circles indicate the two subjet directions.  The discriminating variable $\tau_2/\tau_1$ measures the relative alignment of the jet energy along the open circles compared to the open square.}
  \label{fig:eventDisplaysW}
\end{figure}

Boosted hadronic objects have a fundamentally different energy pattern than QCD jets of comparable invariant mass.  For concreteness, we will consider the case of a boosted $W$ boson as shown in \Fig{fig:eventDisplaysW}, though a similar discussion holds for boosted top quarks or new physics objects.  Since the $W$ decays to two quarks, a single jet containing a boosted $W$ boson should be composed of two distinct---but not necessarily easily resolved---hard subjets with a combined invariant mass of around 80 GeV.  A boosted QCD jet with an invariant mass of 80 GeV usually originates from a single hard parton and acquires mass through large angle soft splittings.  We want to exploit this difference in expected energy flow to differentiate between these two types of jets by ``counting'' the number of hard lobes of energy within a jet.

\subsection{Introducing $N$-subjettiness}

We start by defining an inclusive jet shape called ``$N$-subjettiness'' and denoted by $\tau_N$.  First, one reconstructs a candidate $W$ jet using some jet algorithm.  Then, one identifies $N$ candidate subjets using a procedure to be specified in \Sec{subsec:candidate_subjets}.  With these candidate subjets in hand, $\tau_N$ is calculated via
\begin{equation}
 \tau_N = \frac{1}{d_0} \sum\limits_k p_{T,k} \min \left\lbrace \Delta R _{1,k}, \Delta R _{2,k}, \cdots, \Delta R _{N,k} \right\rbrace.
 \label{eq:tau_N}
\end{equation}
Here, $k$ runs over the constituent particles in a given jet, $p_{T,k}$ are their transverse momenta, and $ \Delta R_{J,k} = \sqrt{(\Delta \eta)^2+(\Delta \phi)^2}$ is the distance in the rapidity-azimuth plane between a candidate subjet $J$ and a constituent particle $k$.  The normalization factor $d_0$ is taken as
\be
d_0 = \sum\limits_k p_{T,k} R_0,
\ee
where $R_0$ is the characteristic jet radius used in the original jet clustering algorithm.

\begin{figure}[tp]
  \begin{center}
    \subfigure[][]{\label{fig:tau1w}\includegraphics[trim = 00mm 0mm 00mm 0mm, clip,height=5cm]{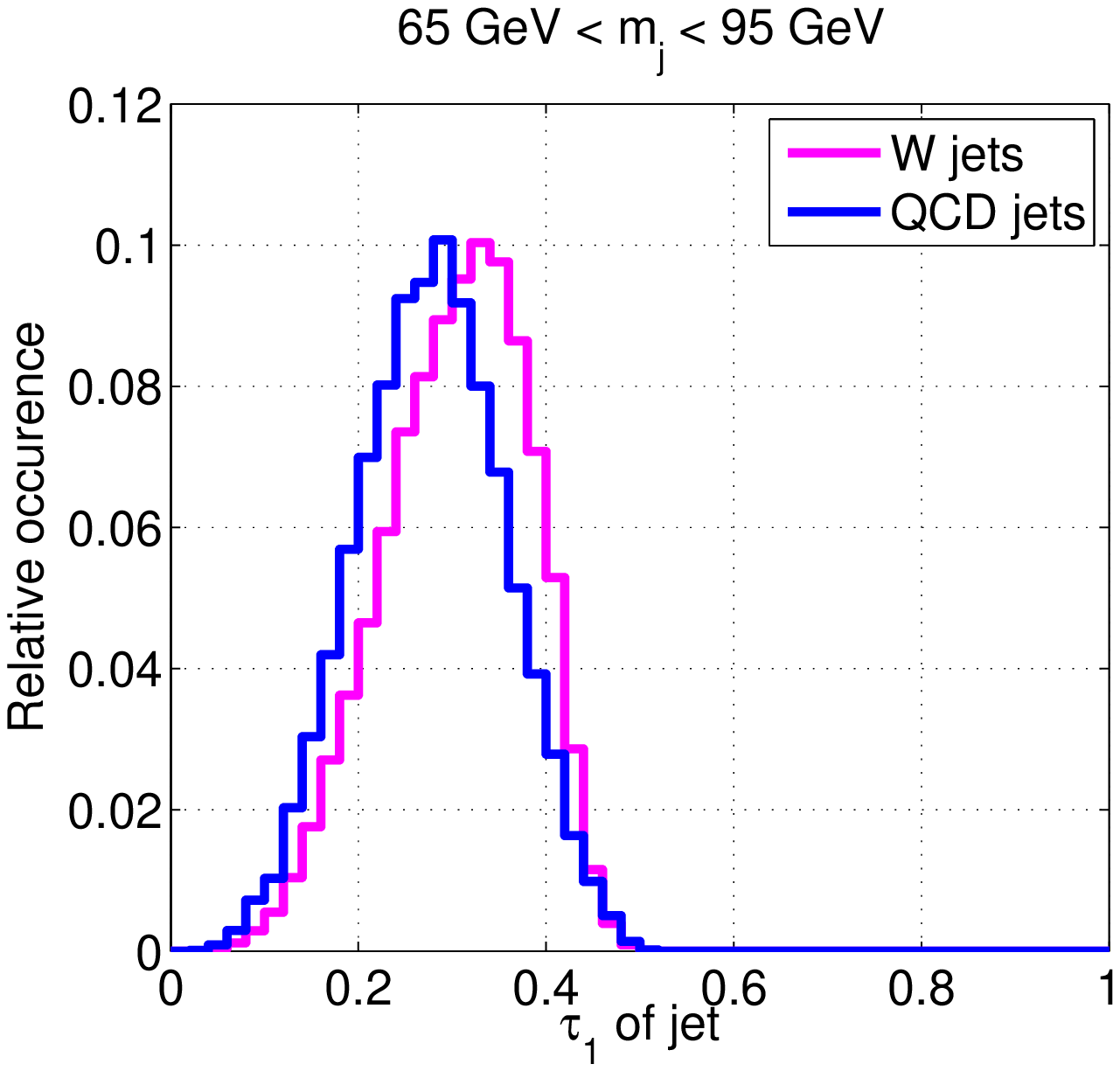}} 
    \subfigure[][]{\label{fig:tau2w}\includegraphics[trim = 00mm 0mm 00mm 0mm, clip,height=5cm]{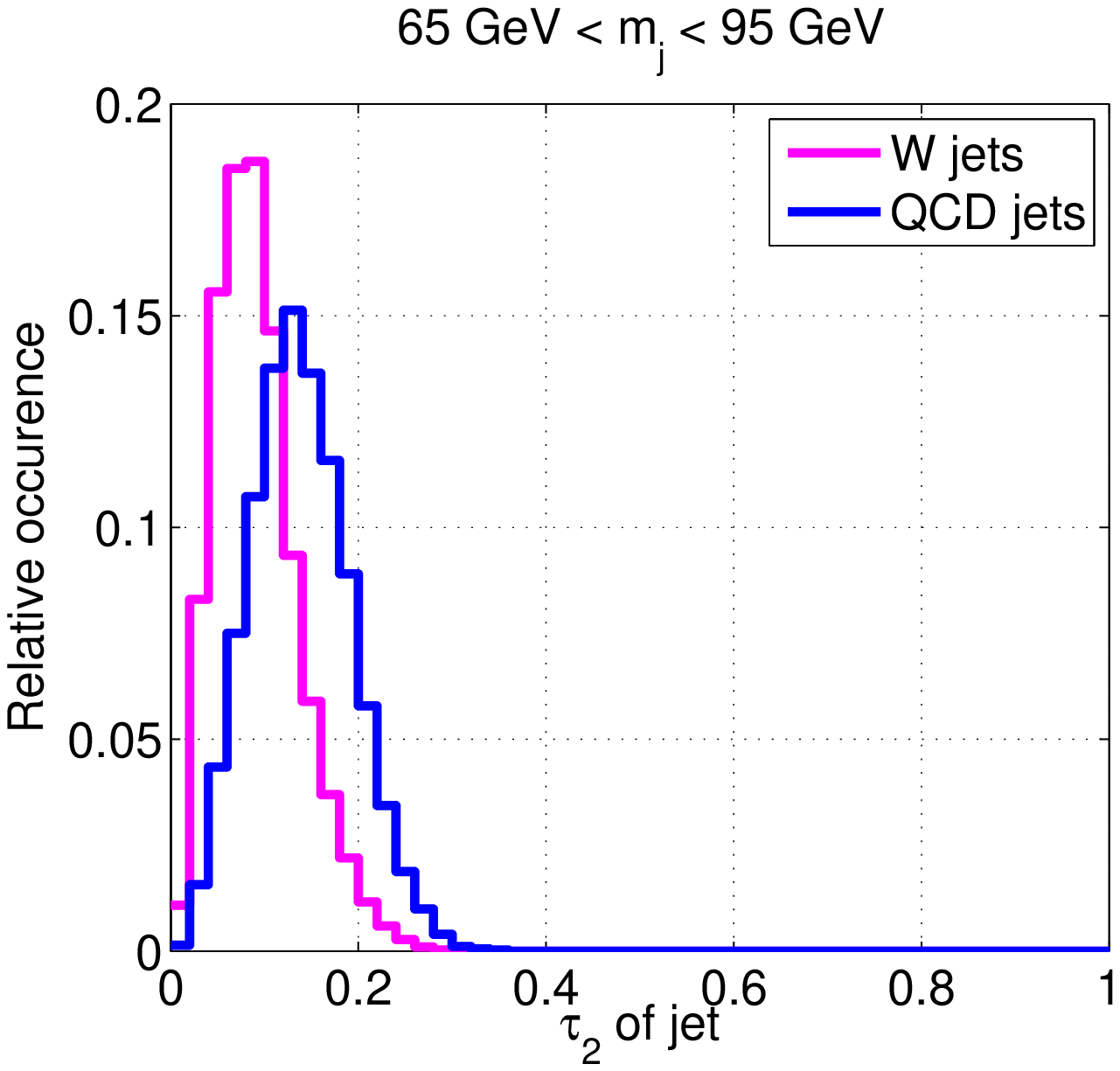}}
  \end{center}
  \vsh
\caption{Distributions of (a) $\tau_1$ and (b) $\tau_2$ for boosted $W$ and QCD jets.  For these plots, we impose an invariant mass window of $65 \text{ GeV} < m_{\text{jet}} < 95 \text{ GeV}$ on jets of $R = 0.6$, $p_T > 300$ GeV, and $|\eta| < 1.3$.  By themselves, the $\tau_N$ do not offer that much discriminating power for boosted objects beyond the invariant mass cut.}
  \label{fig:1Dtau1tau2tau3w}
\end{figure}

\begin{figure}[tp]
  \begin{center}
      \subfigure[][]{\label{fig:tau21w}\includegraphics[trim = 0mm 0mm 0mm 0mm, clip,height=5cm]{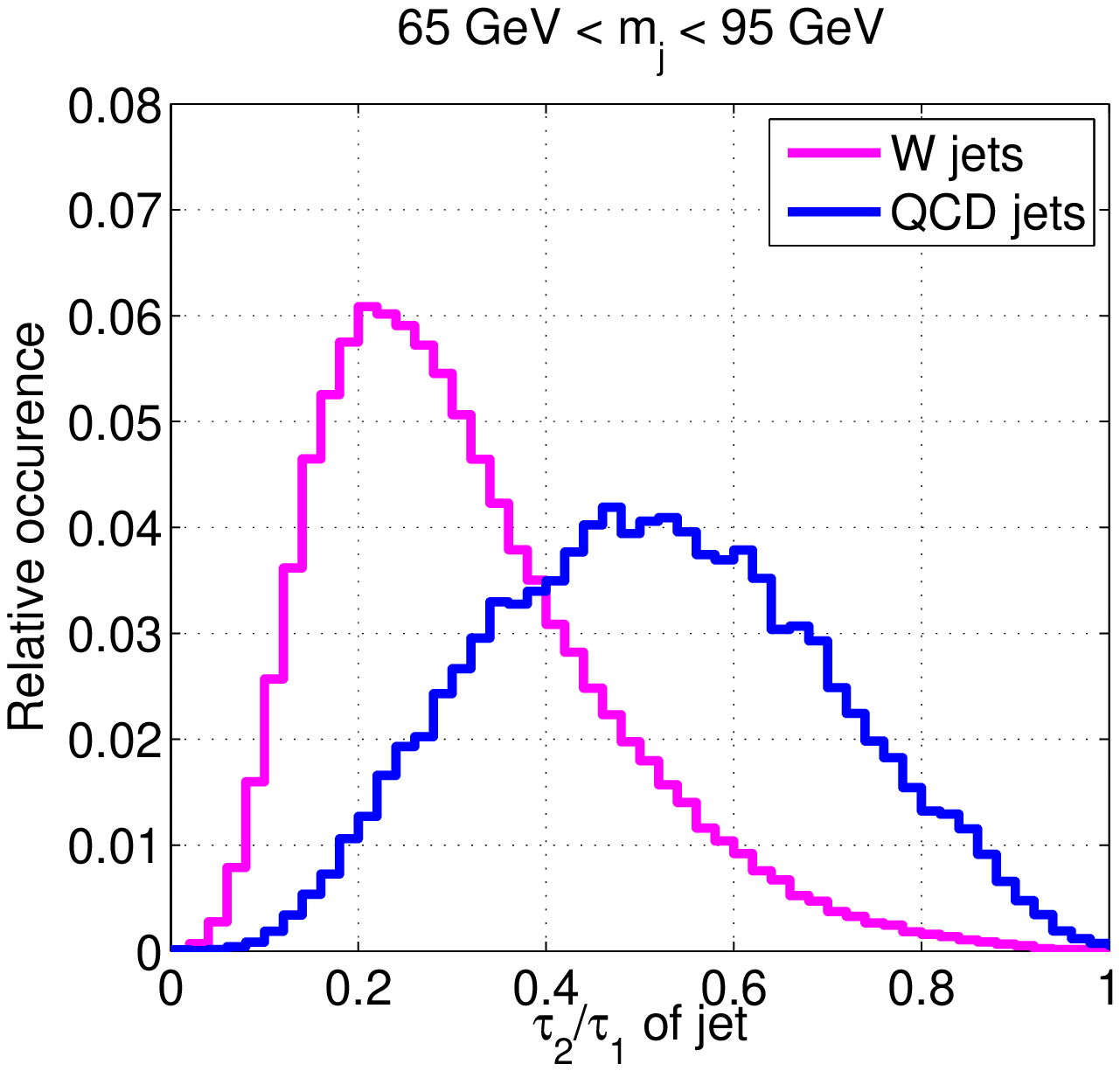}}
      \subfigure[][]{\label{fig:tau21density}\includegraphics[trim = 0mm 0mm 0mm 0mm, clip, height=5cm]{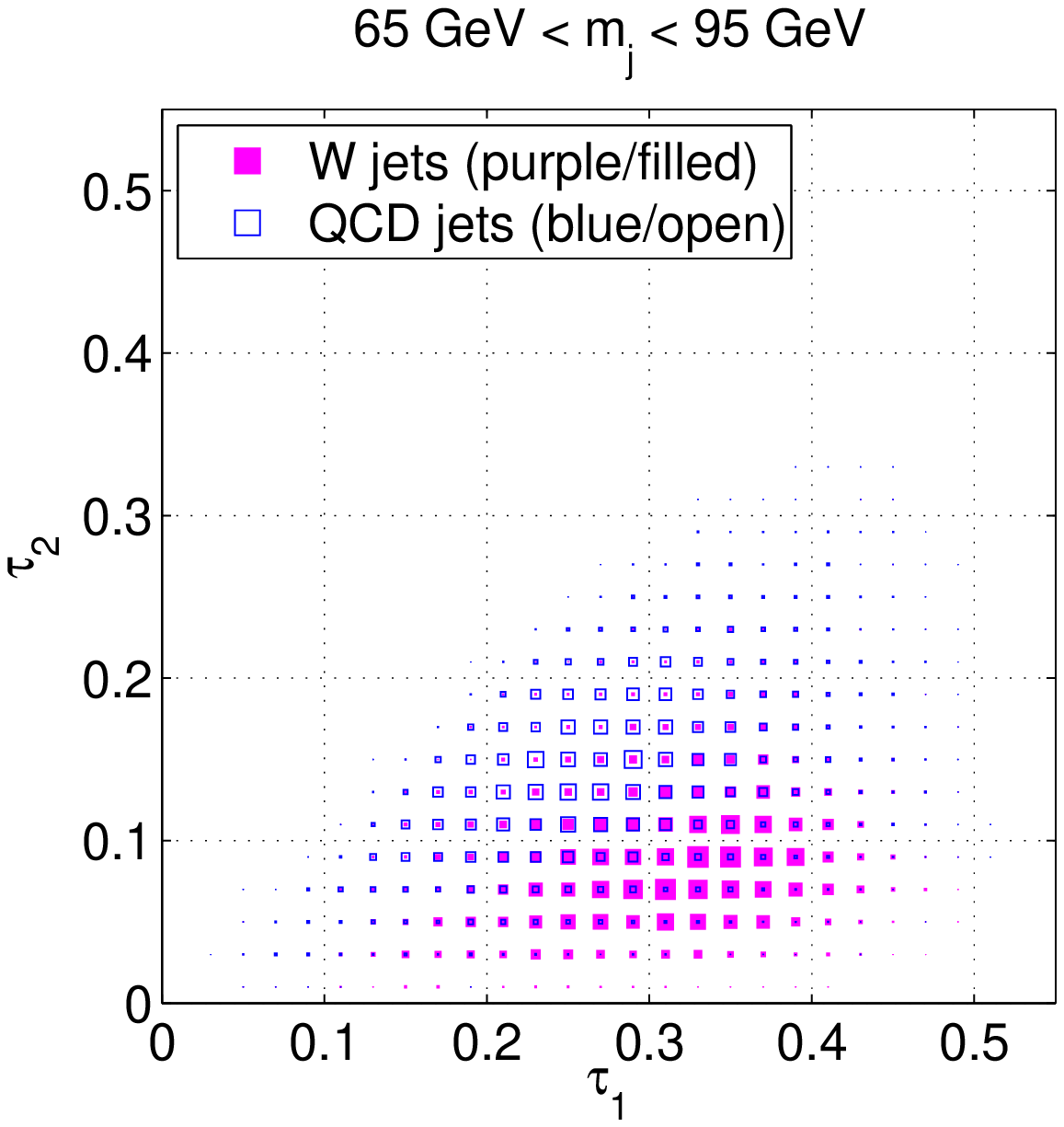}} \\
  \end{center}
  \vsh
  \caption{(a): Distribution of $\tau_2/\tau_1$ for boosted $W$ and QCD jets.  The selection criteria are the same as in \Fig{fig:1Dtau1tau2tau3w}.  One sees that the $\tau_2/\tau_1$ ratio gives considerable separation between $W$ jets and QCD jets beyond the invariant mass cut.  (b): Density plot in the $\tau_1$--$\tau_2$ plane.  Marker sizes are proportional to the number of jets in a given bin. In principle, a multivariate cut in the $\tau_1$--$\tau_2$ plane would give further distinguishing power.}
  \label{fig:W2Dtau12}
\end{figure}

It is straightforward to see that $\tau_N$ quantifies how $N$-subjetty a particular jet is, or in other words, to what degree it can be regarded as a jet composed of $N$ subjets.  Jets with $\tau_N \approx 0$ have all their radiation aligned with the candidate subjet directions and therefore have $N$ (or fewer) subjets.  Jets with $\tau_N \gg 0$ have a large fraction of their energy distributed away from the candidate subjet directions and therefore have at least $N+1$ subjets.  Plots of $\tau_1$ and $\tau_2$ comparing $W$ jets and QCD jets are shown in \Fig{fig:1Dtau1tau2tau3w}.

Less obvious is how best to use $\tau_N$ for identifying boosted $W$ bosons.  While one might naively expect that an event with small $\tau_2$ would be more likely to be a $W$ jet, observe that QCD jet can also have small $\tau_2$, as shown in \Fig{fig:tau2w}.  Similarly, though $W$ jets are likely to have large $\tau_1$, QCD jets with a diffuse spray of large angle radiation can also have large $\tau_1$, as shown in \Fig{fig:tau1w}.  However, those QCD jets with large $\tau_1$ typically have large values of $\tau_2$ as well, so it is in fact the \emph{ratio} $\tau_2/\tau_1$ which is the preferred discriminating variable.  As seen in \Fig{fig:tau21w}, $W$ jets have smaller $\tau_2/\tau_1$ values than QCD jets.  Of course, one can also use the full set of $\tau_N$ values in a multivariate analysis, as suggested by \Fig{fig:tau21density}, and we will briefly explore this possibility in \Sec{sec:optimization}.

As mentioned in the introduction, $N$-subjettiness is adapted from the similar quantity $N$-jettiness introduced in \Ref{Stewart:2010tn}.  There are three important differences:  the sum over $k$ only runs over the hadrons in a particular jet and not over the entire event, we do not have candidate (sub)jets corresponding to the beam directions, and our distance measure is only longitudinally boost invariant and not fully Lorentz invariant.  The definition of $\tau_N$ is by no means unique, and some variations are discussed in \App{app:Nsubdef}, though \Eq{eq:tau_N} appears to be well-suited for boosted object identification.  

\subsection{Finding Candidate Subjets}
\label{subsec:candidate_subjets}

A key step for defining $N$-subjettiness is to appropriately choose the candidate subjet directions.  As also mentioned in \Ref{Stewart:2010tn}, ideally one would determine $\tau_N$ by minimizing over all possible candidate subjet directions, analogously to how the event shape thrust is defined \cite{Farhi:1977sg}.  In that case, $\tau_N$ is a strictly decreasing function of $N$, and $0 < \tau_{N}/\tau_{N-1} < 1$.  

In practice, such a minimization step is computationally intensive, and for the studies presented in this paper, we will determine the candidate subjet directions by using the exclusive-$k_T$ clustering algorithm \cite{Catani:1993hr,Ellis:1993tq}, forcing it to return exactly $N$ jets.   This algorithmic method for finding the candidate subjet momenta is not perfect, and though not displayed in \Fig{fig:tau21w}, a small fraction ($\ll$ 1\%) of jets have $\tau_{N} > \tau_{N-1}$, a ``feature'' which would be eliminated with a proper minimization procedure.  That said, we have not found many cases where the exclusive $k_T$ jet algorithm identifies completely incorrect subjets, and the additional event displays in \App{app:additional_events} show that the exclusive $k_T$ algorithm finds subjets that look reasonable.

Once the candidate subjets are identified, $N$-subjettiness is a proper inclusive jet shape. Since \Eq{eq:tau_N} is linear in each of the constituent particle momenta, $\tau_N$ is an infrared- and collinear-safe observable.  That is, the addition of infinitesimally soft particles does not change $N$-subjettiness (infrared safety), and the linear dependence on the particle momenta combined with the smooth angular dependence ensures that the same $\tau_N$ is obtained for collinear splittings (collinear safety).  Crucially, the candidate subjets used in $N$-subjettiness must be determined via a method that is also  infrared- and collinear-safe, something that is automatic with a minimization procedure or by using $k_T$ declustering.  

\subsection{Summary}

To summarize, $N$-subjettiness is an inclusive jet shape that offers a direct measure of how well jet energy is aligned into subjets, and is therefore an excellent starting point for boosted object identification.  The ratio $\tau_N/\tau_{N-1}$ is an easily adjustable offline parameter which can be varied to adjust signal efficiency/background rejection without having to redo the clustering of the particles in an event.  While there is some residual jet algorithm dependence in the identification of the original seed jet and in the identification of the candidate subjets, this latter effect could be completely removed by using a minimization procedure at the expense of introducing more computational complexity.  Though we will not attempt to do so here, we suspect that $N$-subjettiness will lend itself better to theoretical studies than algorithmic boosted object tagging methods, either in fixed-order or resummed QCD calculations.   As we will see in the next two sections, $N$-subjettiness compares favorably to other boosted object tagging methods in terms of discriminating power.  

\begin{figure}[tp]
  \begin{center}
    \subfigure[]{\label{fig:ttbar}\includegraphics[trim = 0mm 0mm 0mm 10mm, clip, scale=0.25]{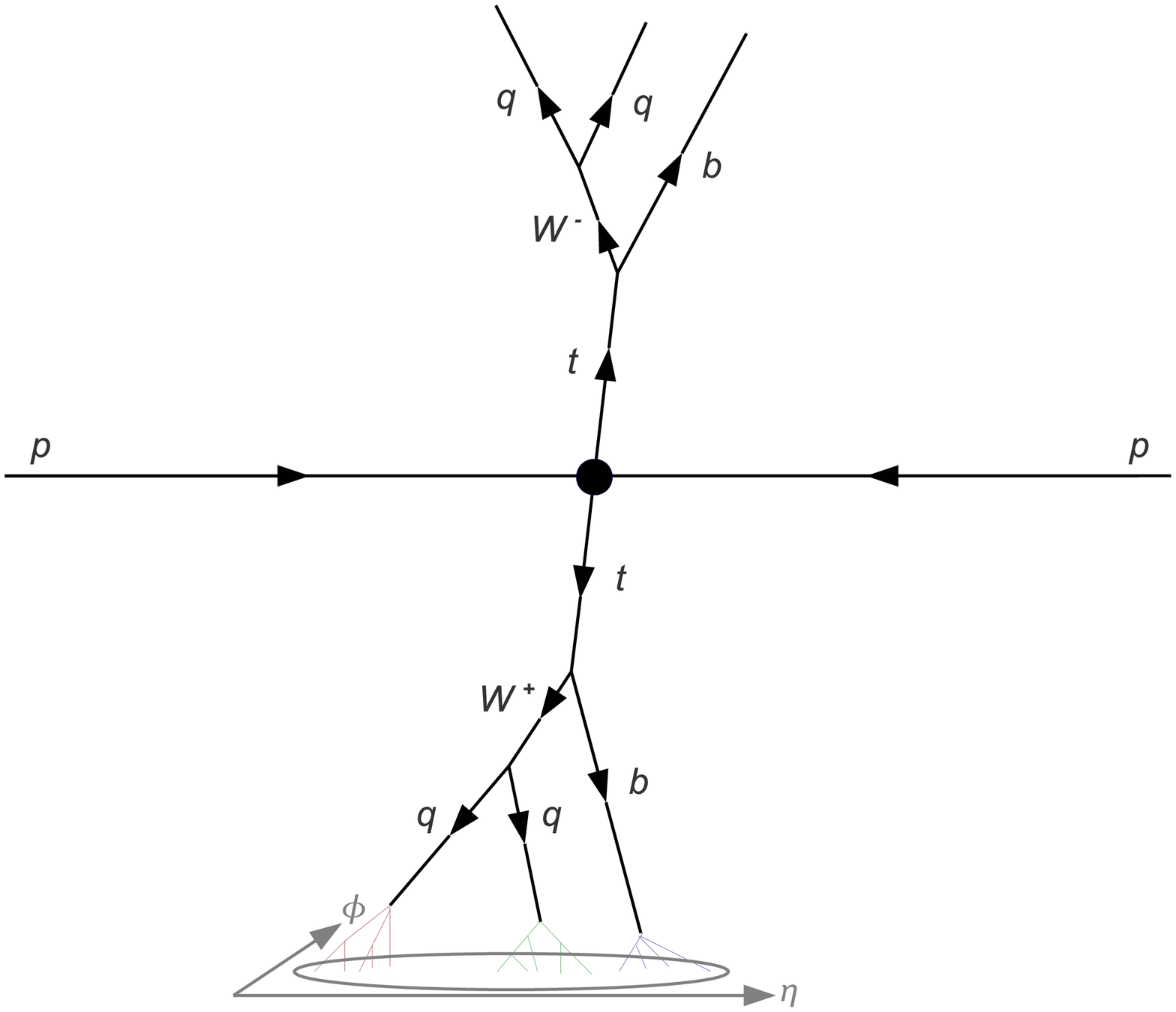}}
    \subfigure[]{\label{fig:eventDisplayTop}\includegraphics[trim = 0mm 0mm 0mm 0mm, clip, scale=0.4]{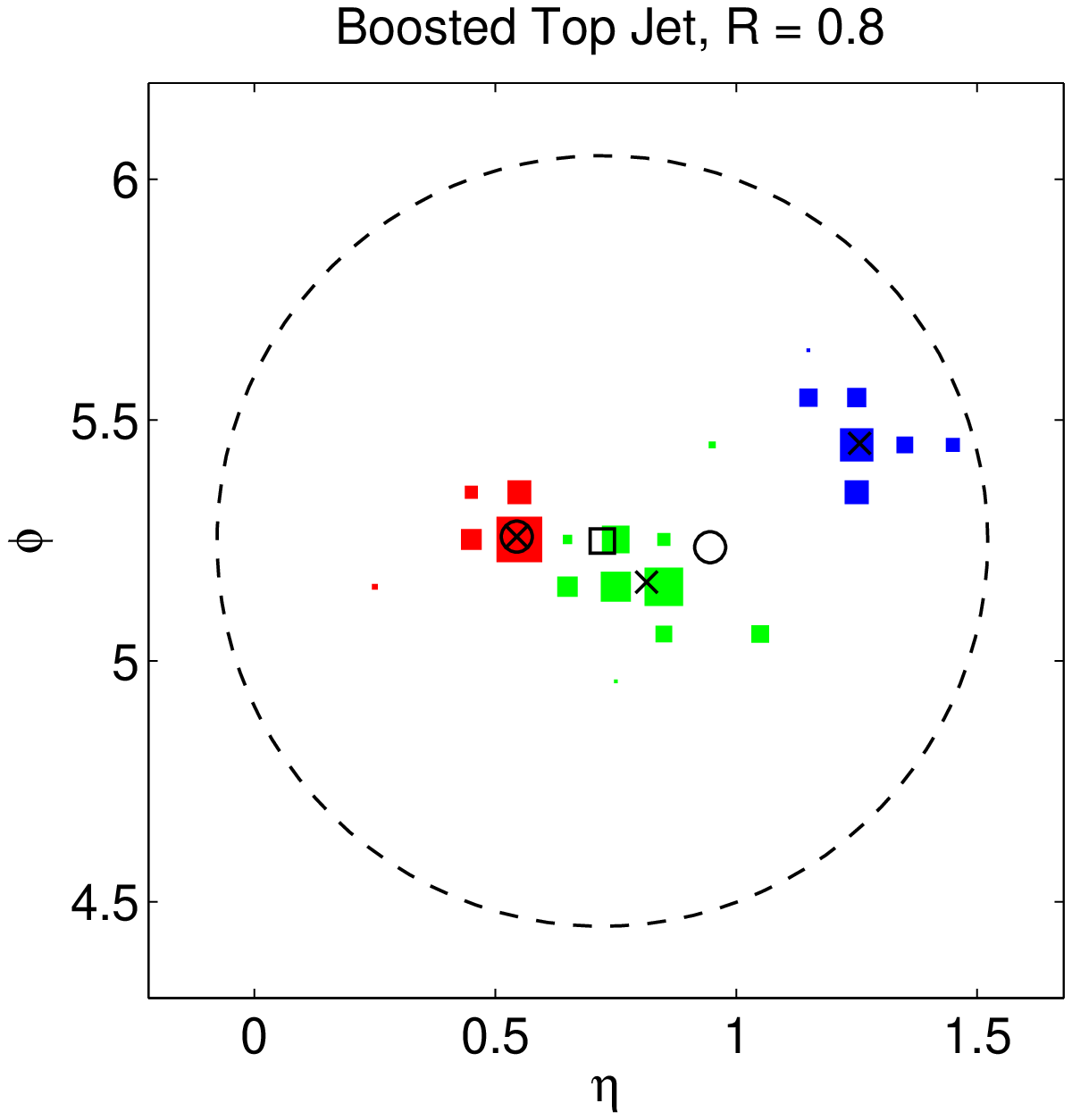}}
    \subfigure[]{\label{fig:qcd}\includegraphics[trim = 0mm 10mm 0mm 0mm, clip, scale=0.25]{figure/qcd.eps}}
    \subfigure[]{\label{fig:eventDisplayQCD}\includegraphics[trim = 0mm 0mm 0mm 0mm, clip, scale=0.4]{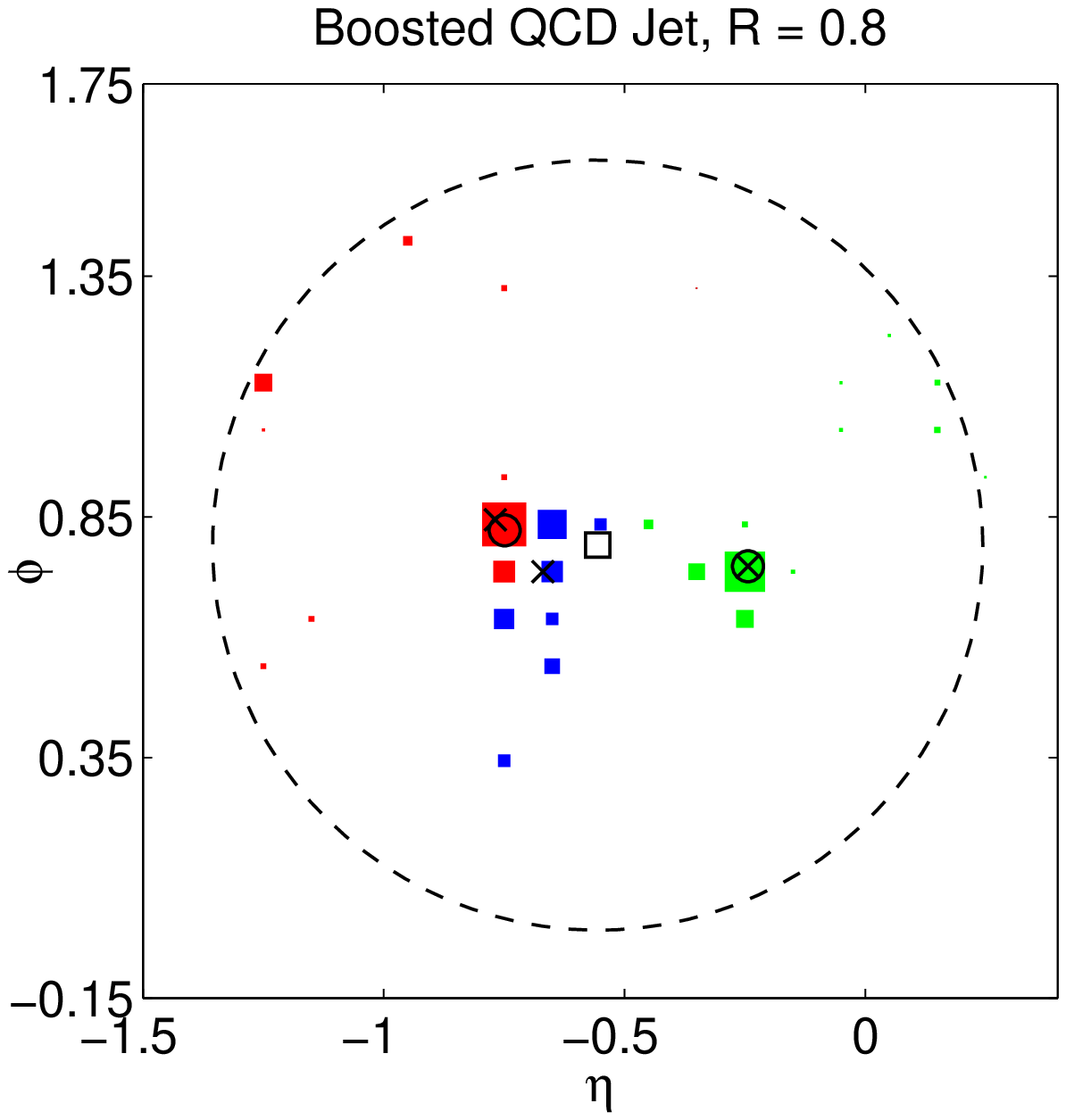}}\\
  \end{center}
  \caption{Left:  Decay sequences in (a) $t\overline{t}$ and (c) dijet QCD events.  Right:  Event displays for (b) top jets and (d) QCD jets with invariant mass near $m_{\rm top}$.  The labeling is similar to \Fig{fig:eventDisplaysW}, though here we take $R=0.8$, and the cells are colored according to how the jet is divided into three candidate subjets.  The open square indicates the total jet direction, the open circles indicate the two subjet directions, and the crosses indicate the three subjet directions.  The discriminating variable $\tau_3/\tau_2$ measures the relative alignment of the jet energy along the crosses compared to the open circles.}
  \label{fig:eventDisplaysTop}
\end{figure}

\begin{figure}[p]
  \begin{center}
    \subfigure[][]{\label{fig:tau1}\includegraphics[trim = 1mm 0mm 6mm 0mm, clip,height=4.5cm]{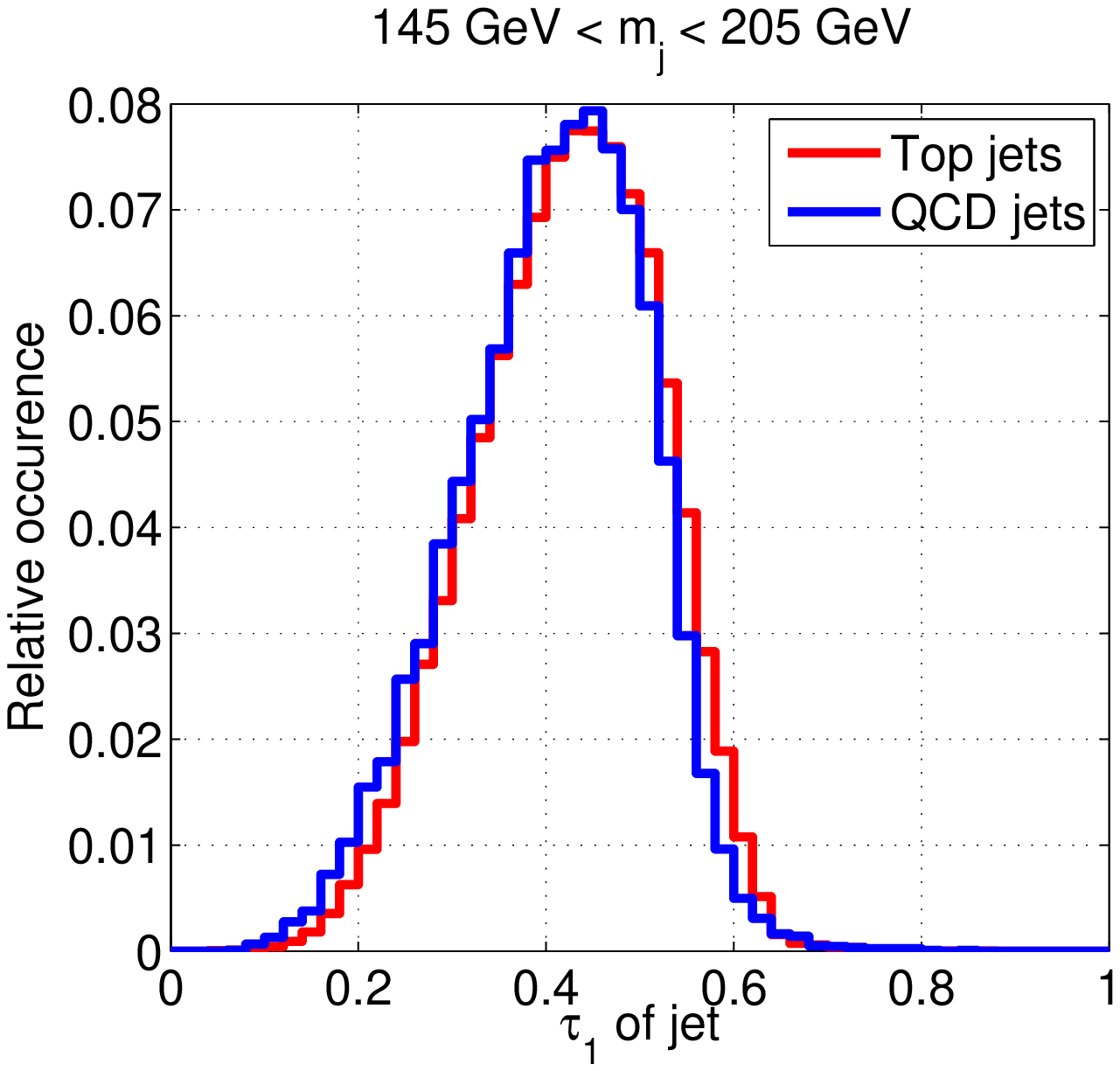}} 
    \subfigure[][]{\label{fig:tau2}\includegraphics[trim = 1mm 0mm 6mm 0mm, clip,height=4.5cm]{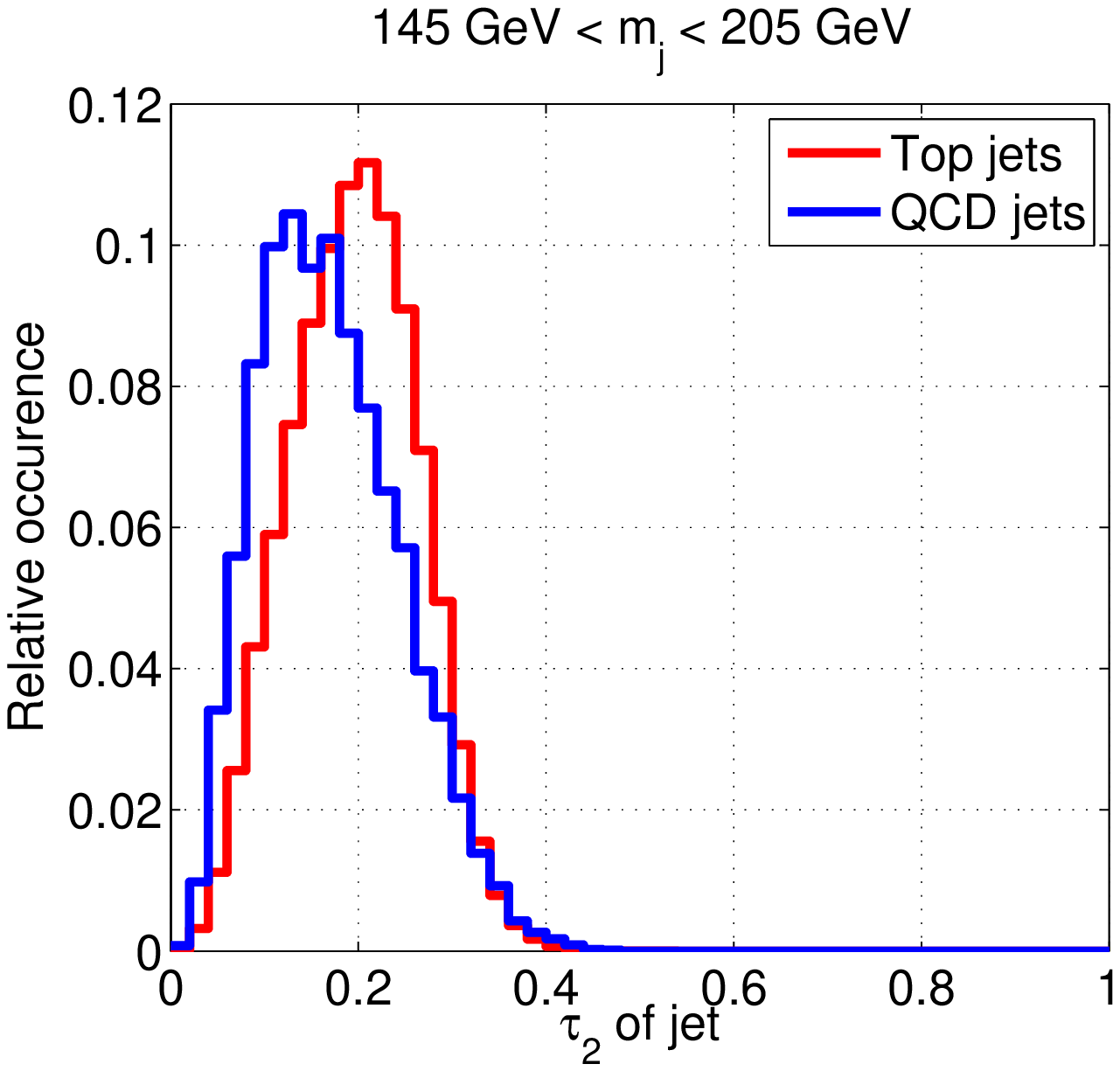}}
    \subfigure[][]{\label{fig:tau3}\includegraphics[trim = 1mm 0mm 6mm 0mm, clip,height=4.5cm]{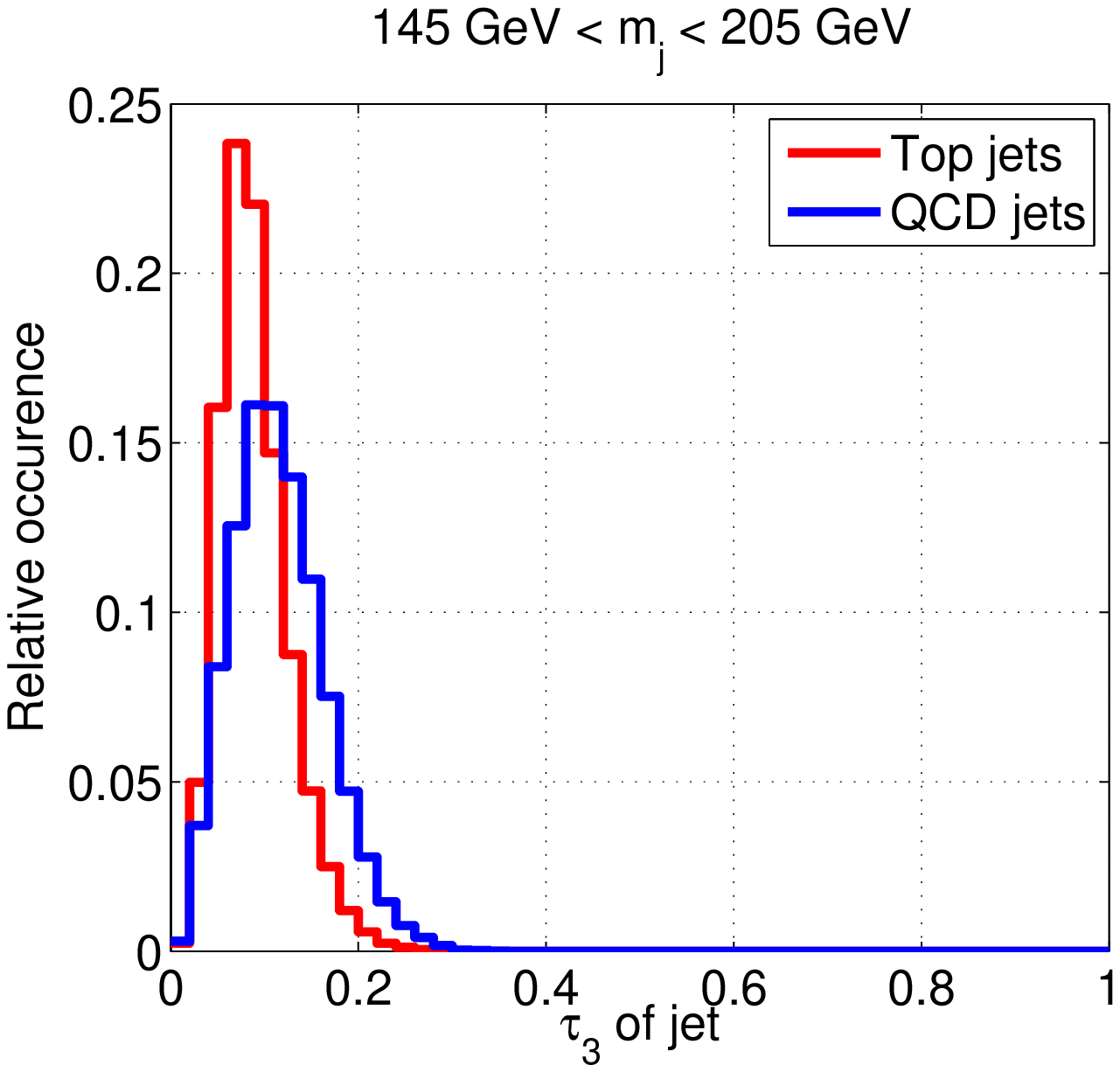}} 
  \end{center}
  \vsh
    \caption{Distributions of (a) $\tau_1$, (b) $\tau_2$ and (c) $\tau_3$ for boosted top and QCD jets.  For these plots, we impose an invariant mass window of $145 \text{ GeV} < m_{\text{jet}} < 205 \text{ GeV} $ on jets with $R = 0.8$, $p_T > 300$ GeV and $|\eta| < 1.3$.}
    \label{fig:1Dtau1tau2tau3}
\end{figure}

\begin{figure}[p]
  \begin{center}
    \subfigure[]{\label{fig:tau12ratio}\includegraphics[trim = 0mm 0mm 0mm 0mm, clip,height=4.5cm]{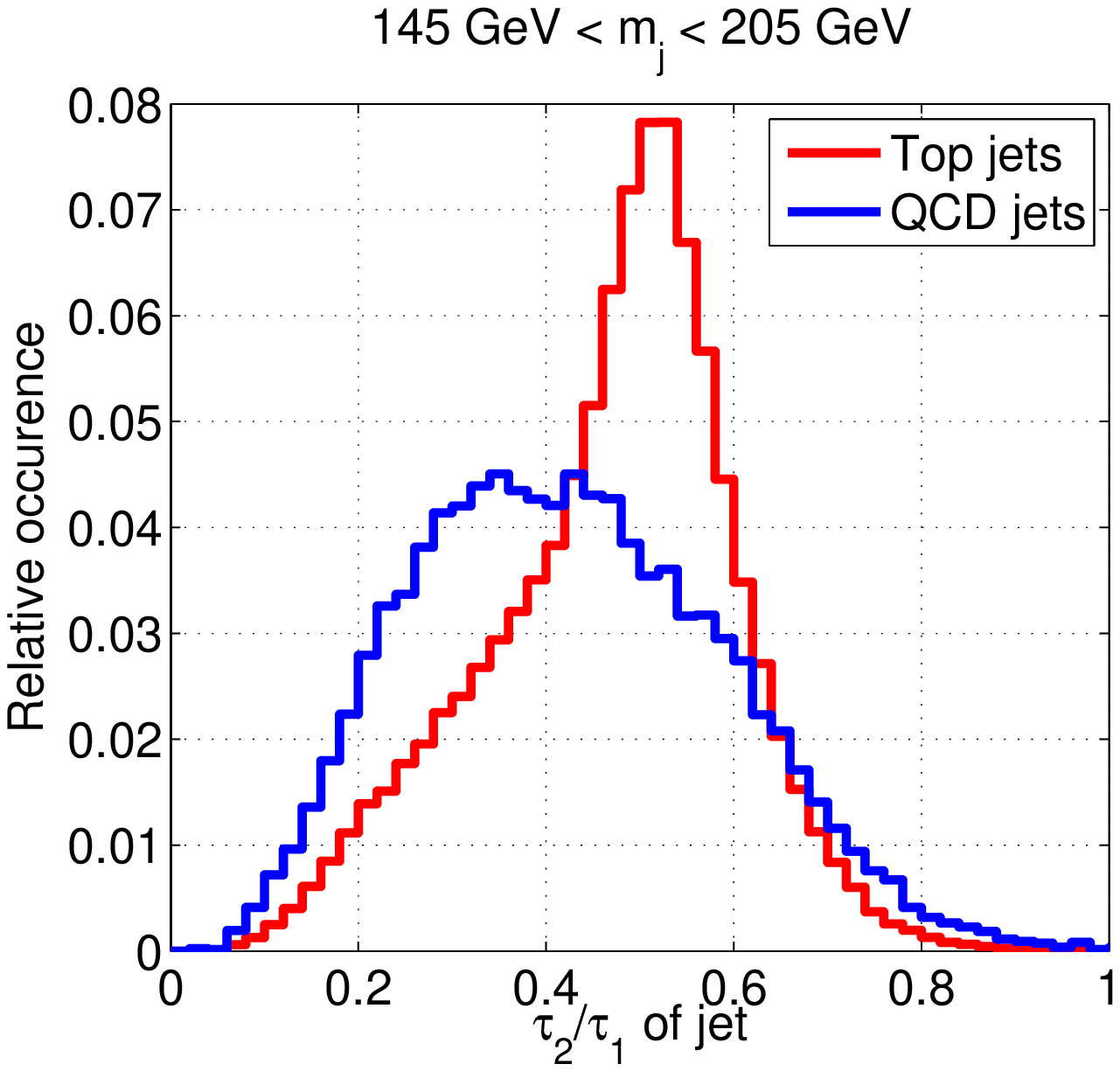}} 
    \subfigure[]{\label{fig:tau23ratio}\includegraphics[trim = 0mm 0mm 0mm 0mm, clip,height=4.5cm]{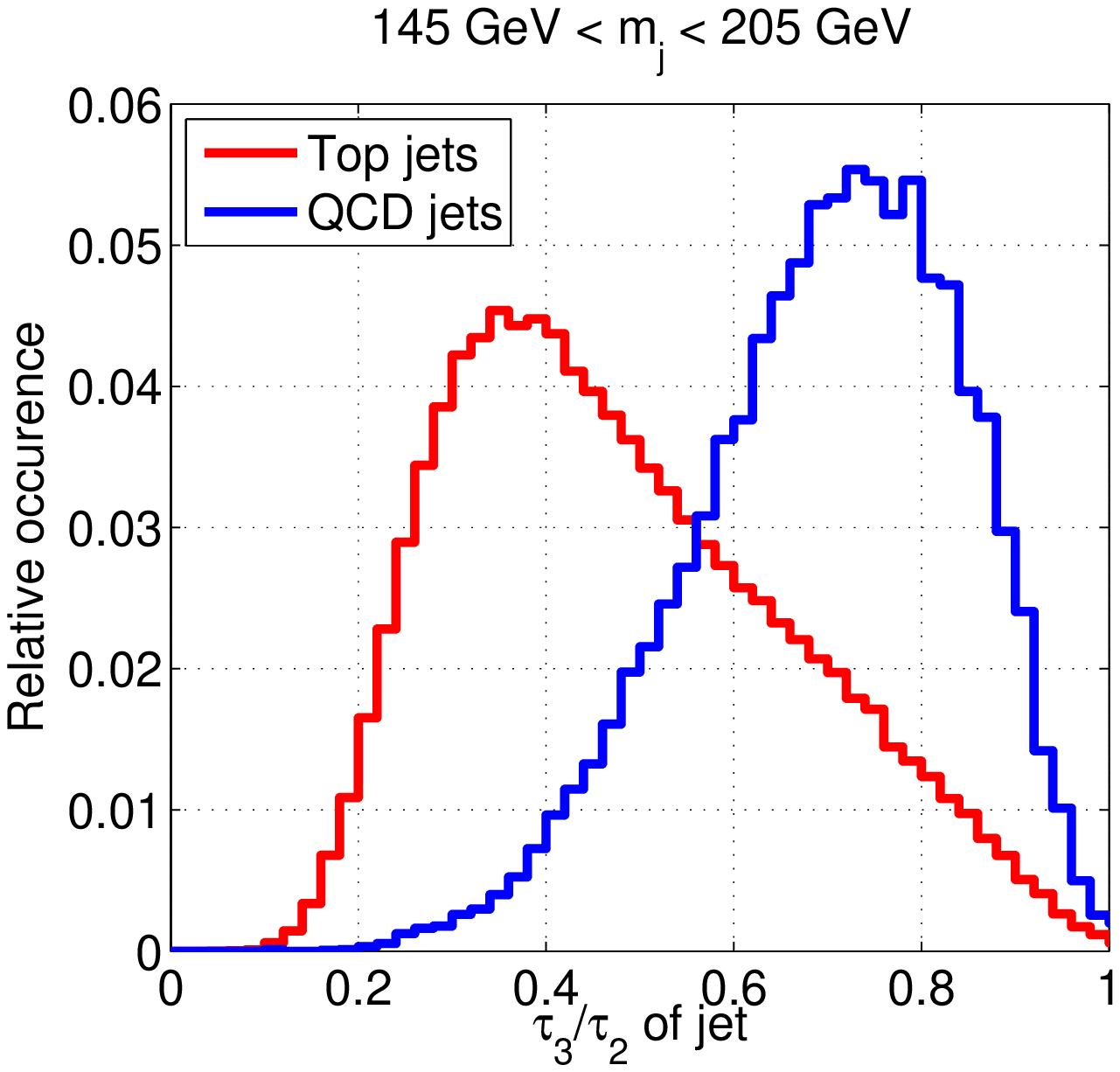}}
  \end{center}
  \vsh
    \caption{Distributions of (a) $\tau_2/\tau_1$ and (b) $\tau_3/\tau_2$ for boosted top and QCD jets.  The selection criteria are the same as in \Fig{fig:1Dtau1tau2tau3}.  We see that $\tau_3/\tau_2$ is a good discriminating variable between top jets and QCD jets.  In this paper, we do not explore $\tau_2/\tau_1$ for top jets, though it does contain additional information.}  \label{fig:1Dtau123ratios}  
\end{figure}

\begin{figure}[p]
  \begin{center}
    \subfigure[]{\label{fig:Top2Dtau12}\includegraphics[trim = 0mm 0mm 0mm 0mm, clip,height=5.0cm]{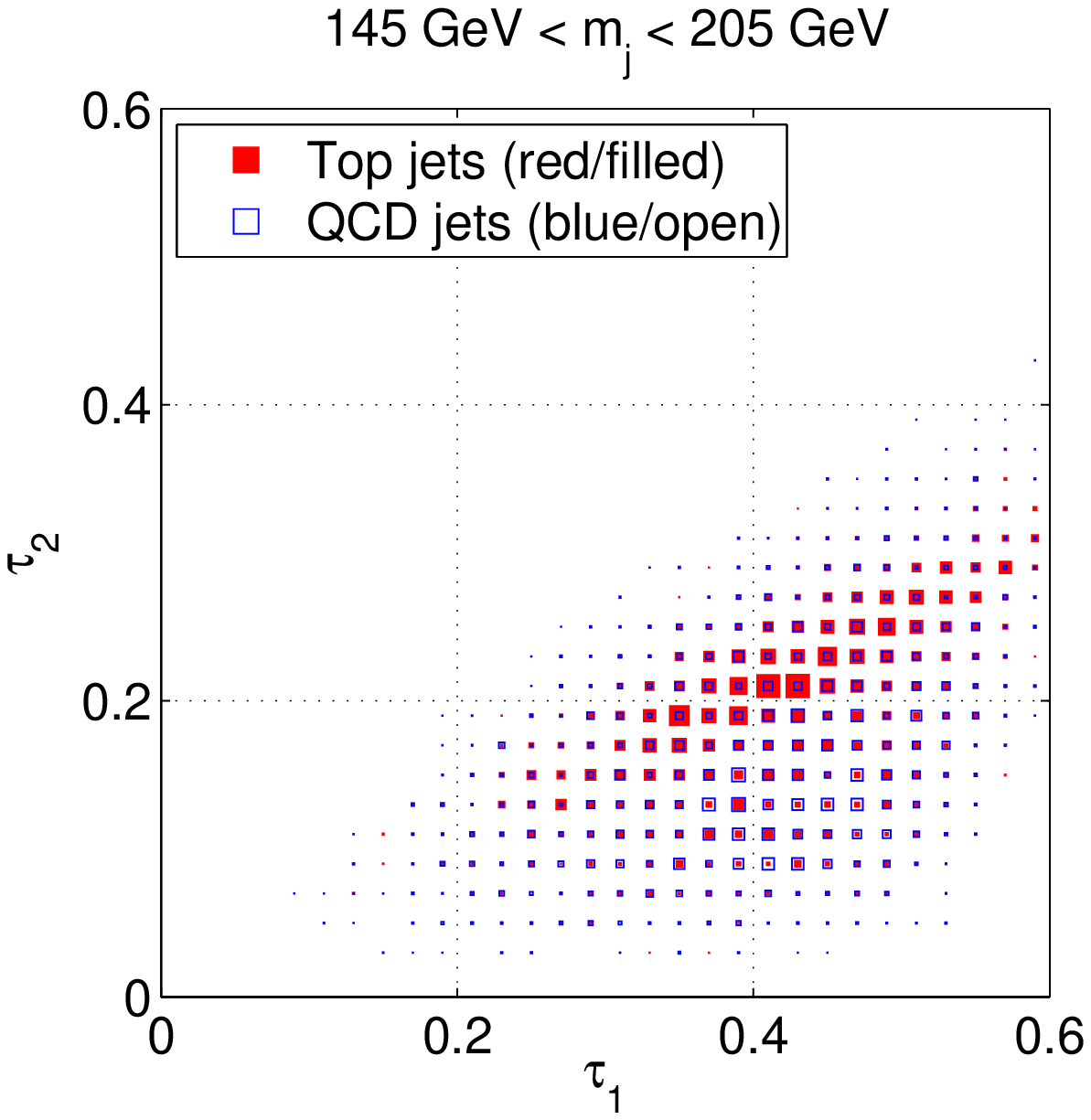}}
    \subfigure[]{\label{fig:Top2Dtau23}\includegraphics[trim = 0mm 0mm 0mm 0mm, clip,height=5.0cm]{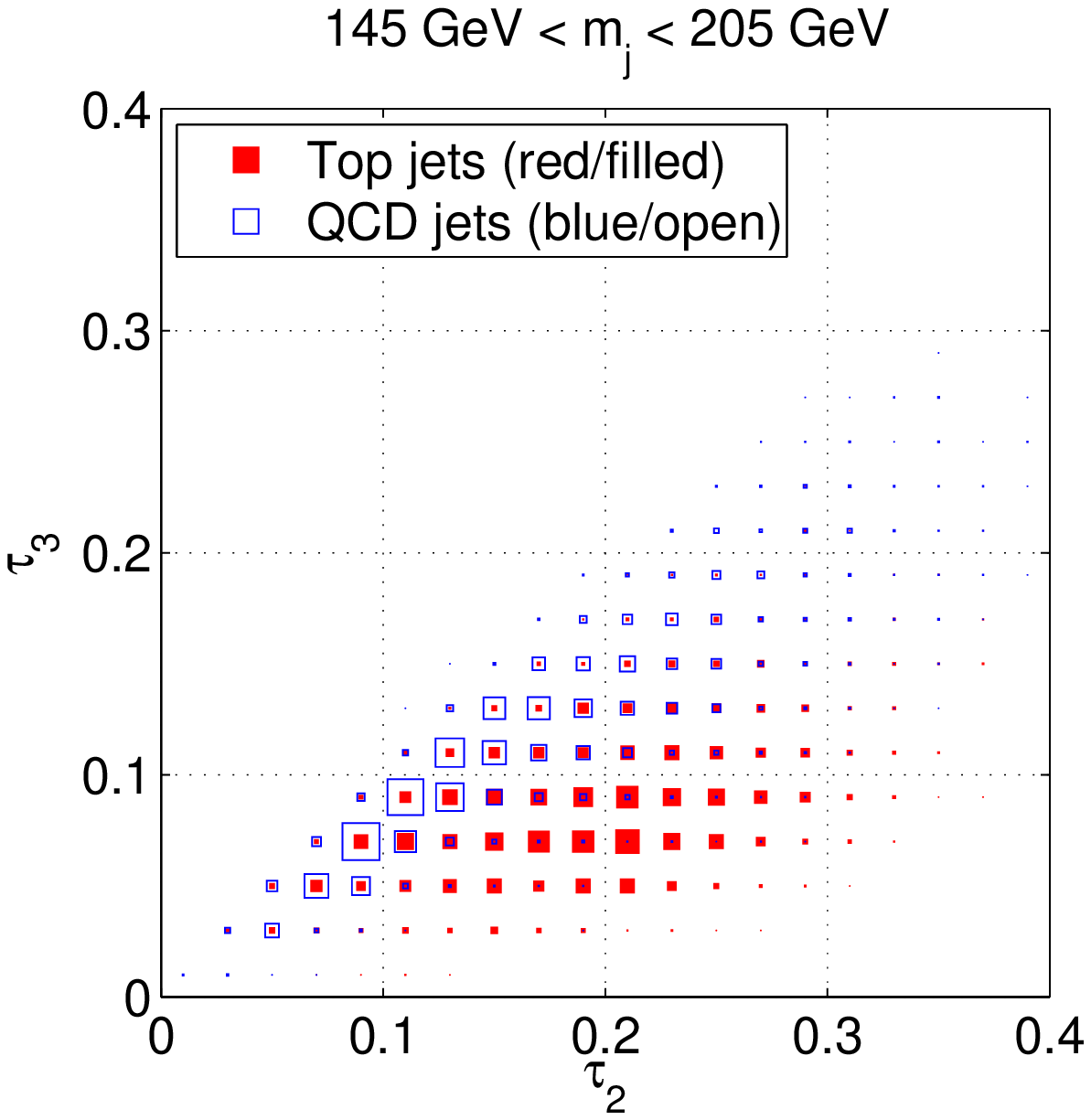}}
  \end{center}
  \vsh
  \caption{Density plots in the (a) $\tau_1$--$\tau_2$ plane and (b) $\tau_2$--$\tau_3$ plane for boosted top and QCD jets.  The selection criteria are the same as in \Fig{fig:1Dtau1tau2tau3}.  These plots suggest further improvement in boosted top identification is possible with a multivariate method.}
  \label{fig:Top2Dtau123}
\end{figure}

Finally, in the above discussion, we used boosted $W$ bosons just as an example, and a similar discussion holds for $Z$ bosons and Higgs bosons.  Also, $N$-subjettiness will be effective for identifying boosted top quarks.   A top quark with mass of 175 GeV decays to a $b$ jet and a $W$ boson, and if the $W$ boson decays hadronically into two quarks, the top jet will have three lobes of energy.  Thus, instead of $\tau_2/\tau_1$, one expects $\tau_3/\tau_2$ to be an effective discriminating variable for top jets.  This is indeed the case, as sketched in Figs.~\ref{fig:eventDisplaysTop}, \ref{fig:1Dtau1tau2tau3}, \ref{fig:1Dtau123ratios}, and \ref{fig:Top2Dtau123}.

\clearpage

\clearpage

\section{Efficiency Studies}
\label{sec:efficiency}

In this section, we investigate the tagging efficiencies for individual $W$ jets and top jets and the mistagging rates for QCD jets.  The $W$ boson case is representative of hadronically decaying $Z$ bosons and Higgs bosons, though for simplicity we only show $W$ jet results in this paper.  Here, we select candidate boosted objects using an invariant mass cut augmented with an $N$-subjettiness criterium and compare our results to the YSplitter method \cite{Butterworth:2002tt,YSplitter,Brooijmans:2008} as well as to the Johns Hopkins Top Tagger \cite{Kaplan:2008ie}.  This study will lay the groundwork for the case study in \Sec{sec:case} for reconstructing a new physics resonance which decay to pairs of gauge bosons or top quarks.  

\subsection{Analysis Overview}
\label{sec:eff_overview}

The basic criteria for tagging a boosted $W$ boson or top quark is that the jet invariant mass should fall near $m_W$ or $m_{\rm top}$, respectively.  For concreteness, we consider the mass windows of $65~\GeV < m_{\rm jet}< 95~\GeV$ for $W$ jets and $145~\GeV < m_{\rm jet} < 205~\GeV$ for top jets.  We then apply a cut on the $\tau_2 / \tau_1$ ratio (for $W$ jets) or the $\tau_3 / \tau_2$ ratio (for top jets), where the cut is adjusted to change the relative signal tagging efficiency and background mistagging rate.  

For all of our studies, we generate events for $pp$ collisions at a center-of-mass energy of $\sqrt{s} = 7$ TeV with \texttt{Pythia 8.135} \cite{Sjostrand:2006za,Sjostrand:2007gs} and perform jet clustering with \texttt{FastJet 2.4.2} \cite{FastJet,Cacciari:2005hq}.  For the QCD background of light quarks and gluons, we use the default QCD dijet production routines.   For the boosted $W$ boson signal we use standard model $W^+ W^-$ production, and for the boosted top signal we use standard model $t \overline{t}$ production.   In both signal samples, we force the $W$ bosons to decay entirely hadronically, eliminating the leptonic decays of the $W$ bosons for which our method is not applicable. Everywhere we include multiple interactions, initial- and final-state radiation (ISR/FSR), and hadron level decay, though not the effects of event pileup.  

We apply a global analysis cut to isolate boosted central jets, namely jets with transverse momentum $p_T > 300$ GeV and rapidity $|\eta| < 1.3$), and we take only the hardest jet in each event.\footnote{To reduce computing time for the simulations, we utilize a parton-level momentum phase space cut of $p_{T} > 200$ GeV on all partons studied in this paper.   This cut applies both to gluons and light quarks, as well as $W$ bosons and top quarks.  Cross-checks show that outside this kinematic regime, a negligible number of events contain jets with $p_T > 300$ GeV, so no significant bias is introduced.}  In some of our subsequent analyses, we apply harder cuts beyond the $p_T > 300$ GeV restriction, because for more moderately boosted jets, our method becomes less effective.

To partially simulate detector effects and to speed up jet reconstruction, observable final-state particles with $|\eta| < 4$ are collected into ``calorimeter'' cells arranged in a rectangular lattice with 80 rapidity ($\eta$) and 64 azimuth ($\phi$) bins (corresponding to approximately $0.1 \times 0.1$ sized cells).  The calorimeter momenta are interpreted as massless pseudo-particles with energy given by the calorimeter energy.  

The seed jets for analysis are determined using the anti-$k_T$ jet algorithm \cite{Cacciari:2008gp}, with various jet radii $R$.  To compute $\tau_N$, the seed jets are reclustered with the exclusive-$k_T$ algorithm \cite{Catani:1993hr,Ellis:1993tq} into exactly $N$ candidate subjets, and these subjets are used as input to \Eq{eq:tau_N}.  Note that \Eq{eq:tau_N} only depends on the three-momenta of the candidate subjets.  

Finally, in our study, we used the default \texttt{Pythia} showering algorithm to describe ISR/FSR and did not include possible matrix element corrections.  While a proper modeling of perturbative radiation would certainly affect the specific values of $N$-subjettiness as well as the jet invariant mass distribution, we found that the ratio $\tau_N/\tau_{N-1}$ is reasonably robust to changes in the shower model.  Thus, we suspect that using matched multi-jet samples \cite{Alwall:2007fs} would not significantly affect the tagging efficiencies of our $N$-subjettiness cuts, though a detailed study is beyond the scope of this work.

\subsection{Comparing to Other Methods}

To evaluate the performance of $N$-subjettiness, we want to compare it to previous jet substructure methods.   The most natural comparison is to the YSplitter technique \cite{Butterworth:2002tt,YSplitter,Brooijmans:2008}.  In YSplitter, a jet is declustered using the exclusive-$k_T$ jet algorithm, and the $y_{N,N+1}$ scale is the square-root of the $k_T$ distance measure at which the jet declusters from $N$ subjets to $N+1$ subjets.  Especially since we are using the  exclusive-$k_T$ jet algorithm in this paper to define $\tau_N$, one might naively think that $\tau_N/\tau_{N-1}$ and $y_{N,N+1}/y_{N-1,N}$ should have similar discriminating power.  However, we will see that $\tau_N$ does provide additional information beyond the $y_{N,N+1}$ variable.  This makes sense, as $y_{N,N+1}$ measures the ``scale'' at which subjets decompose, whereas $\tau_N/\tau_{N-1}$ measures the degree to which jet radiation is collimated around the candidate subjet directions.  In other words, $\tau_N$ is a more direct measure of how $N$-subjetty a jet really is.  

While a detailed comparison of $N$-subjettiness and YSplitter is beyond the scope of this work, we find that $N$-subjettiness does compare favorably to YSplitter in a simple, non-optimized test.  For this naive comparison with the YSplitter method, we use the same anti-${k_T}$ jet clustering parameters and the same invariant mass windows for $W$ jets and top jets.  We then place cuts on the ratio $y_{23} / y_{12}$ (for $W$ jets) and $y_{34} / y_{23}$ (for top jets).  This is certainly not an ideal use of the YSplitter variables, and we consider potential optimizations in \Sec{sec:optimization}.  Plots of various $y_{N,N+1}$ distributions appear in \App{app:ysplitter}.

Finally, for top jets, we also do one benchmark comparison to the Johns Hopkins Top Tagging (JHTT) method, using settings similar to those in \Ref{Kaplan:2008ie}.  Here, jet clustering is performed with the Cambridge-Aachen algorithm \cite{Dokshitzer:1997in,Wobisch:1998wt} combined with the \texttt{FastJet} JHTT plugin \cite{FastJet} with a jet radius of $R = 0.8$, $\delta_p = 0.10$, and $\delta_r = 0.19$.  As above, we only consider the hardest jet in each event, provided it has $p_T > 300$ GeV and $|\eta| < 1.3$.  We impose the top mass window ($145 \text{ GeV} < m_{\text{jet}} < 205 \text{ GeV}$) on the entire jet and also demand that any two subjets in the JHTT declustering sequence have an invariant mass near $m_W$ (65 GeV to 95 GeV).  Finally, we selected on the $W$ helicity angle by requiring $\cos\theta_h < 0.7$.  Again, this is not an optimal comparison, but serves as a useful point of reference to gauge the effectiveness of $N$-subjettiness.

\clearpage

\subsection{Boosted $W$ and Top Results}
\label{sec:efficiencyt}

\begin{figure}[tp]
  \begin{center}
    \subfigure[]{\label{fig:invMW}\includegraphics[trim =    10mm 0mm 15mm 0mm, clip, height=5.5cm]{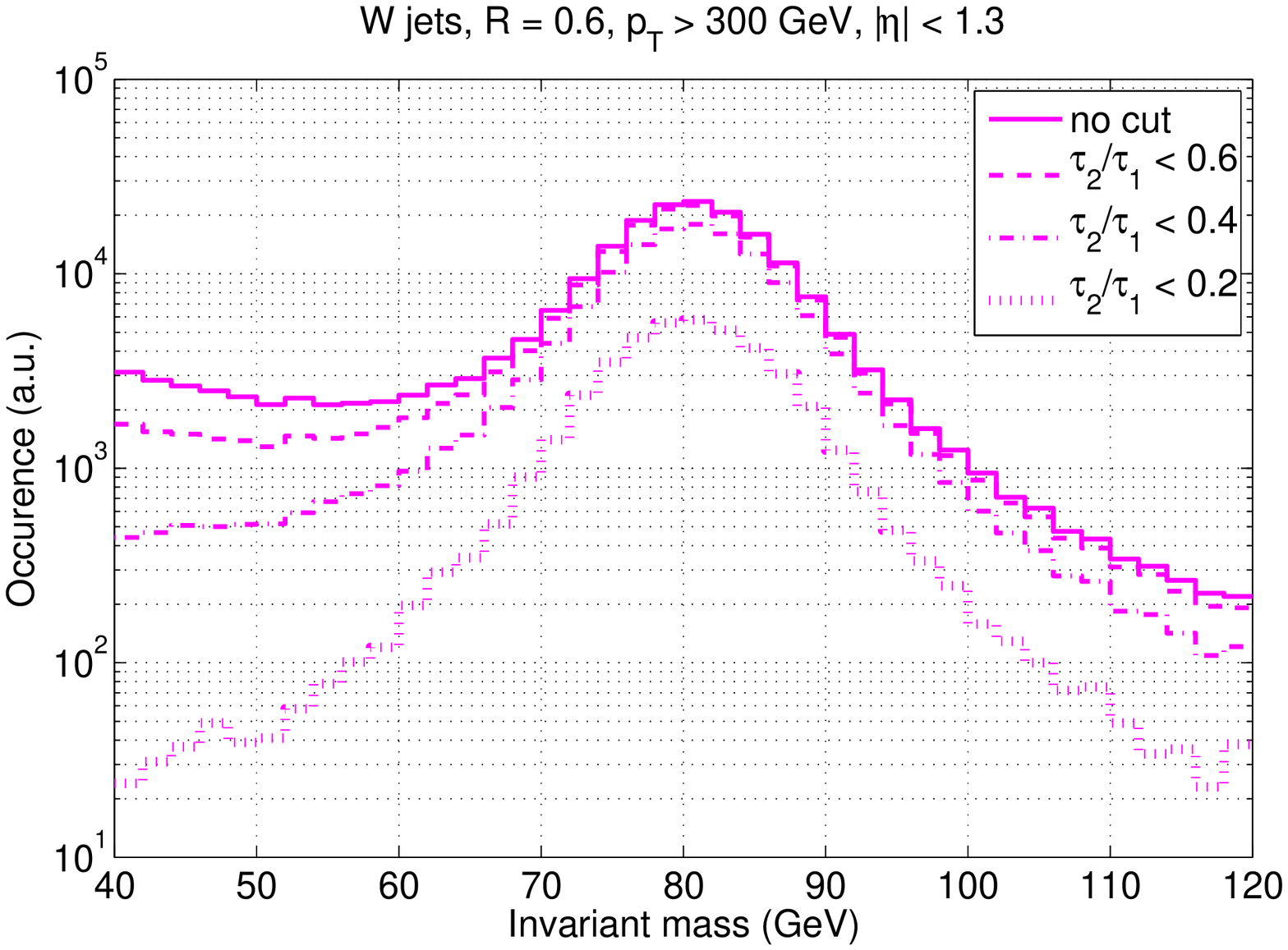}}   
    \subfigure[]{\label{fig:invMQCDW}\includegraphics[trim = 10mm 0mm 15mm 0mm, clip, height=5.5cm]{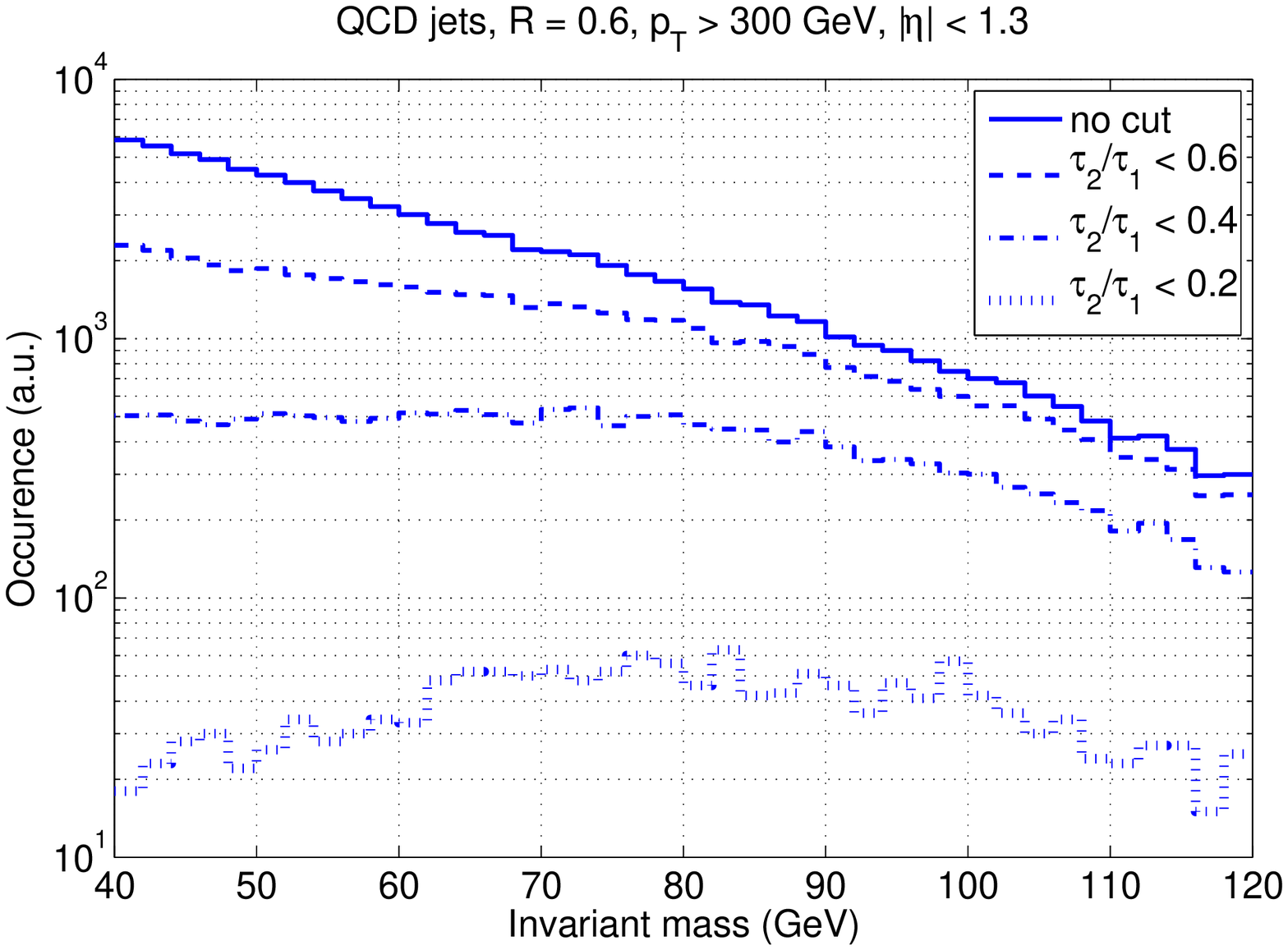}}     
  \end{center}
  \vsh
  \caption{Invariant mass distributions for (a) $W$ jets and (b) QCD jets as the $N$-subjettiness cut on $\tau_2/\tau_1$ is varied.   Here, the jet radius is $R=0.6$ and the jets satisfy $p_T > 300 \text{ GeV}$ and $|\eta| < 1.3$.  Since the QCD jet rate decreases faster than the $W$ jet rate as the $\tau_2/\tau_1$ cut is tightened, $\tau_2/\tau_1$ is an effective discriminating variable.}
  \label{fig:WInvM}
\end{figure}

We are now ready to investigate the efficiency of the $N$-subjettiness method applied to $W$ jets and top jets.  We start with $W$ jets.  In \Fig{fig:WInvM}, the invariant mass distributions for boosted $W$ jets and QCD jets are shown for boosted central jets with $R = 0.6$, $p_T > 300 \text{ GeV}$ and $|\eta| < 1.3$.  As the cut on $\tau_2/\tau_1$ is tightened, the boosted $W$ jet signal experiences a modest decrease, but the peak of the distribution stays well-centered on the $W$ mass.  The QCD background decreases by a much greater degree, with no significant sculpting of the distribution.  Thus we confirm that $\tau_2/\tau_1$ is effective as a $W$ jet discriminating variable.

\begin{table}[t]
  \begin{center}
    \begin{tabular}{ r |r@{ : }lr@{ : }lr@{ : }lr@{ : }lr@{ : }l}
     \hline
     \hline
     $p_T$ range (GeV) & \multicolumn{2}{c}{300--400} & \multicolumn{2}{c}{400--500} & \multicolumn{2}{c}{500--600} & \multicolumn{2}{c}{600--700} & \multicolumn{2}{c}{700--800} \\ 
     \hline
     No $\tau_2/\tau_1$ cut & .62 & .14 & .72 & .19 & .73 & .21 & .71 & .23 & .69 & .25 \\
     $\tau_2/\tau_1 < 0.5$ & .56 & .072 & .61 & .077 & .59 & .077 & .55 & .084 & .51 & .085 \\
     $\tau_2/\tau_1 < 0.3$ & .36 & .019 & .35 & .020 & .33 & .020 & .31 & .019 & .30 & .024 \\
     $\tau_2/\tau_1 < 0.2$ & .16 & .0044 & .16 & .0056 & .16 & .0052 & .15 & .0036 & .16 & .0034 \\
     \hline
     1\% mistag rate & .26 & .010 & .24 & .010 & .23 & .010 & .24 & .010 & .26 & .010 \\
     40\% tag efficiency & .40 & .025 & .40 & .025 & .40 & .028 & .40 & .036 & .40 & .045 \\
     \hline
     \hline
    \end{tabular}
    \caption{Tagging efficiencies vs. mistagging rates for $W$ jets : QCD jets with $R= 0.6$.  The top row corresponds to just applying the $m_W$ invariant mass window (65 GeV to 95 GeV) criteria, and the subsequent rows include an additional $\tau_2/\tau_1$ cut.  The bottom two rows indicate the tagging efficiencies achievable with  a fixed mistagging rate of 1\%, and the mistagging rate achievable with a fixed tagging efficiency of 40\%.}
      \label{tab:WSigEff06}
  \end{center}
\end{table}

\begin{figure}[tp]
  \begin{center}
    \subfigure[]{\label{fig:WSigEffa}\includegraphics[trim = 2mm 0mm 10mm 0mm, clip, height=4.5cm]{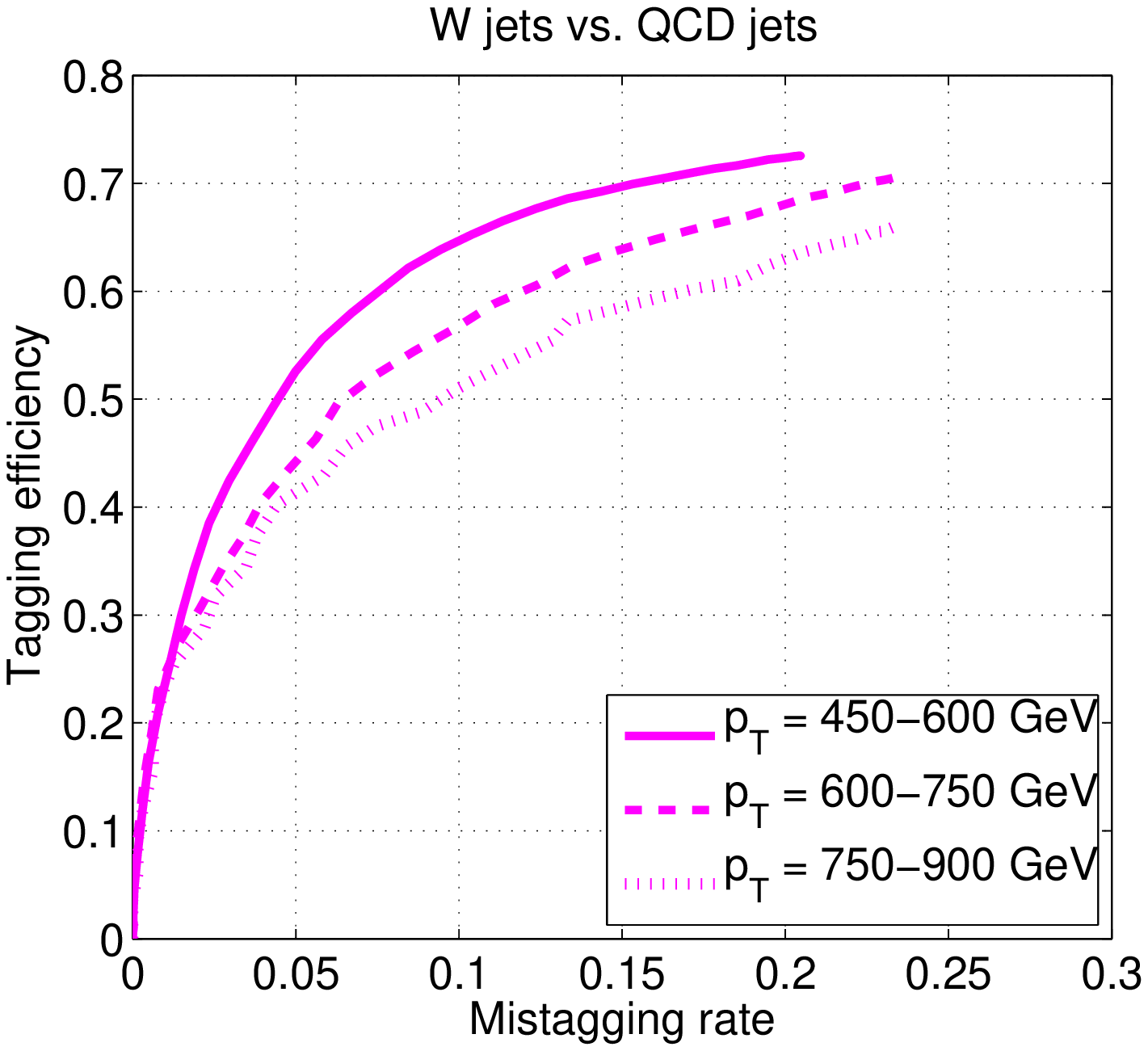}}
    \subfigure[]{\label{fig:WSigEffb}\includegraphics[trim = 2mm 0mm 10mm 0mm, clip, height=4.5cm]{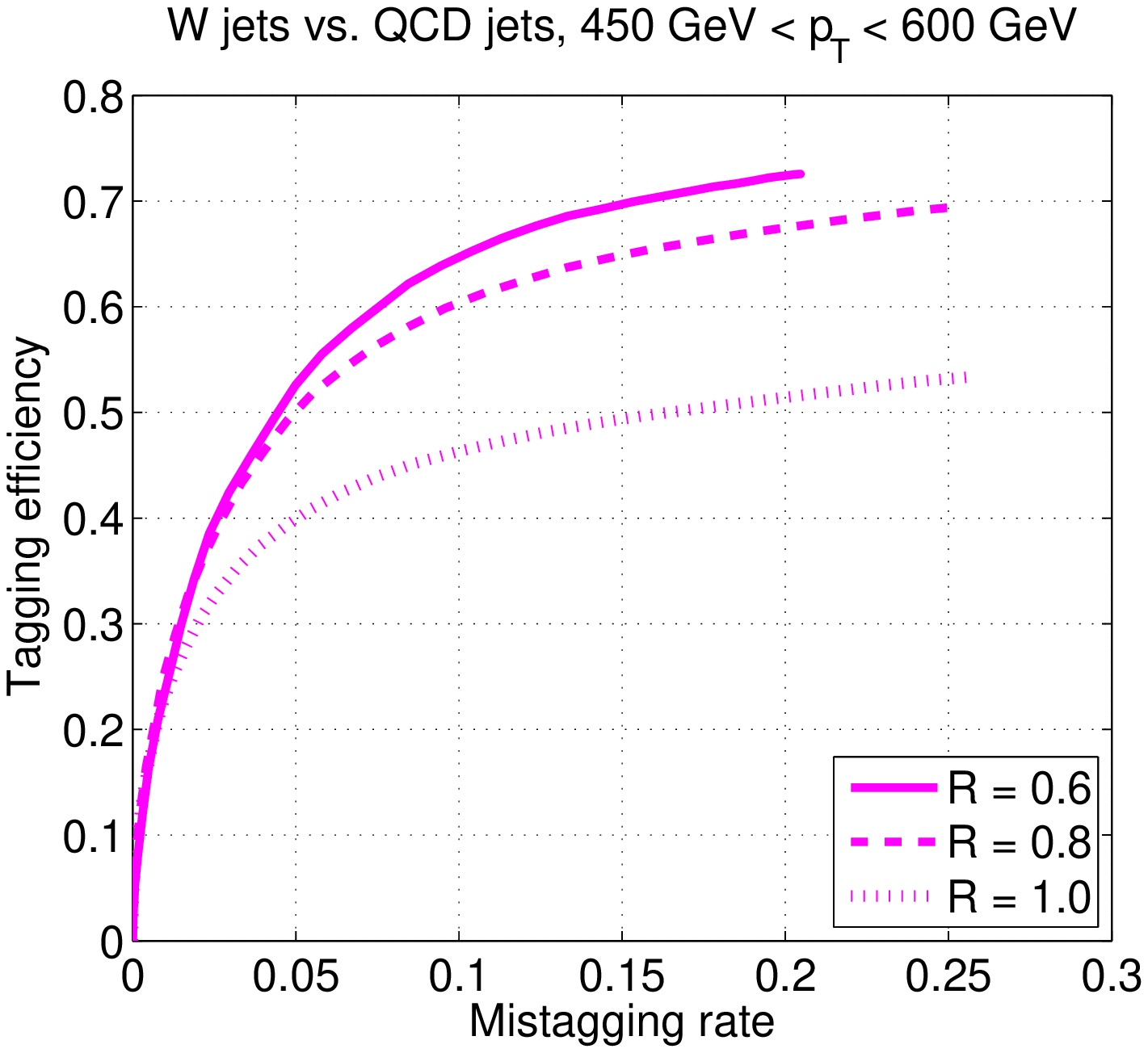}}   
    \subfigure[]{\label{fig:WSigEffc}\includegraphics[trim = 2mm 0mm 10mm 0mm, clip, height=4.5cm]{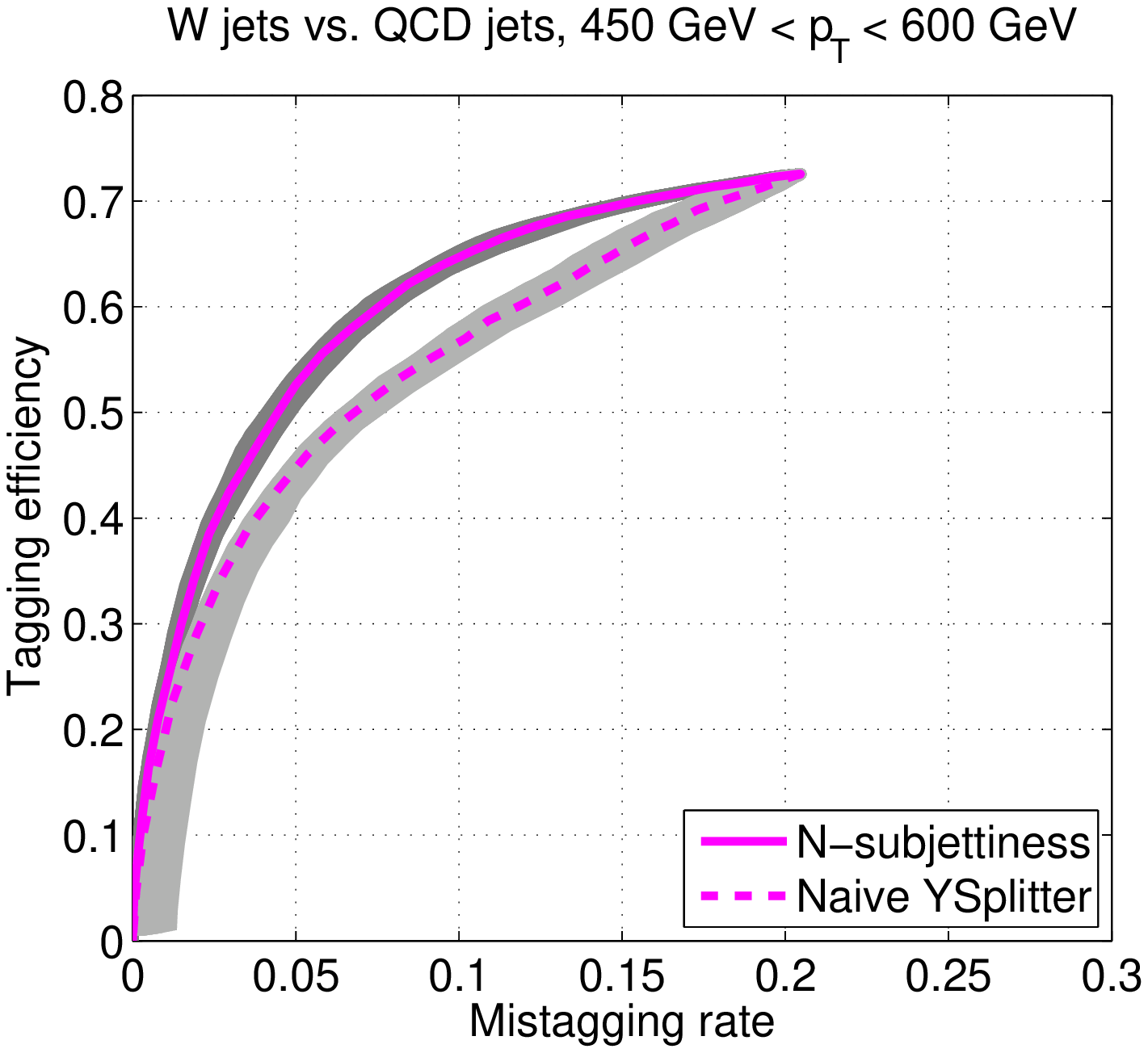}}
  \end{center}
  \vsh
  \caption{$W$ jet signal efficiency/background rejection plots.  Here, the $\tau_2/\tau_1$ cut is varied for different (a) jet transverse momenta and (b) jet radii.  Unless otherwise indicated, the jets have $R= 0.6$ and $450 \text{ GeV} < p_T < 600 \text{ GeV}$.  The rightmost points in each plot are for the $m_W$ invariant mass window criterium alone, and points to the left of these are obtained with additional cuts on the $\tau_2/\tau_1$ ratio. Figure (c) shows a naive comparison against the YSplitter method, where the purple lines indicate varying cuts on $\tau_2/\tau_1$ and $y_{23}/y_{12}$, respectively, while the shaded bands indicate the modified linear cuts of \Sec{sec:optimization}.  In this non-optimized test, $N$-subjettiness compares favorably to YSplitter.}
  \label{fig:WSigEff}
\end{figure}

\begin{figure}[tp]
  \begin{center}
      \subfigure[]{\label{fig:WSigImprova}\includegraphics[trim = 0mm 0mm 10mm 0mm, clip, height=4.5cm]{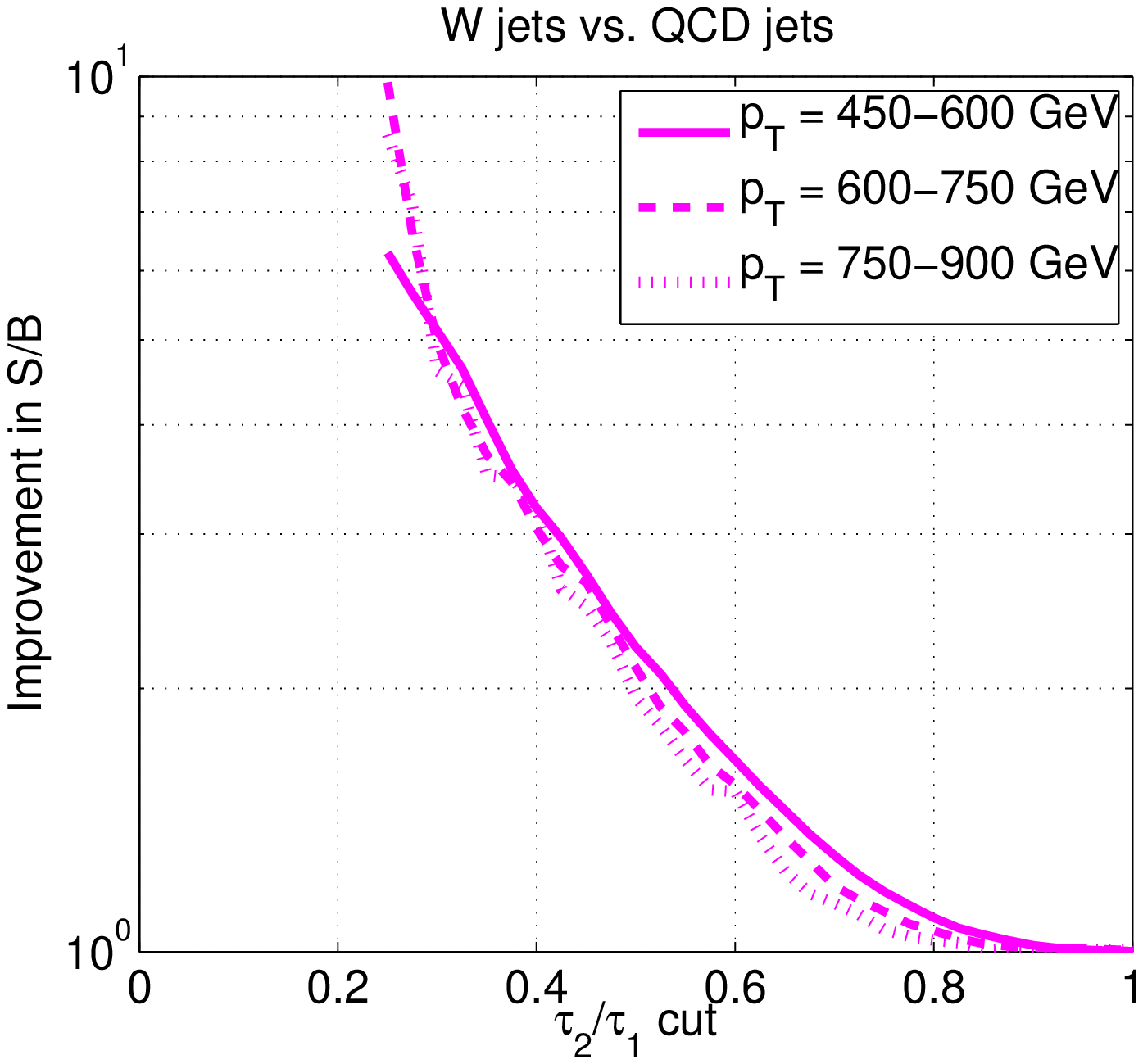}}
      \subfigure[]{\label{fig:WSigImprovb}\includegraphics[trim = 0mm 0mm 10mm 0mm, clip, height=4.5cm]{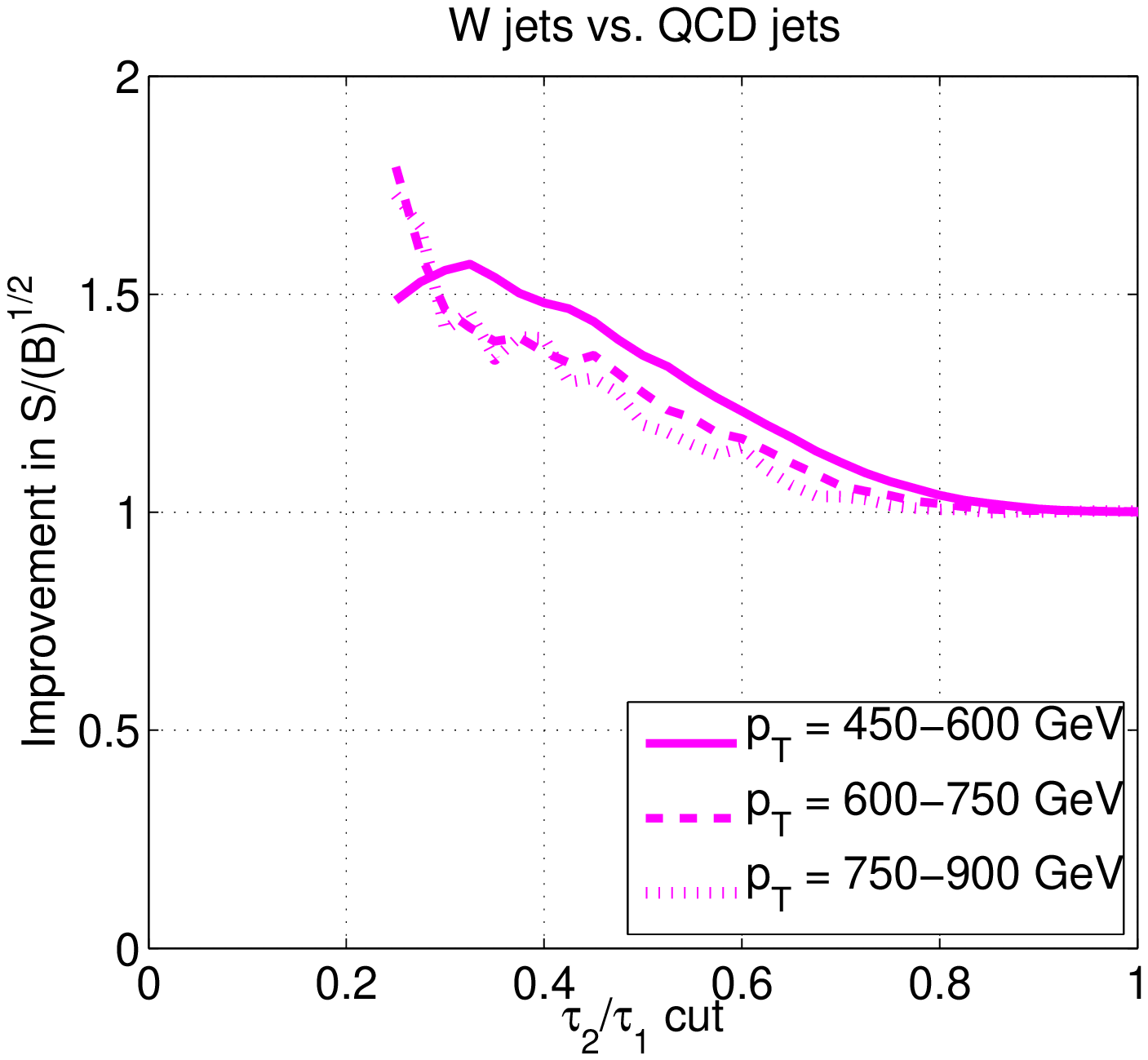}}\\
  \end{center}
  \vsh
  \caption{Improvement compared to an invariant mass cut alone for $W$ jets in (a) signal over background ($S/B$) and (b) signal over square-root background ($S/\sqrt{B}$) by using the $\tau_2/\tau_1$ cut. The value of unity on the right corresponds to using only the $m_W$ invariant mass window.}
  \label{fig:WSigImprov}
\end{figure}

In \Tab{tab:WSigEff06}, we show the quantitative effect of various $\tau_2/\tau_1$ cuts for different jet $p_T$ bins.  First, note that the $m_W$ invariant mass window ($65~\GeV < m_{\rm jet} < 95~\GeV$) already acts as a useful discriminating variable, with a tagging efficiency of around 70\% for  mistagging rates between 14\% and 25\% across the whole kinematic range.  (The higher mistagging rate for more boosted jets occurs because the invariant mass of QCD jets increases with their transverse momenta.)  By including a modest cut on $\tau_2/\tau_1$, we can maintain nearly the same signal efficiency while halving the mistagging rate.  By tightening the $\tau_2/\tau_1$ cut, we can control the degree of signal efficiency and background rejection, and across the whole $p_T$ range, a tagging efficiency of 25\% can be achieved for a mistagging rate of only 1\%.

In \Fig{fig:WSigEff}, we plot tagging efficiency curves.  These show the effects of the $\tau_2/\tau_1$ cut in different kinematical regimes and for different jet radii.   In \Fig{fig:WSigEffb}, the signal efficiency is lower for $R = 1.0$ as compared to $R = 0.8$, because a larger fraction of $W$ jets fall outside of the mass window due to fact that more ISR is captured in jets with larger radii.  As a result, the signal efficiency using only a mass cut is significantly reduced.  Beyond this mass cut, however, our method is relatively insensitive to changes in jet radius, as the slope of the efficiency curves does not change considerably with changes in $R$.  In \Fig{fig:WSigEffc}, we compare a $\tau_2/\tau_1$ cut to a $y_{23} / y_{12}$ cut, which shows that an $N$-subjettiness cut compares favorably to a naive YSplitter cut.  In \Fig{fig:WSigImprov}, we show how the signal to background ratio ($S/B$) and signal to square-root background ratio ($S/\sqrt{B}$) improve as the $\tau_2/\tau_1$ cut is tightened.  Compared to only having an invariant mass cut, the $S/B$ ratio can improve by as much as an order of magnitude with an improvement of $S/\sqrt{B}$ of around 50\%.  These improvements will be relevant for the resonance study in \Sec{sec:case}.

\begin{figure}[tp]
  \begin{center}
    \subfigure[]{\label{fig:invMTop}\includegraphics[trim = 10mm 0mm 15mm 0mm, clip, height=5.5cm]{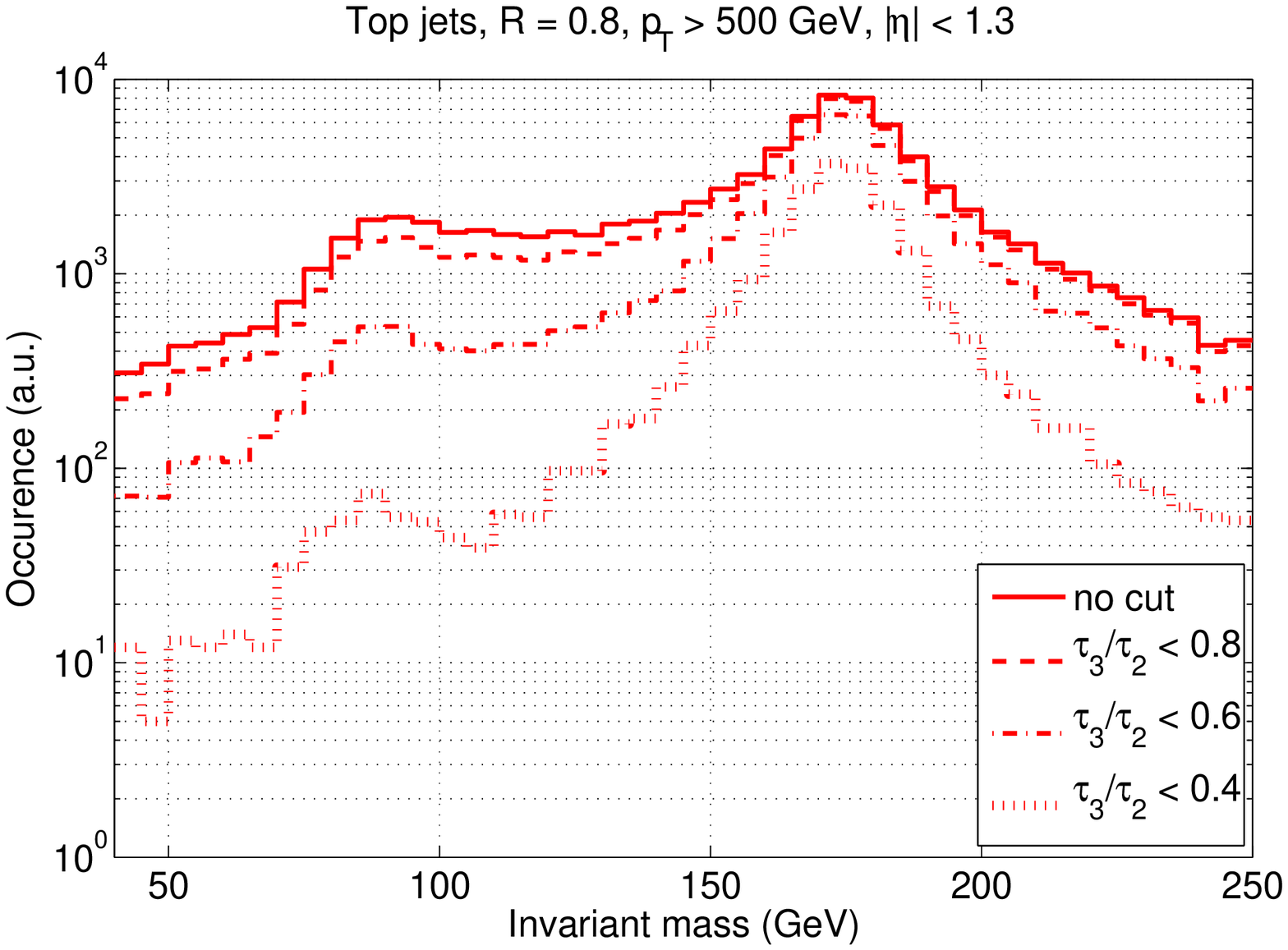}}
    \subfigure[]{\label{fig:invMQCD}\includegraphics[trim = 10mm 0mm 15mm 0mm, clip, height=5.5cm]{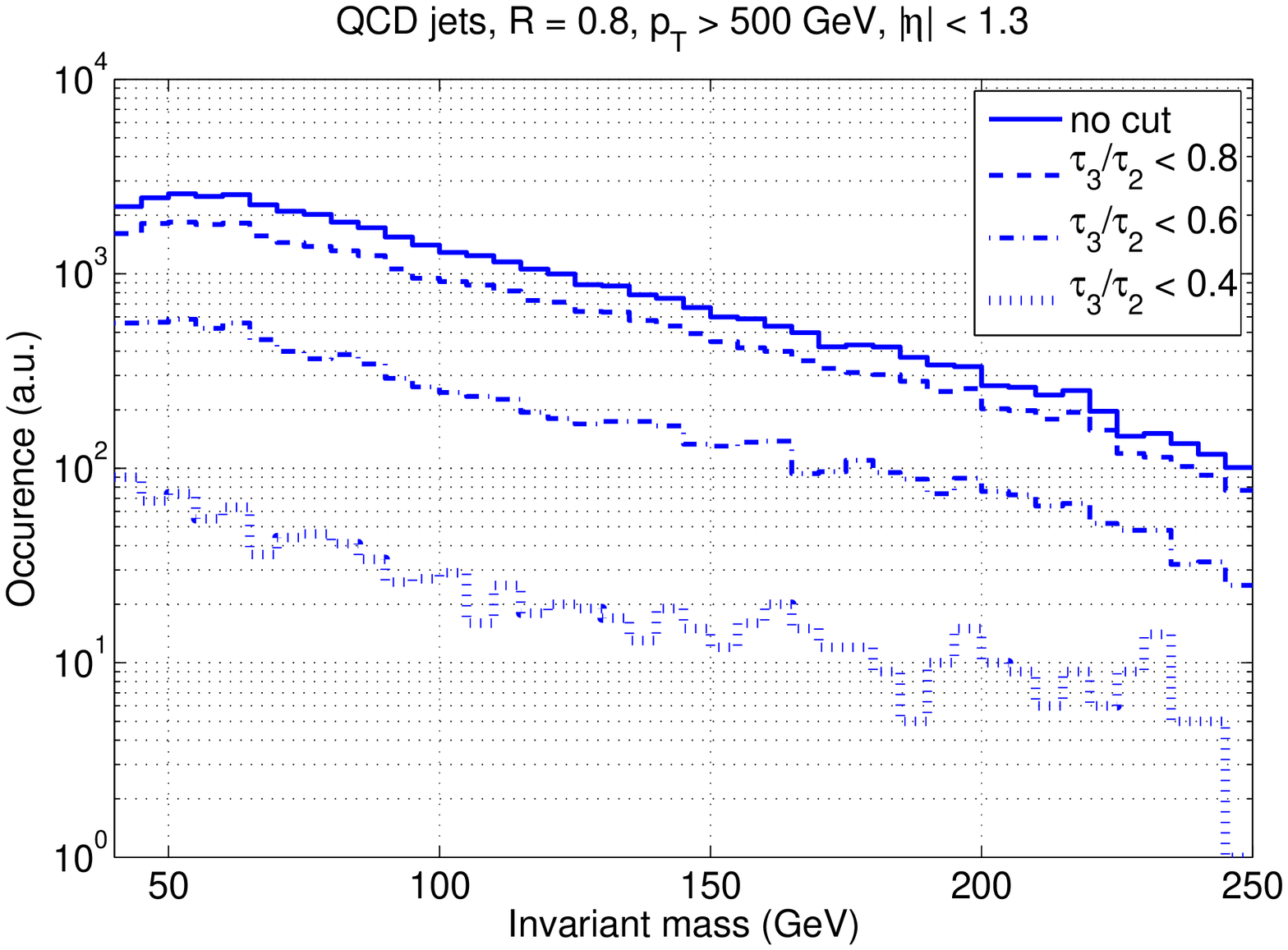}} 
  \end{center}    
    \vsh
  \caption{Invariant mass distributions for (a) top jets and (b) QCD jets as the $N$-subjettiness cut on $\tau_3/\tau_2$ is varied.   Here, the jet radius is $R=0.8$ and the jets satisfy $p_T > 500 \text{ GeV}$ and $|\eta| < 1.3$.  Since the QCD jet rate decreases faster than the top jet rate as the $\tau_3/\tau_2$ cut is tightened, $\tau_3/\tau_2$ is an effective discriminating variable.  Note also that the secondary peak at $m_W$ decreases when the $\tau_3/\tau_2$ cut is applied.}
  \label{fig:TopInvM}
\end{figure}

We now consider the analogous tables and plots for top jets, where we use a jet radius of $R = 0.8$, the $m_{\rm top}$ invariant mass window ($145~\GeV < m_{\rm jet} < 205~\GeV$), and the discriminating variable $\tau_3/\tau_2$. In \Fig{fig:TopInvM}, we show the invariant mass distributions for top jets and QCD jets, considering $p_T > 500 \text{ GeV}$ for both.  Note that the $\tau_3/\tau_2$ cut decreases the background rate faster than the signal rate.  In addition to the top mass peak, there is a secondary peak at $m_W$ which gets less prominent as the $\tau_3/\tau_2$ cut is tightened.   This secondary mass peak would be even higher if we loosened the transverse momentum cut to $p_T > 300 \text{ GeV}$, because for moderately boosted top quarks, it is less likely that a single jet could capture all three of the subjets coming from the top quark decay.  

\begin{figure}[tp]
  \begin{center}
    \subfigure[]{\label{fig:TopSigEffa}\includegraphics[trim = 2mm 0mm 10mm 0mm, clip, height=4.5cm]{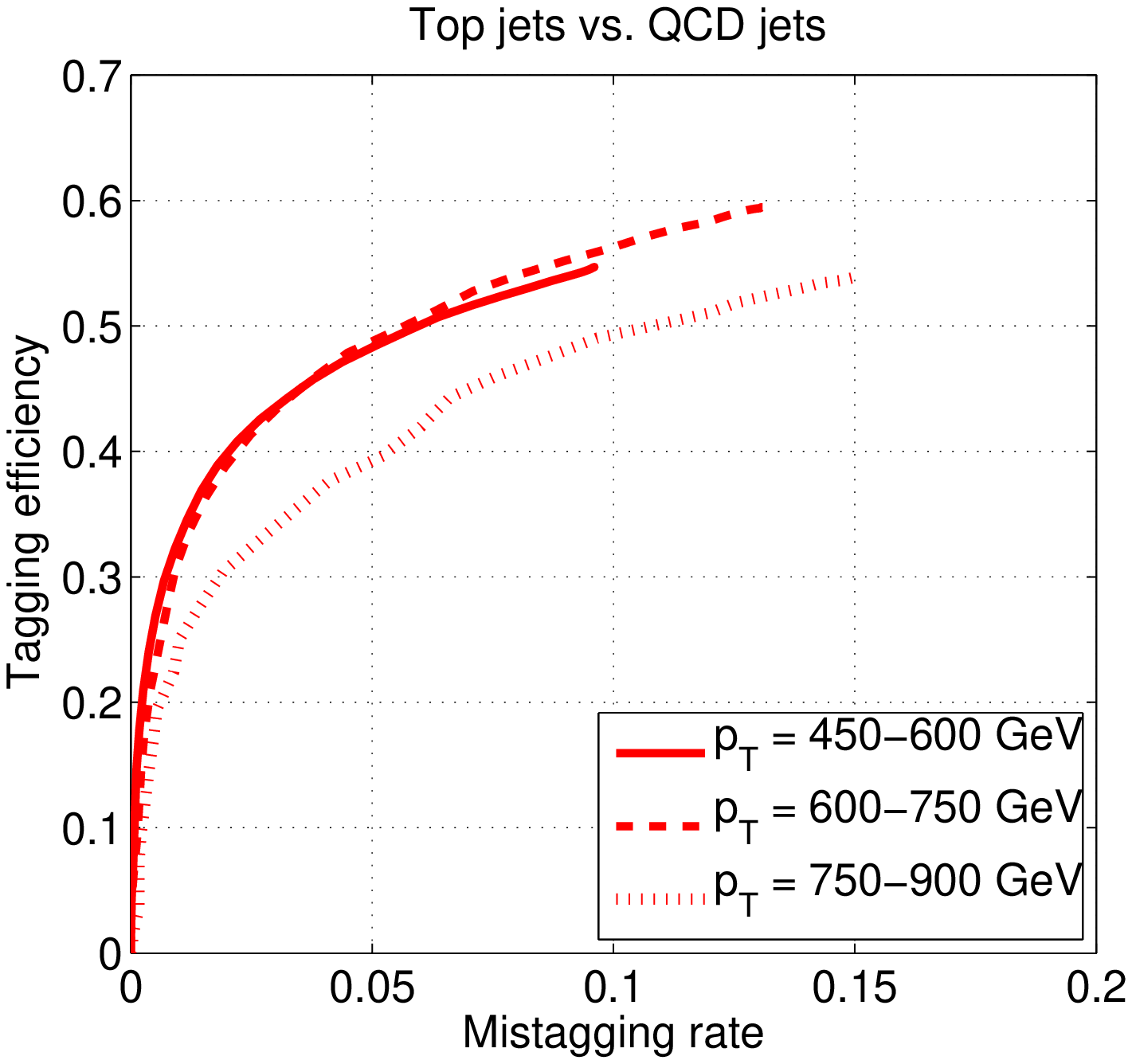}}
    \subfigure[]{\label{fig:TopSigEffb}\includegraphics[trim = 2mm 0mm 10mm 0mm, clip, height=4.5cm]{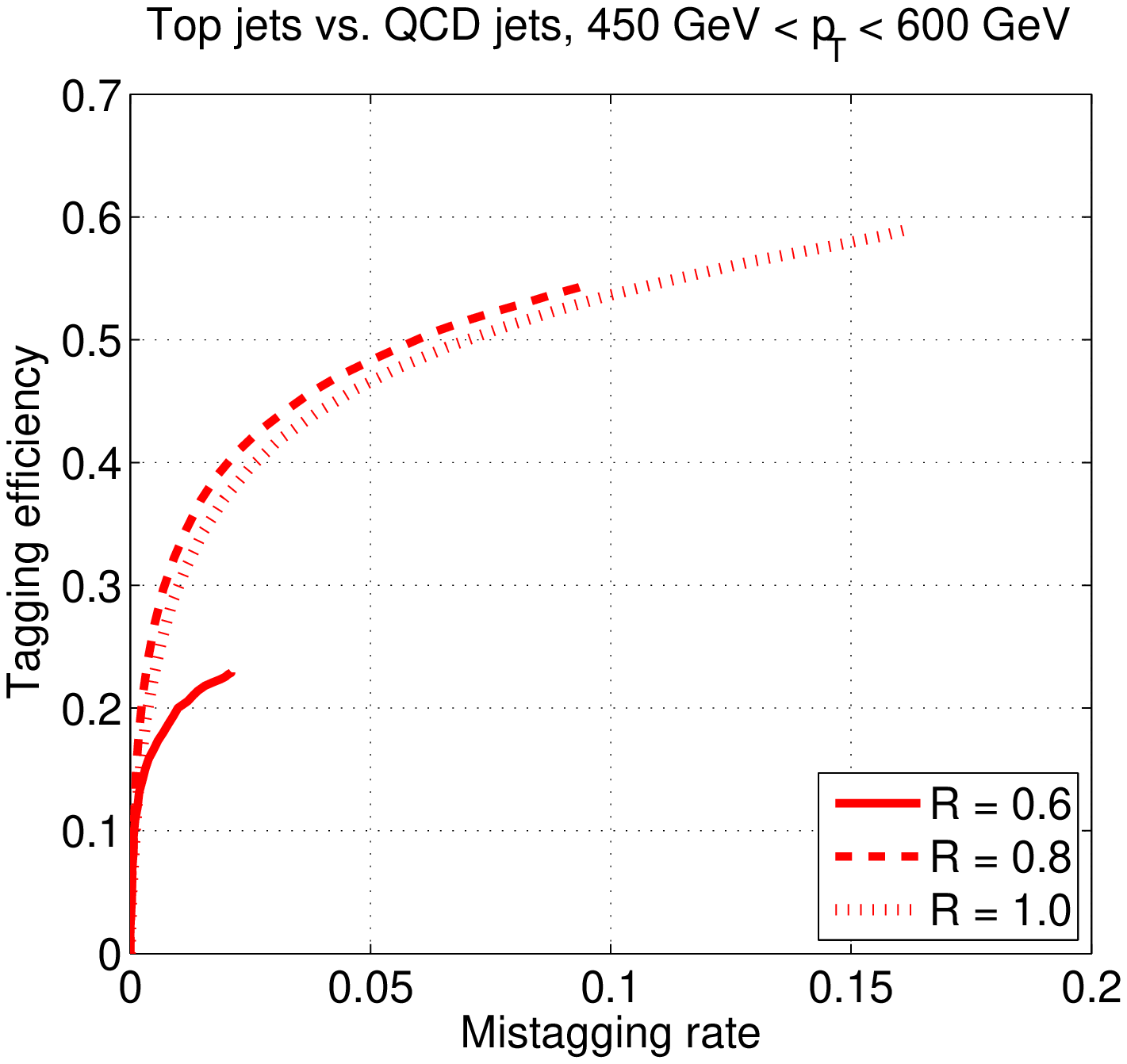}}
    \subfigure[]{\label{fig:TopSigEffc}\includegraphics[trim = 2mm 0mm 10mm 0mm, clip, height=4.5cm]{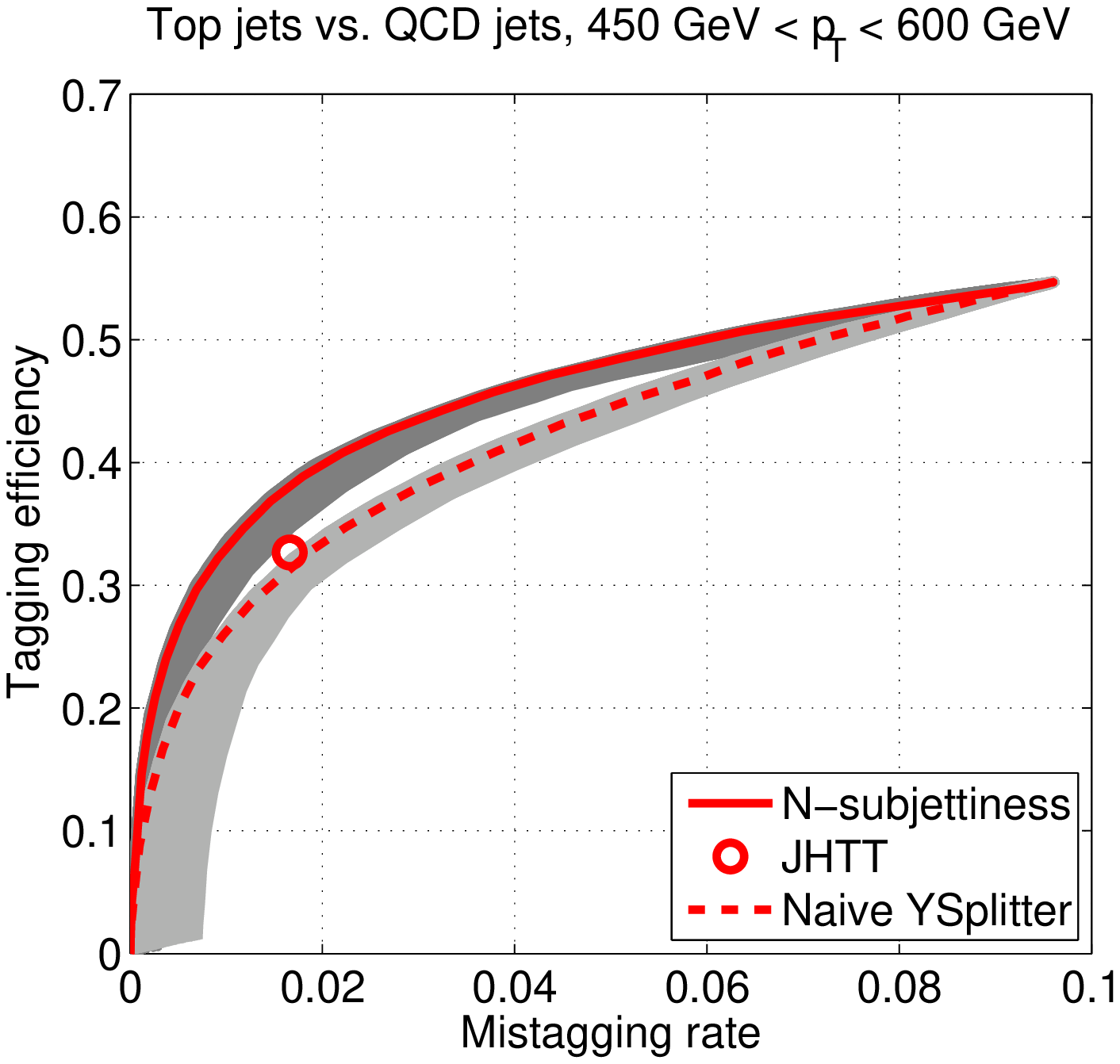}}
  \end{center}
  \vsh
  \caption{Top jet signal efficiency/background rejection plots.  Here, the $\tau_3/\tau_2$ cut is varied for different (a) jet transverse momenta and (b) jet radii.  Unless otherwise indicated, the jets have $R= 0.8$ and $450 \text{ GeV} < p_T < 600 \text{ GeV}$.  The rightmost points in each plot are for the $m_{\rm top}$ invariant mass window criterium alone, and points to the left of these are obtained with additional cuts on the $\tau_3/\tau_2$ ratio.  Figure (c) shows a naive comparison against the YSplitter method, where the red lines indicate varying cuts on $\tau_3/\tau_2$ and $y_{34}/y_{23}$, respectively, while the shaded bands indicate the modified linear cuts of \Sec{sec:optimization}.  The circle indicates a benchmark JHTT point.  In this non-optimized test, $N$-subjettiness  compares favorably to previous top tagging methods.  Note the different horizontal scale in (c).}
  \label{fig:TopSigEff}
\end{figure}

\begin{figure}[tp]
  \begin{center}
      \subfigure[]{\label{fig:TopSigImprova}\includegraphics[trim = 0mm 0mm 0mm 0mm, clip, height=4.5cm]{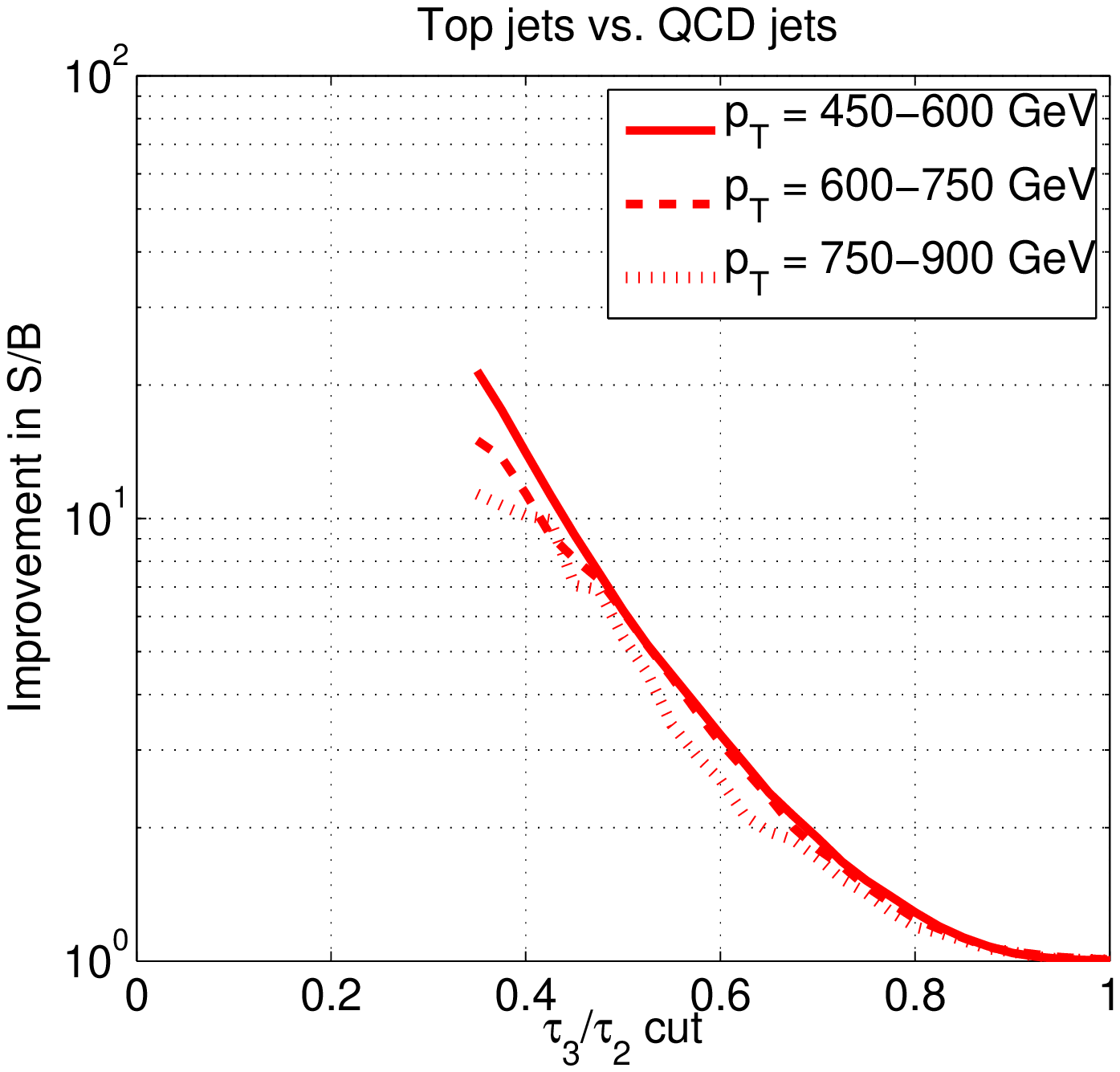}}
      \subfigure[]{\label{fig:TopSigImprovb}\includegraphics[trim = 0mm 0mm 0mm 0mm, clip, height=4.5cm]{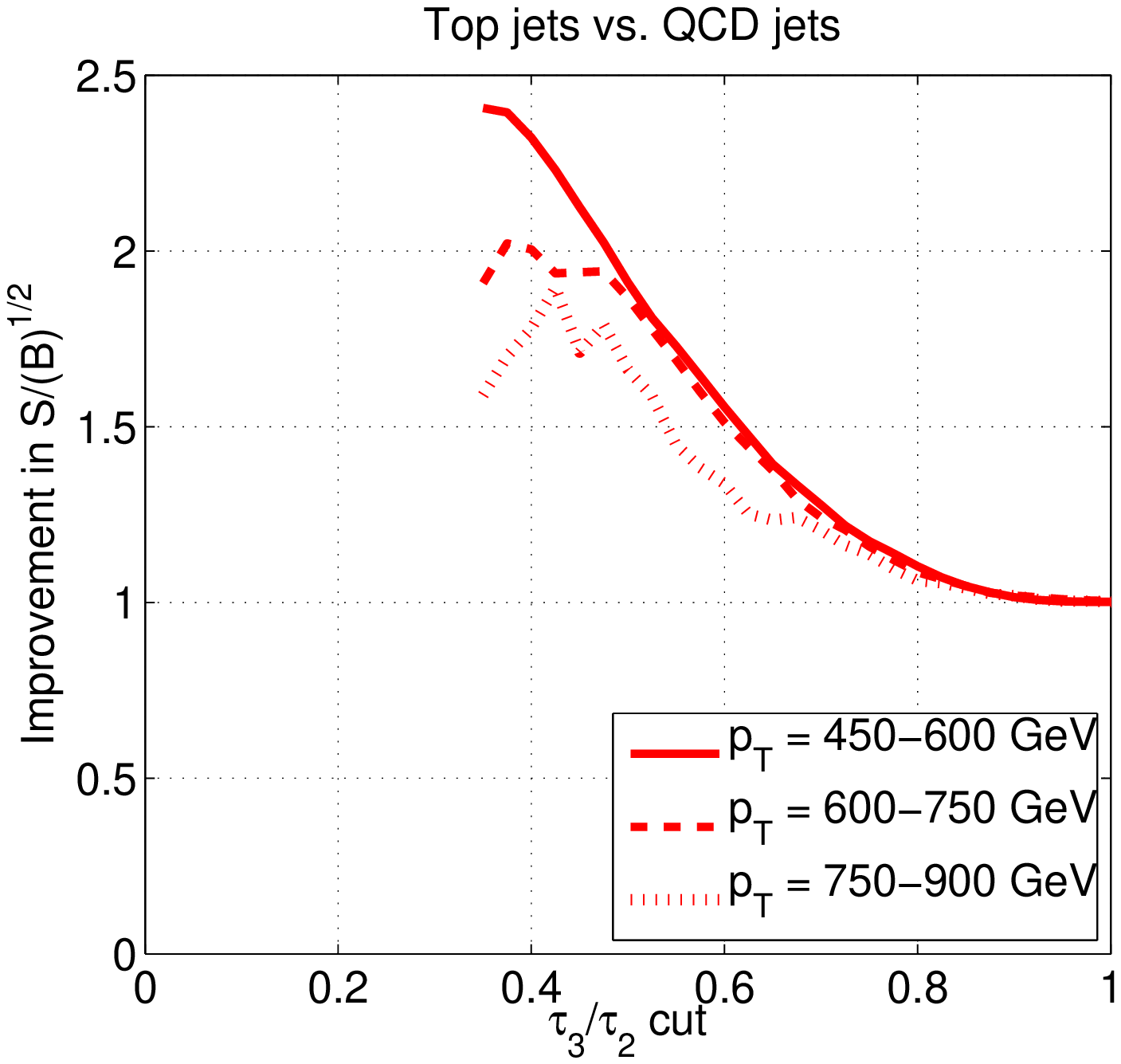}}\\
  \end{center}
  \vsh
  \caption{Improvement compared to an invariant mass cut alone for top jets in (a) signal over background ($S/B$) and (b) signal over square-root background ($S/\sqrt{B}$) by using the $\tau_3/\tau_2$ cut. The value of unity on the right corresponds to using only the $m_{\rm top}$ invariant mass window.}
  \label{fig:TopSigImprov}
\end{figure}

In \Tab{tab:TopSigEff08} and \Fig{fig:TopSigEff}, we see that except for the highest $p_T$ range and a too small jet radius, a tagging efficiency of around 30\% is achievable for a mistagging rate of 1\%.  Note that we do not display results for $300 \text{ GeV} < p_T < 400 \text{ GeV}$ in \Tab{tab:TopSigEff08} nor for $300 \text{ GeV} < p_T < 450 \text{ GeV}$ in \Fig{fig:TopSigEff}, as in this kinematic range, often not all subjets from the decay products of the top quark are captured by the anti-$k_T$ algorithm, leading to low signal efficiencies.  This is already noticeable in the 400-500 GeV $p_T$ range, where only 45\% of the jets pass the mass criterion, whereas the same is true for 58\% of the jets in the 500-600 GeV $p_T$ range.  For the same reason, the tagging efficiency is much lower for $R=0.6$ as compared to other jet radii, as seen in \Fig{fig:TopSigEffb}.  Our method is most effective for top jets with $500 \text{ GeV} < p_T < 700 \text{ GeV}$.  The worse efficiencies for $p_T > 700 \text{ GeV}$ are due to several factors.  A larger portion of top jets have a mass above their mass window due to additional FSR at high $p_T$, while more light quark and gluon jets will acquire masses high enough to pass the mass window cut.  Our $N$-subjettiness cuts are also less effective at very high transverse momenta, since the distinction between 2-subjettiness and 3-subjettiness becomes less clear when the three subjets have considerable overlap.  \Fig{fig:TopSigEffc} compares the $N$-subjettiness $\tau_3/\tau_2$ cut to a $y_{34} / y_{23}$ cut as well as to a benchmark JHTT point, and we see that $N$-subjettiness compares favorably to these methods for boosted top identification.  In \Fig{fig:TopSigImprov}, we see that a factor of 20 improvement in $S/B$ and a factor of 2 improvement in $S/\sqrt{B}$ is possible using $\tau_3/\tau_2$ for top jets.

\begin{table}[t]
  \begin{center}
    \begin{tabular}{r |r@{ : }lr@{ : }lr@{ : }lr@{ : }lr@{ : }l}
     \hline
     \hline
     $p_T$ range (GeV) & \multicolumn{2}{c}{400--500} & \multicolumn{2}{c}{500--600} & \multicolumn{2}{c}{600--700} & \multicolumn{2}{c}{700--800} & \multicolumn{2}{c}{800--900}\\
     \hline
     No $\tau_3/\tau_2$ cut & .45 & .067 & .58 & .11 & .60 & .13 & .58 & .15 & .53 & .15 \\
     $\tau_3/\tau_2 < 0.8$ & .43 & .048 & .55 & .080 & .56 & .094 & .54 & .12 & .50 & .12 \\
     $\tau_3/\tau_2 < 0.6$ & .34 & .016 & .43 & .025 & .44 & .028 & .41 & .039 & .36 & .041 \\
     $\tau_3/\tau_2 < 0.4$ & .18 & .0019 & .21 & .0028 & .21 & .0035 & .19 & .0038 & .16 & .0069 \\
    \hline
     1\% mistag rate & .30 & .010 & .34 & .010 & .32 & .010 & .28 & .010 & .21 & .010\\
     40\% tag efficiency & .40 & .035 & .40 & .018 & .40 & .020 & .40 & .035 & .40 & .061\\
     \hline
     \hline
    \end{tabular}
  \end{center}

    \caption{Tagging efficiencies vs. mistagging rates for top jets : QCD jets with $R= 0.8$.  The top row corresponds to just applying the $m_{\rm top}$ invariant mass window (145 GeV to 205 GeV) criteria, and the subsequent rows include an additional $\tau_3/\tau_2$ cut.  The bottom two rows indicate the tagging efficiencies achievable with  a fixed mistagging rate of 1\%, and the mistagging rate achievable with a fixed tagging efficiency of 40\%.}
  \label{tab:TopSigEff08}
\end{table}

\subsection{Potential Optimization}
\label{sec:optimization}

In the above efficiency studies, we only considered cuts on the $\tau_{N}/\tau_{N-1}$ ratio to tag $N$-prong boosted objects.  One could certainly generalize this approach to include more complicated cuts in the $\tau_{N-1}$--$\tau_{N}$ plane or to use a multivariate analysis on the full set of $\tau_N$ values.  Such studies are beyond the scope of the present work, but as a small step towards optimization, we tested the next simplest generalization of $\tau_{N}/\tau_{N-1}$: a general linear cut in the $\tau_{N-1}$--$\tau_{N}$ plane.

For $N$-subjettiness we tested a cut of $\tau_{N}/(\tau_{N-1} - \tau_{N-1;0}) < s$ with the slope parameter $s \in [0,5]$.  For YSplitter, we tested the analogous cut on $y_{N,N+1}/(y_{N-1,N} - y_{N-1,N;0}) < s$.  These are essentially the same linear cuts as in the previous subsection except with a shift in one of the two $N$-subjettiness (or YSplitter) values.  In \Fig{fig:WSigEffc} for $W$ jets, the dark grey envelope corresponds to the range $\tau_{1;0} \in [-0.075, 0.075]$ with the solid purple curve for the special case $\tau_{2;0} = 0$.  Similarly, the light grey envelope corresponds to the range $y_{12;0} \in [-25 \text{ GeV}, 25 \text{ GeV}]$ with the dashed purple curve for the special case $y_{23,0} = 0$.  The same logic holds for top jets in \Fig{fig:TopSigEffc} with the ranges $\tau_{2;0} \in [-0.05, 0.05]$ and $y_{23;0} \in [-10 \text{ GeV}, 10 \text{ GeV}]$.  For both methods, small improvements are possible, warranting a further exploration of multivariate methods.

\section{Applications to New Physics Searches}
\label{sec:case}

Having seen gains in single $W$ jet and top jet identification in the previous section, we now apply $N$-subjettiness to the task of heavy resonance reconstruction.  Consider a hypothetical $Z'$ spin-$1$ resonance which decays to pairs of boosted objects as $Z' \rightarrow W^+ W^-$ or $Z' \rightarrow t \overline{t}$.  We will show that an $N$-subjettiness cut can lower the minimum detectable cross section for this resonance by up to an order of magnitude with 1 fb$^{-1}$ of 7 TeV LHC data.  This illustrates the versatility and power of the $N$-subjettiness tagging method in the search for new physics.  

\subsection{Analysis Overview}
\label{sec:analysisz}

We use the same basic \texttt{Pythia} settings as in \Sec{sec:eff_overview}, simulating $pp$ collisions at $\sqrt{s} = 7$ TeV using the same virtual calorimeter setup.  The jet clustering procedure is exactly as before, though for all samples we set the anti-$k_T$ jet radius at $R=0.8$, which slightly degrades the $Z' \rightarrow W^+ W^-$ reach.  We perform our selection criteria on events with at least two boosted central jets ($p_T > 300$ GeV and $|\eta|<1.3$).  We impose the same $W$ mass and top mass windows and use $\tau_2/\tau_1$ and $\tau_3/\tau_2$ as the respective discriminating variables. 

The signal source is a hypothetical ``sequential'' $Z'$ boson of the ``extended gauge model'' \cite{Altarelli:1989ff}, one of the reference spin-$1$ resonance options available in \texttt{Pythia}.  We choose this type of $Z'$ for convenience since it has couplings both to $W$ bosons and to top quarks.\footnote{Of course, for a true sequential $Z'$, one would likely use the $Z' \rightarrow \ell^+ \ell^-$ mode for discovery.}  At the end of the day, we will report the reach in terms of $\sigma \times \text{Br}$, so our analysis is roughly independent of signal source, apart from parton distribution functions and angular correlations, which somewhat affect the signal acceptance. We consider $Z'$ masses between 750 GeV and 2 TeV, and consider the signal region to be where the combined invariant mass of the boosted jets are within 100 GeV of the $Z'$ boson mass.   

There are two main types of backgrounds:  reducible backgrounds from QCD dijets and irreducible backgrounds from standard model $W^+W^-$ and $t\bar{t}$ production.  For simplicity, we will only consider QCD dijets for discussing the reach, as the $W^+W^-$ and $t\bar{t}$ processes have lower cross sections than the achievable reach with 1 fb$^{-1}$ of 7 TeV LHC data.\footnote{For $t\bar{t}$, we will see that this statement is borderline, so strictly speaking the standard model $t\bar{t}$ background should not be ignored.}    There is a potentially important background from $W/Z$+jets, where the $W/Z$ decays hadronically to form a real boosted $W/Z$ and a jet fakes a boosted $W$.  However, while the $W/Z$+jets background (with only one jet faking a $W$) is reduced less by our tagging method than the QCD dijet background (with both jets faking $W$'s), the contribution from $W/Z$+jets still ends up being about an order of magnitude smaller than that from QCD dijets after our optimal cuts.\footnote{Note that if the mistagging rate is reduced even further with more aggressive $N$-subjettiness cuts or with an additional selection criteria, it may no longer be possible to ignore the background from $W/Z$+jets.}  There is also an interesting background from $W$-strahlung that can mimic the top jet signal \cite{Rehermann:2010vq} which we will not include.

\subsection{Di-W and Di-Top Resonance Results}
\label{sec:caseresults}

\begin{figure}[tp]
  \begin{center}
    \subfigure[]{\label{fig:ZprimeInvMw}\includegraphics[trim = 0mm 0mm 0mm 0mm, clip, height=5.5cm]{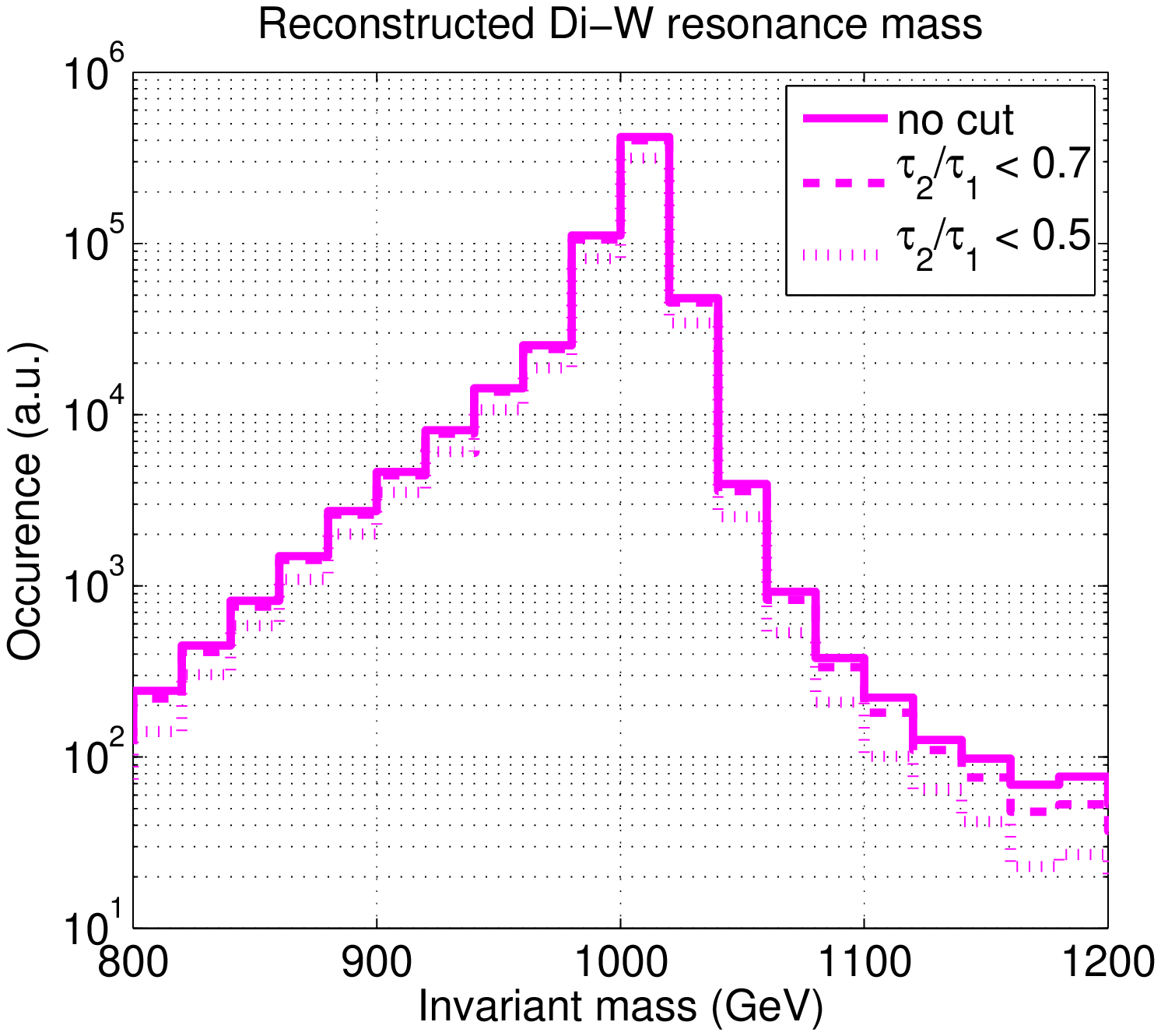}} 
    \subfigure[]{\label{fig:ZprimeInvMqw}\includegraphics[trim = 0mm 0mm 0mm 0mm, clip, height=5.5cm]{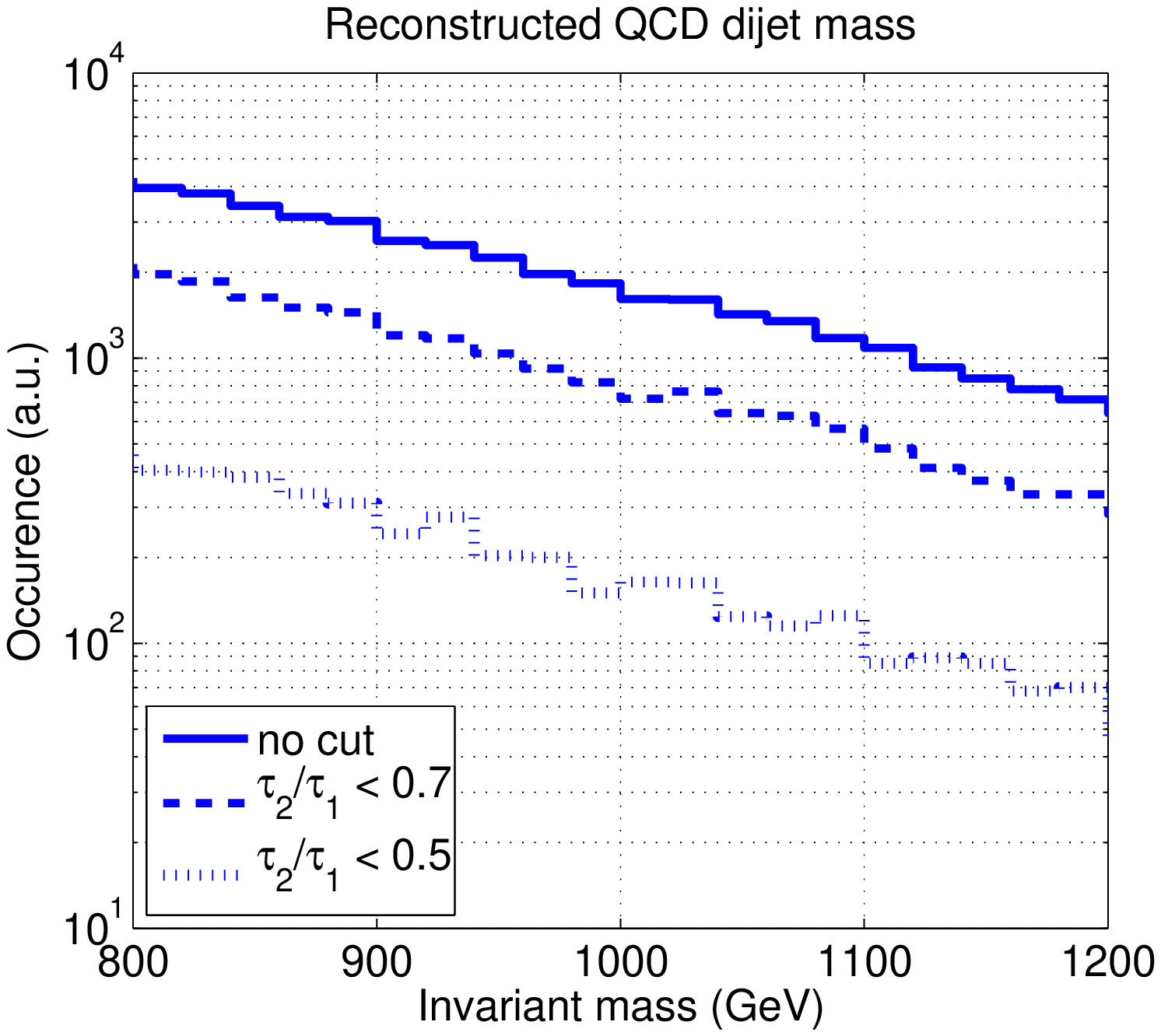}}
  \end{center}
    \vsh
    \caption{Reconstructed invariant mass distributions for (a) a narrow 1 TeV $Z'$ decaying to two $W$ jets and (b) dijet QCD events faking $Z' \rightarrow W^+W^-$.  In both cases, the hardest two jets in the event were required to satisfy $p_T > 300~\GeV$, $|\eta|<1.3$, and $65~\GeV < m_{\text{jet}} < 95~\GeV$, and the $\tau_2/\tau_1$ criterium is applied to both jets.  Note that we can use milder $N$-subjettiness cuts than in \Sec{sec:efficiencyt}, since the cuts are applied to both jets.}
    \label{fig:ZprimeInvM2}
\end{figure}

\begin{figure}[tp]
  \begin{center}
    \subfigure[]{\label{fig:ZprimeInvM}\includegraphics[trim = 0mm 0mm 0mm 0mm, clip, height=5.5cm]{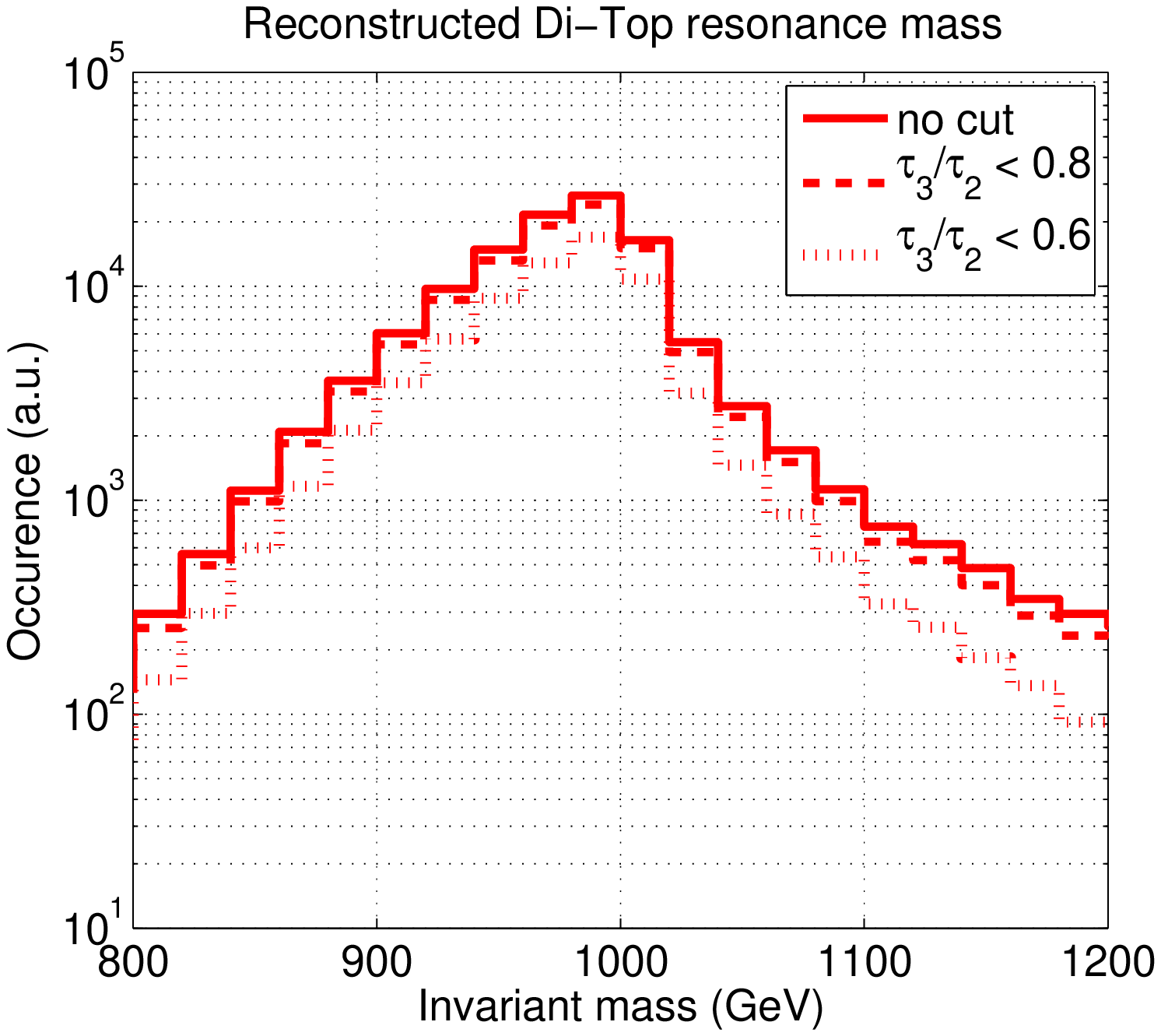}} 
    \subfigure[]{\label{fig:ZprimeInvMq}\includegraphics[trim = 0mm 0mm 0mm 0mm, clip, height=5.5cm]{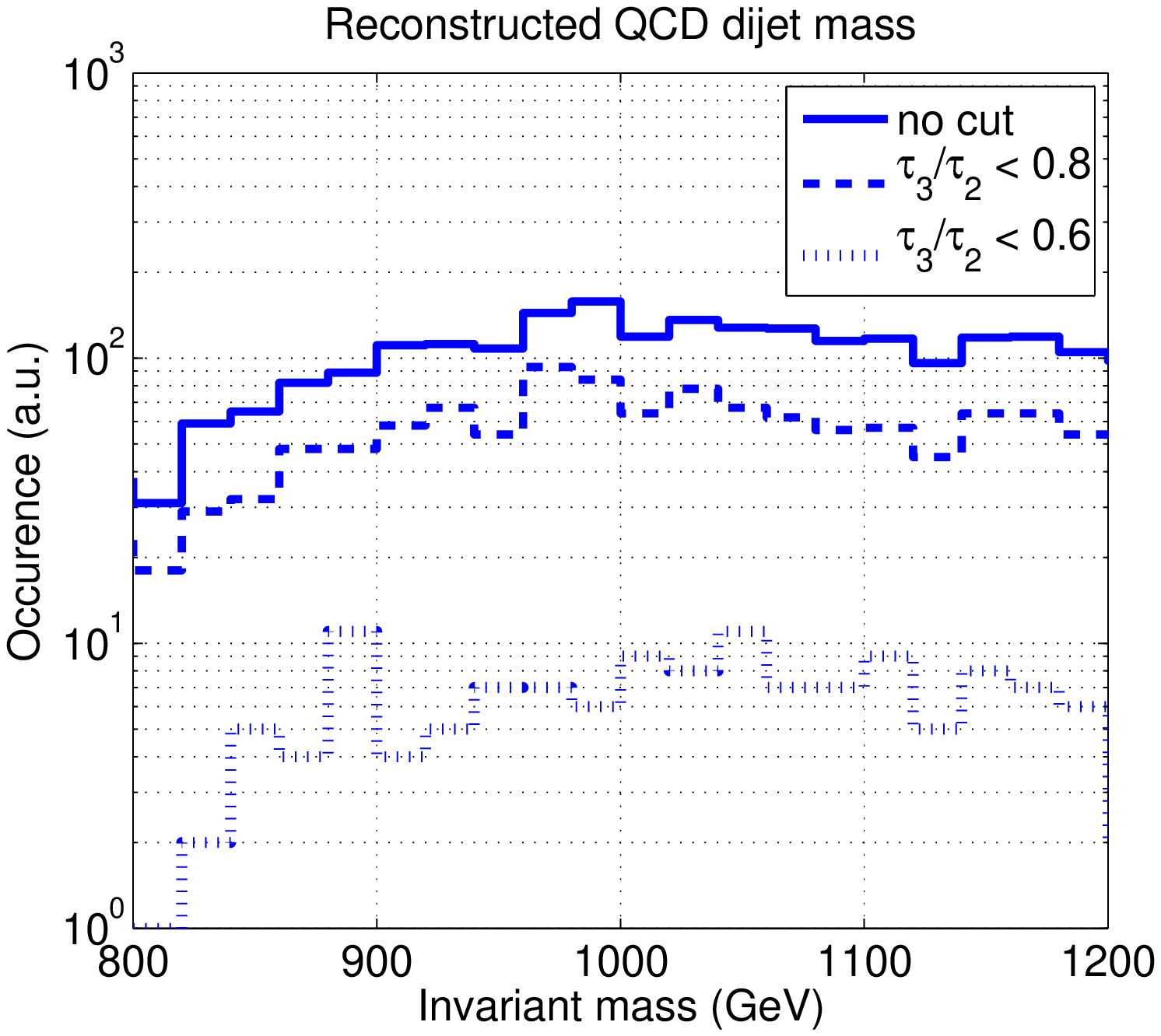}}\\
  \end{center}
  \vsh
    \caption{Reconstructed invariant mass distributions for (a) a narrow 1 TeV $Z'$ decaying to two top jets and (b) dijet QCD events faking $Z' \rightarrow \bar{t}t$.  In both cases, the hardest two jets in the event were required to satisfy $p_T > 300~\GeV$, $|\eta|<1.3$, and $145~\GeV < m_{\text{jet}} < 205~\GeV$, and the $\tau_3/\tau_2$ criterium is applied to both jets.  Again note the milder $N$-subjettiness cuts compared to \Sec{sec:efficiencyt}.}
    \label{fig:ZprimeInvM1}
\end{figure}

To see the effect of the $N$-subjettiness cut on resonance reconstruction, consider \Fig{fig:ZprimeInvM2} for $Z' \rightarrow W^+ W^-$ and \Fig{fig:ZprimeInvM1} for $Z' \rightarrow t \overline{t}$, where both jets have been tagged as boosted hadronic objects.  Even for a moderate cut on $\tau_{N}/\tau_{N-1}$, there is a substantial decrease in the QCD dijet background with only a slight decrease in the resonance signal.  

\begin{figure}[tp]
  \begin{center}
    \subfigure[]{\label{fig:ZpimreSigEffWb}\includegraphics[trim = 0mm 0mm 0mm 0mm, clip, height=4.5cm]{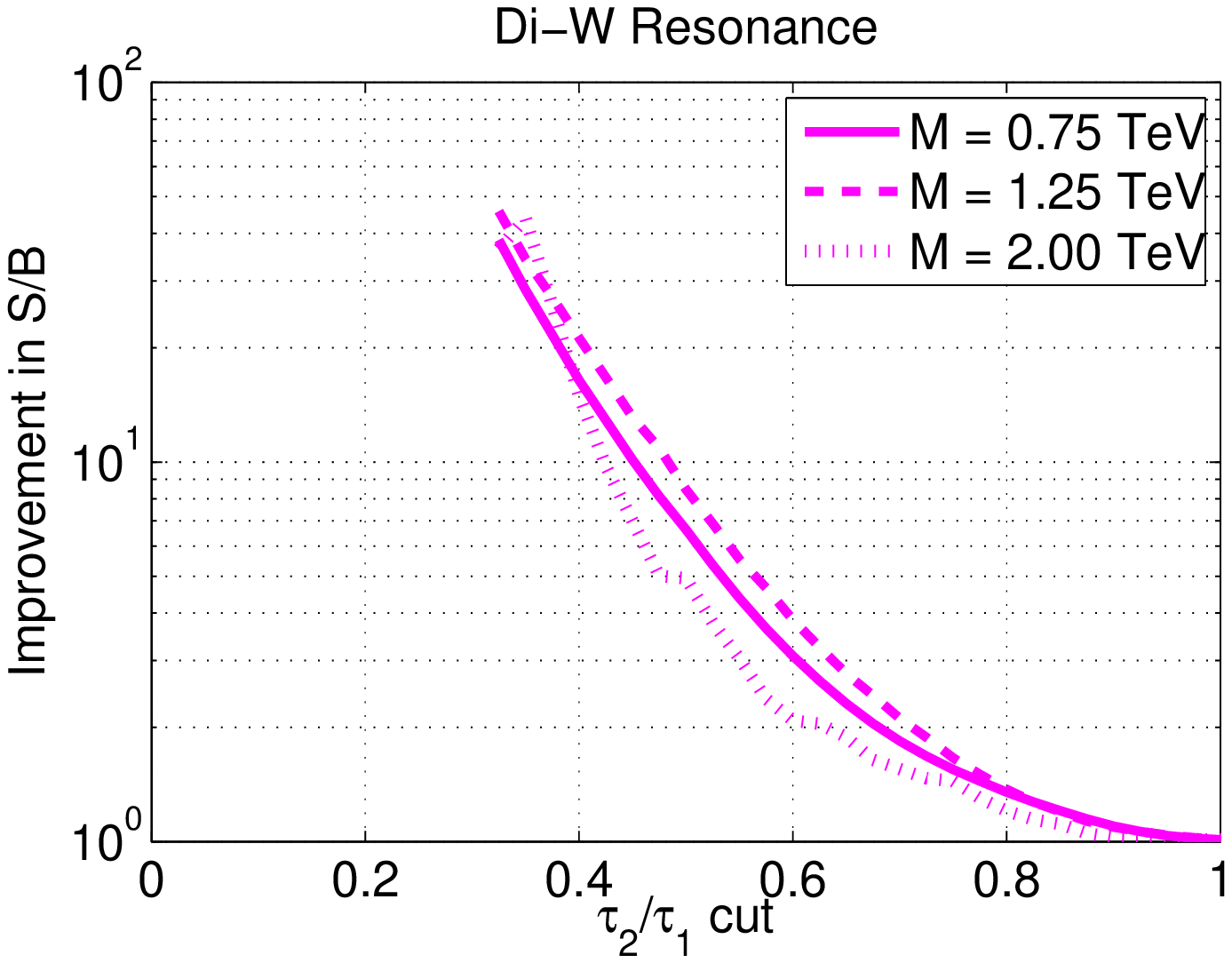}}
    \subfigure[]{\label{fig:ZpimreSigEffWc}\includegraphics[trim = 0mm 0mm 0mm 0mm, clip, height=4.5cm]{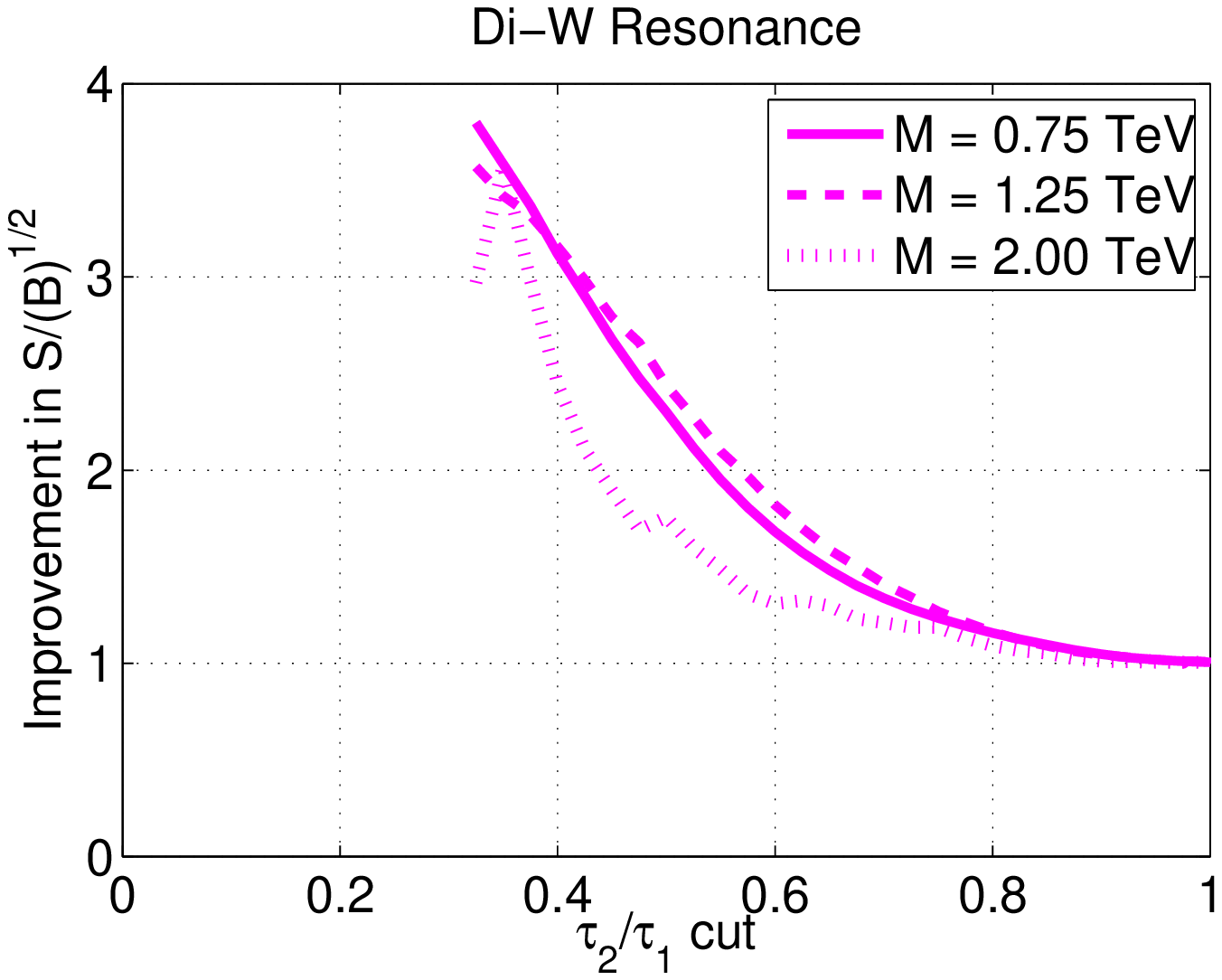}}
  \end{center}
  \vsh
  \caption{Improvement in (a) $S/B$ and (b) $S/\sqrt{B}$ for identifying a $Z'$ boson decaying to $W$ jet pairs.   Compared to the $m_W$ invariant mass cut alone (rightmost points on the plot), cuts on $N$-subjettiness can significantly improve signal efficiency/background rejection.}
  \label{fig:ZprimeEffW}
\end{figure}

\begin{figure}[tp]
  \begin{center}
    \subfigure[]{\label{fig:ZpimreSigEffb}\includegraphics[trim = 0mm 0mm 0mm 0mm, clip, height=4.5cm]{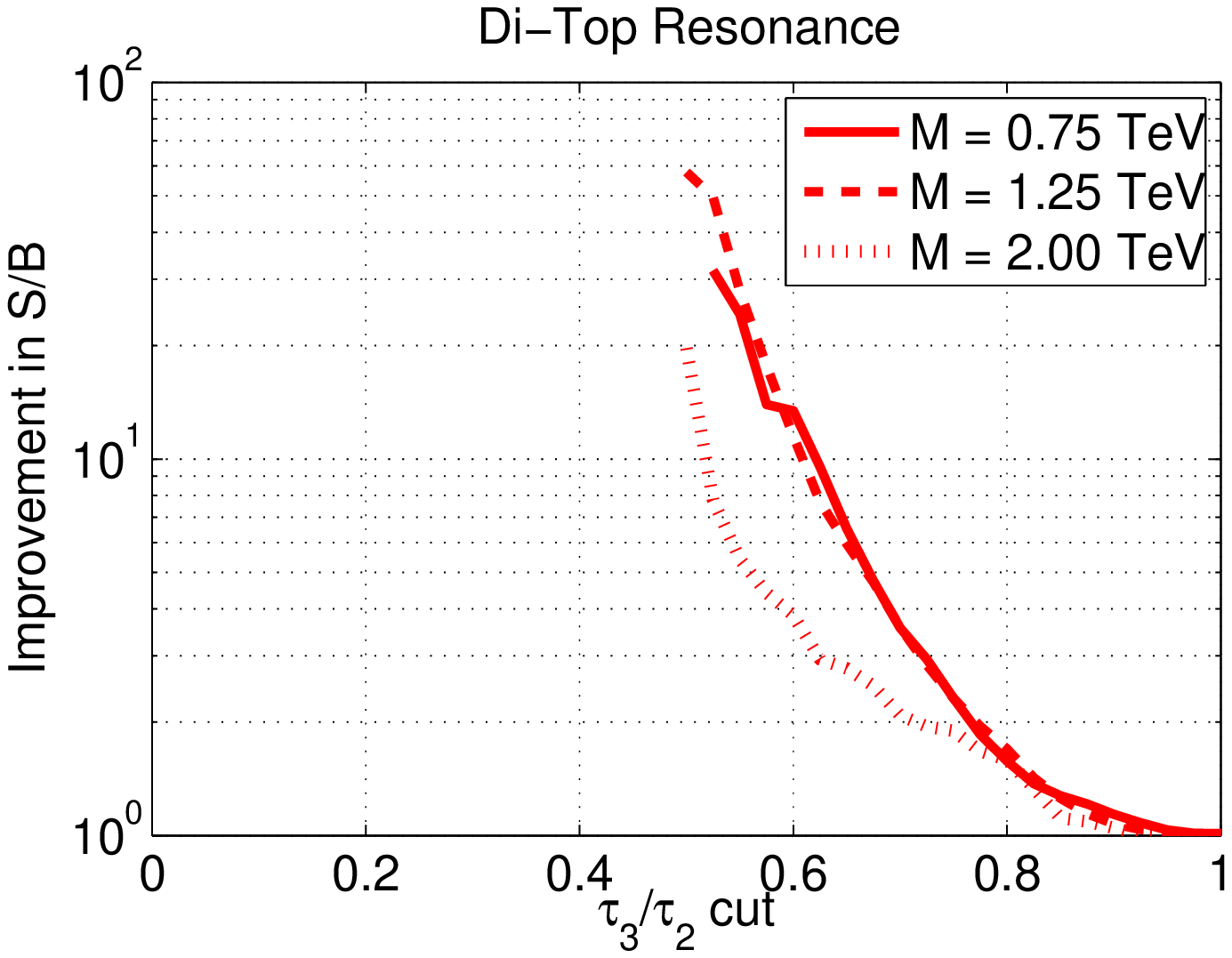}}
    \subfigure[]{\label{fig:ZpimreSigEffc}\includegraphics[trim = 0mm 0mm 0mm 0mm, clip, height=4.5cm]{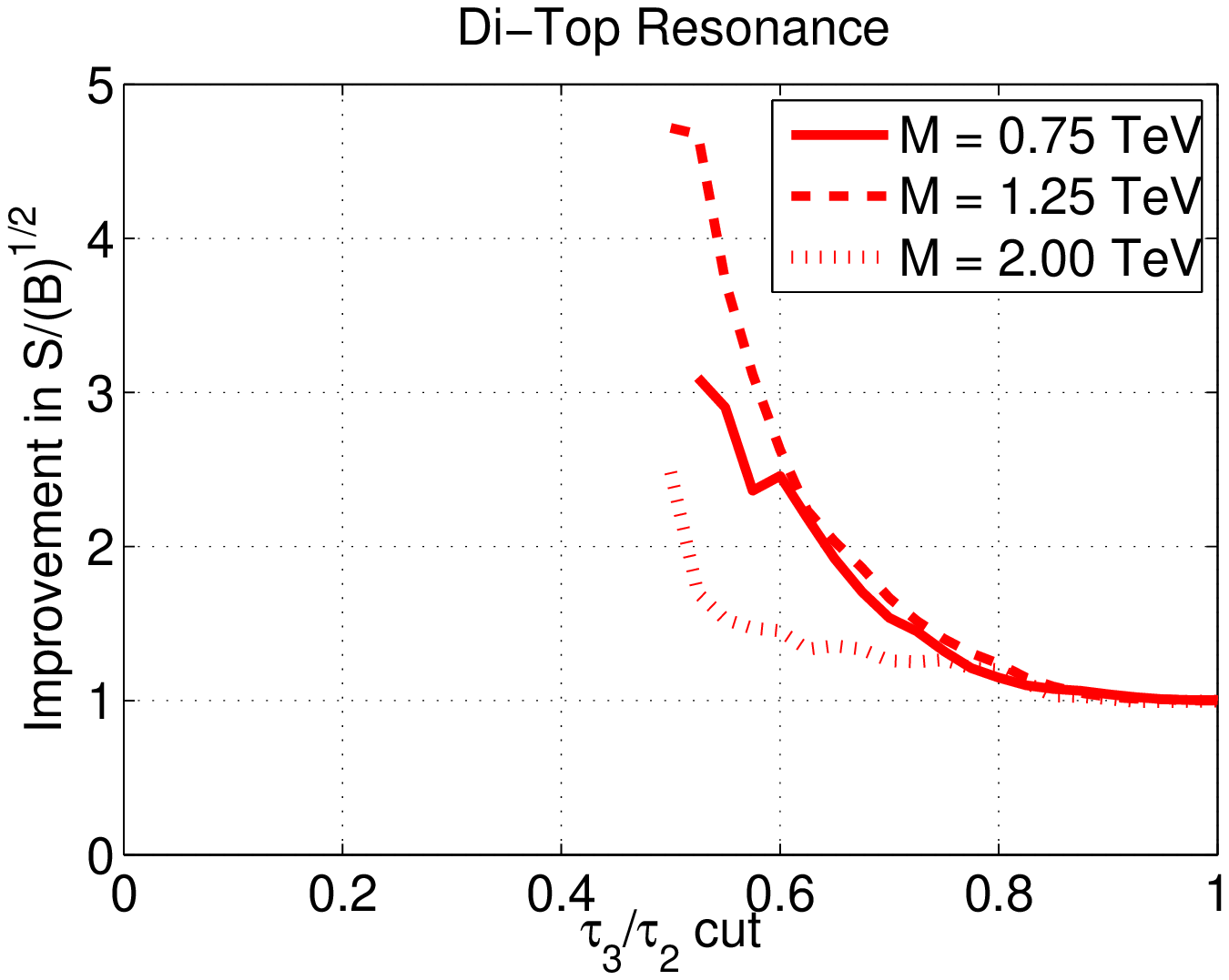}}
  \end{center}
  \vsh
  \caption{Improvement in (a) $S/B$ and (b) $S/\sqrt{B}$ for identifying a $Z'$ boson decaying to top jet pairs.   Compared to the $m_{\rm top}$ invariant mass cut alone (rightmost points on the plot), cuts on $N$-subjettiness can significantly improve signal efficiency/background rejection.}
  \label{fig:ZprimeEff}
\end{figure}

We can isolate the $Z'$ resonance region by considering dijet invariant masses satisfying $|m_{jj} - m_{Z'}| < 100~\GeV$.  In \Fig{fig:ZprimeEffW} and \Fig{fig:ZprimeEff}, we plot the improvements in $S/B$ and $S/\sqrt{B}$ for di-$W$ and di-top resonances for three values of the $Z'$ boson mass.  Compared to the single object efficiencies from \Sec{sec:efficiency}, the improvement seen in resonance reconstruction approximately factorizes.  That is, the $\tau_{N}/\tau_{N-1}$ values for the two hardest jets in any event are roughly independent of each other, such that an $S/B$ improvement of $\epsilon$ for a given $\tau_{N}/\tau_{N-1}$ cut in \Sec{sec:efficiencyt} yields an $S/B$ improvement of roughly $\epsilon^2$ for the $Z'$ resonance.

To see how these improvements in $S/B$ and $S/\sqrt{B}$ work in practice, we now calculate how much an $N$-subjettiness cut would lower the minimum cross section for detecting a $Z'$ boson in the boosted $W^+W^-$ or boosted $\bar{t}t$ channel.  For this purpose, we define the fiducial $Z'$ detection criteria to be twofold.  First, we require at least 10 reconstructed candidate $Z'$ resonances ($S > 10$) at a luminosity of 1 fb$^{-1}$.  Second, we require $S/\text{Err}(B) > 5$ (``five sigma discovery''), where the combined statistical and systematic error on the QCD background is estimated as $\text{Err}(B) = \sqrt{(\sqrt{B})^2 + (0.1 B)^2}$.\footnote{A 10\% systematic on a jet measurement is likely optimistic, but fine as a benchmark for comparison.}  Both constraints restrict the detectable physical cross section values for the $Z'$ resonance, and we optimize the $N$-subjettiness cuts to satisfy both constraints.

\begin{figure}[tp]
  \begin{center}
    \subfigure[]{\label{fig:crossBoundWa}\includegraphics[trim = 10mm 0mm 10mm 0mm, clip, height=5.0cm]{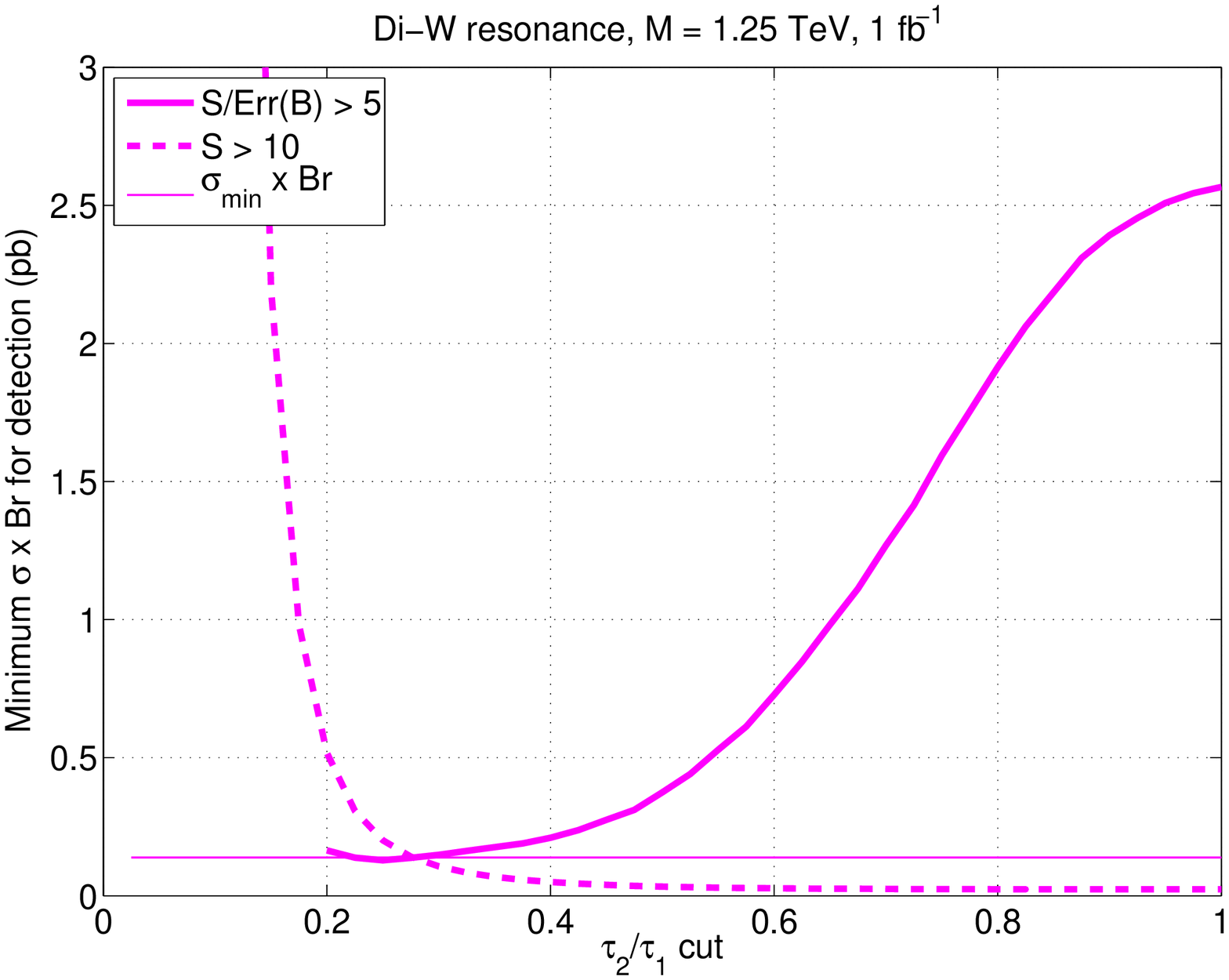}}
    \subfigure[]{\label{fig:crossBoundWb}\includegraphics[trim = 10mm 0mm 10mm 0mm, clip, height=5.0cm]{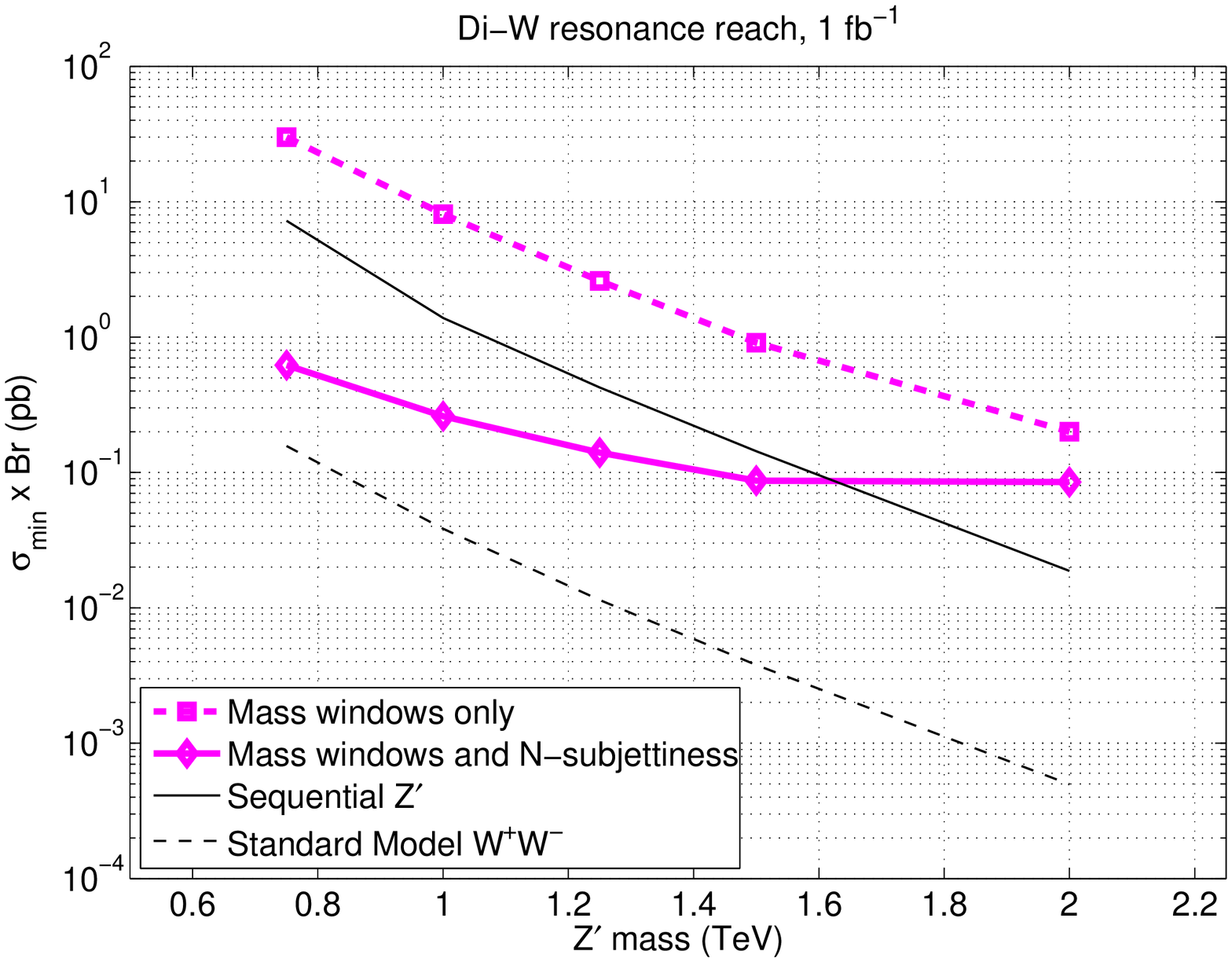}}
  \end{center}
\vsh
    \caption{(a): Optimizing the $\tau_2/\tau_1$ cut to determine the minimum physical cross section for detection of $Z' \rightarrow W^+ W^-$ with a $Z'$ mass of 1.25 TeV.  The value of the cut is chosen such that both $S > 10$ and $S/\text{Err}(B) > 5$.  (b): Lower bounds on the detectable $Z'$ cross section times the branching ratio to hadronically-decaying $W$ bosons as a function of the mass of the $Z'$.  Compared to a jet invariant mass cut alone, the $N$-subjettiness method to tag boosted $W$s gives a substantial improvement in the reach.  Shown for reference are the $\sigma \times \text{Br}$ for the sequential $Z'$ model and standard model $W^+W^-$ production where the di-$W$ invariant mass is within 100 GeV of the fiducial $Z'$ mass.}
    \label{fig:crossBoundW}
\end{figure}

\begin{figure}[tp]
  \begin{center}
    \subfigure[]{\label{fig:CrossBounda}\includegraphics[trim = 10mm 0mm 10mm 0mm, clip, height=5.0cm]{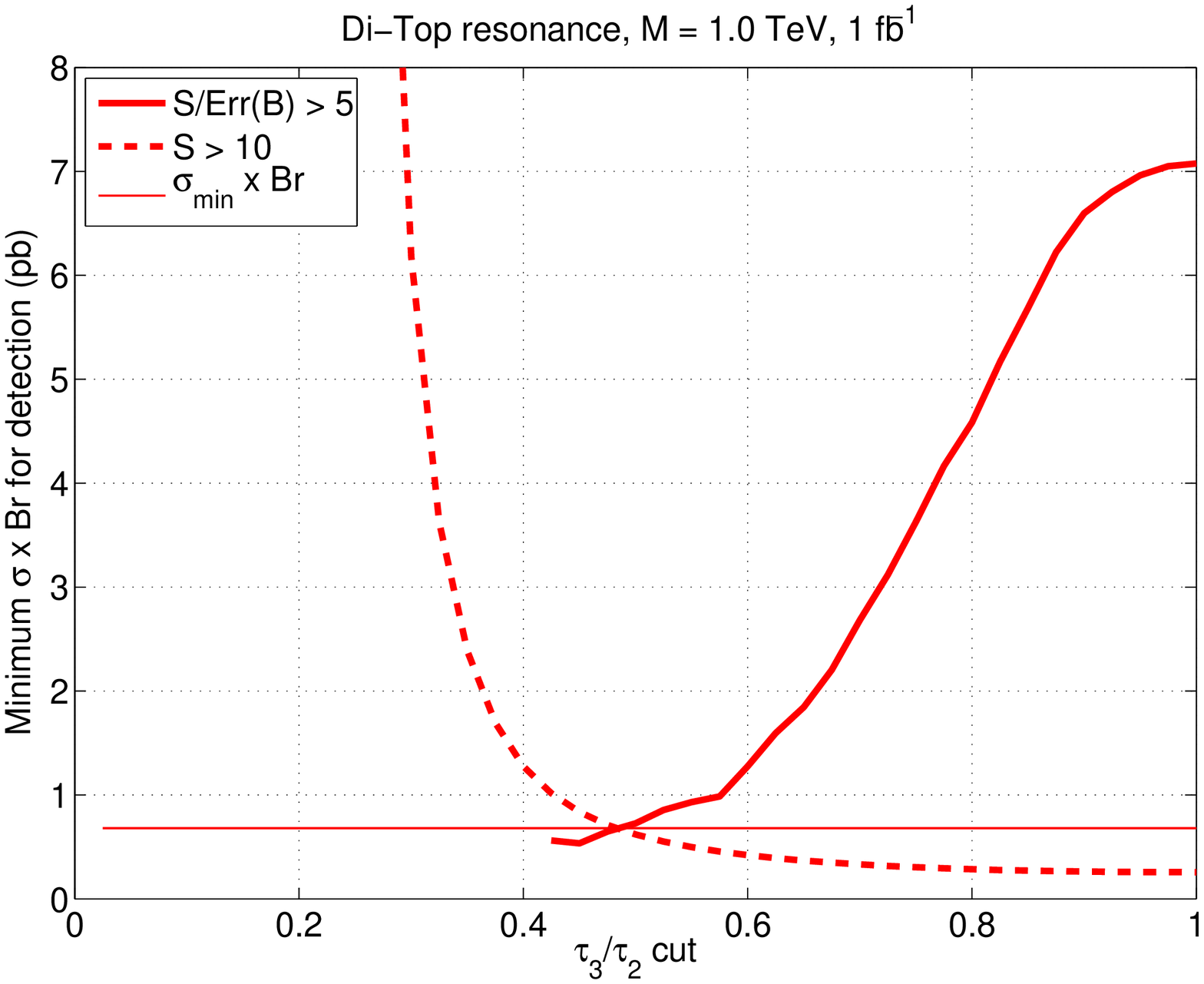}}
    \subfigure[]{\label{fig:CrossBoundb}\includegraphics[trim = 10mm 0mm 10mm 0mm, clip, height=5.0cm]{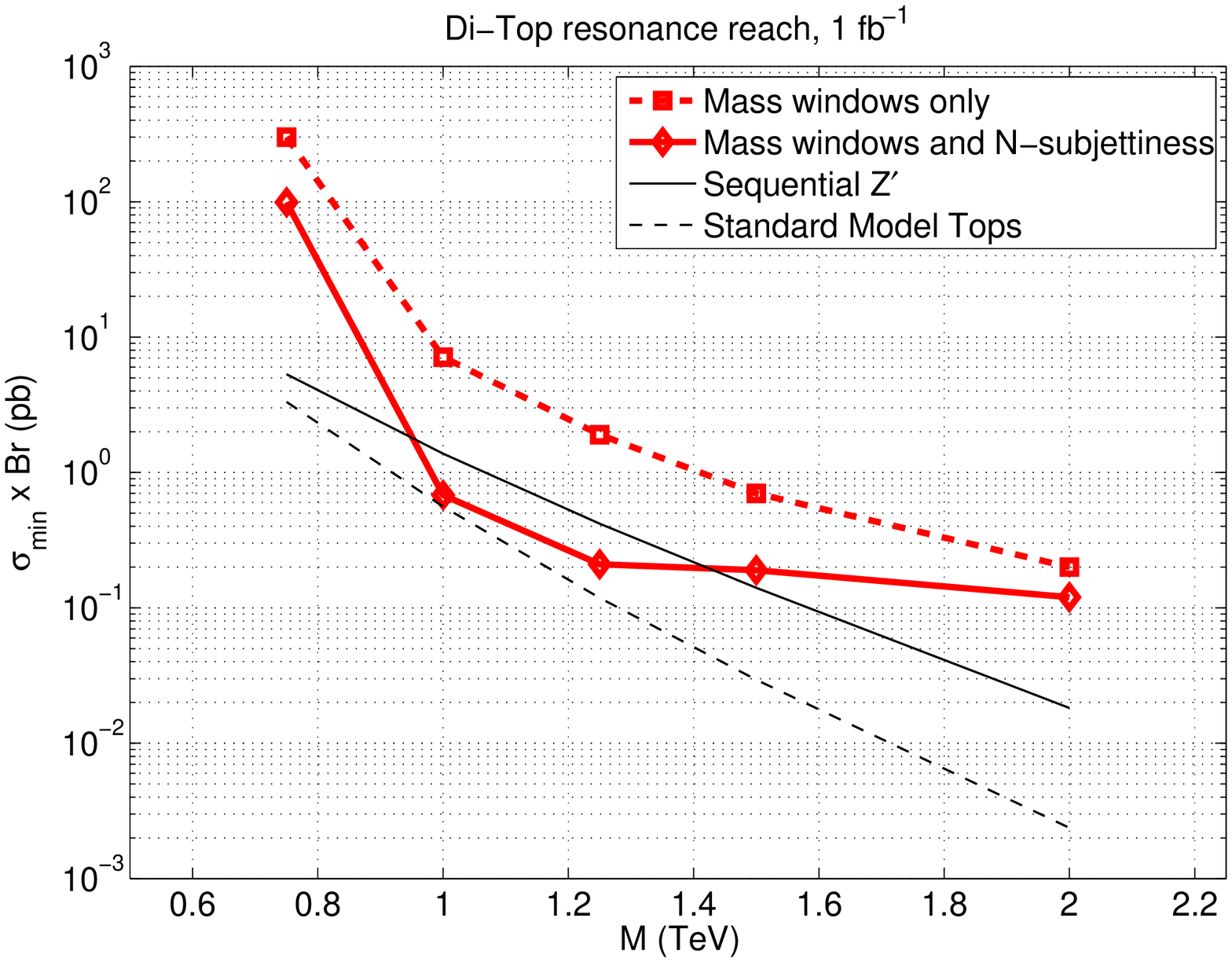}}
  \end{center}
\vsh
    \caption{(a): Optimizing the $\tau_3/\tau_2$ cut to determine the minimum physical cross section for detection of $Z' \rightarrow \bar{t} t$ with a $Z'$ mass of 1 TeV.  The value of the cut is chosen such that both $S > 10$ and $S/\text{Err}(B) > 5$.  (b): Lower bounds on the detectable $Z'$ cross section times the branching ratio to hadronically-decaying top quark as a function of the mass of the $Z'$.  Compared to a jet invariant mass cut alone, the $N$-subjettiness method to tag boosted tops gives a substantial improvement in the reach.  Shown for reference are the $\sigma \times \text{Br}$ for the sequential $Z'$ model and standard model $\bar{t}t$ production where the di-top invariant mass is within 100 GeV of the fiducial $Z'$ mass.}
    \label{fig:crossBound}
\end{figure}

In \Fig{fig:crossBoundWa}, we demonstrate the interplay between the $S > 10$ and $S/\text{Err}(B) > 5$ requirements for selecting the optimal $\tau_2/\tau_1$ cut on the $W$ jets.  In \Fig{fig:crossBoundWb}, we plot the lower bound on the detectable cross section for $Z'$, given in terms of $\sigma(Z') \times \text{Br}(Z' \rightarrow W^+ W^-) \times \text{Br}(W \rightarrow jj)^2$.  For reference, we also plot the $\sigma \times \text{Br}$ for the sequential $Z'$ and standard model $W^+ W^-$ production.  The $\tau_2/\tau_1$ cut improves the discovery reach upwards of an order of magnitude compared to only using invariant mass to tag $W$ jets.  The analogous bounds for $Z' \rightarrow \bar{t}t$ appear in \Fig{fig:crossBound} where the reach is given in terms of $\sigma(Z') \times \text{Br}(Z' \rightarrow \bar{t}t) \times \text{Br}(t \rightarrow bjj)^2$. 
The $\tau_3/\tau_2$ cut can bring $Z'$ signal detection threshold down almost to the irreducible $t\bar{t}$ background level.\footnote{The reach for the 750 GeV resonance could likely be improved by using a larger initial jet radius.}  Thus, $N$-subjettiness provides substantial gains for discovering new physics with boosted hadronic objects.

\section{Conclusion}
\label{sec:conclusions}

In this paper, we introduced an inclusive jet shape $N$-subjettiness designed to tag boosted hadronic objects.  We found that the ratio $\tau_2 / \tau_1$ is an effective discriminating variable to isolate boosted hadronic $W$, $Z$, and Higgs bosons from the background of QCD jets with large invariant mass.  Similarly, $\tau_3 / \tau_2$ is an effective variable for identifying boosted top quarks.   As a case study, we observed that $N$-subjettiness offers significant improvements in the detection sensitivity of hypothetical heavy resonances decaying to pairs of electroweak bosons or top quarks.  Overall, $N$-subjettiness selection methods are at least as good as other commonly used discriminating methods for identification of boosted objects.

$N$-subjettiness exhibits several desirable properties which warrant further experimental and theoretical investigations.  On the experimental side, $\tau_N$ can be calculated on a jet-by-jet basis and thereby offers considerable flexibility of application.  While we focused just on ratios of $\tau_N$ as discriminating variables, multivariate optimization along the lines of \Sec{sec:optimization} could improve signal efficiency and background rejection.  In addition, some of the $N$-subjettiness variations mentioned in \App{app:Nsubdef} might also be effective discriminating variables by themselves or in combination.   On the theoretical side, $\tau_N$ is an infrared and collinear safe inclusive jet shape which in principle can be defined without the need for an algorithmic subjet finding method.  Thus, the prospects for theoretical calculations involving $N$-subjettiness look promising both using fixed-order perturbative calculations and using resummation techniques.

With the first LHC data on the books, the search for new physics is already underway.  New phenomena may be revealed in the production of highly boosted electroweak bosons and top quarks, and we expect that $N$-subjettiness will prove to be a useful variable for exploring such extreme kinematic regimes.

~\\

\noindent  \textbf{Note Added}:  While this paper was being completed, \Ref{Kim:2010uj} appeared which defines a Lorentz-invariant version of $N$-subjettiness and uses $\tau_2$ for boosted Higgs identification. 

\acknowledgments
We thank David Krohn, Iain Stewart, Frank Tackmann, and Wouter Waalewijn for helpful conversations.  J.T. is supported by the U.S. Department of Energy (D.O.E.) under cooperative research agreement DE-FG02-05ER-41360.  K.V.T is supported by the MIT Undergraduate Research Opportunities Program (UROP).

\appendix

\newpage

\section{Definition of $N$-subjettiness}
\label{app:Nsubdef}

The definition of $N$-subjettiness in \Eq{eq:tau_N} is not unique, and different choices for $\tau_N$ can be used to give different weights to the emissions within a jet.  These generalizations of $N$-subjettiness are similar to different ``angularities'' \cite{Berger:2003iw} used in $e^+ e^- \rightarrow \text{hadrons}$ measurements.  

Analogously to \Ref{Stewart:2010tn}, a general $N$-subjettiness measure is
\be
\label{eq:tauNgen}
\tau_N^{\rm gen} = \frac{1}{d_0}\sum_k \min_{J} \left\{d(p_J,p_k) \right\},
\ee
where $d_0$ is a normalization factor, $J$ runs over the $N$ candidate subjets, and $d(p_J,p_k)$ is distance measure between a candidate subjet $p_J$ and a jet constituent $p_k$.  Like in \Sec{subsec:candidate_subjets}, one needs a method to figure out the candidate subjet directions, which could be achieved through a separate subjet finding algorithm or by minimizing $\tau_N$ over possible candidate subjets $p_J$.

There are many choices for $d(p_J,p_k)$, but a nice two-parameter, boost-invariant choice for the distance measure is
\be
d^{\alpha, \beta}(p_J,p_k) = p_{T,k}  \left(p_{T,J}\right)^\alpha \left(\Delta R_{J,k} \right)^\beta.
\ee
If desired, one could replace $p_{T,J}$ with $E_{T,J} = \sqrt{p_{T,J}^2 + m_J^2}$ to include information about the subjet mass.\footnote{Obviously, one could also use $E_{T,k}$ to include the mass of the jet constituent, though in our studies, the four-vectors of the calorimeter cells were massless by definition.}  For $e^+ e^-$ applications, one would replace the transverse momentum $p_T$ with the total momentum $|\vec{p}|$ (or the energy $E$) and $\Delta R$ with the opening angle $\Delta \Omega$.  A natural choice for the normalization factor to keep $0< \tau_N < 1$ is
\be
d_0 = \max_{J} \left\{ \left(p_{T,J}\right)^\alpha  \right\} \left(R_0 \right)^\beta \sum_k p_{T,k}  ,
\ee 
where $R_0$ is the characteristic jet radius.

By making $d(p_J,p_k)$ linear in $p_{T,k}$, $\tau_N$ is automatically an infrared-safe observable. Collinear-safety requires linearity in $p_{T,k}$ as well, but imposes the addition requirement that $\beta \geq 0$.  The value of $\alpha$ is unconstrained.  Of course, we are assuming that the candidate subjet finding method is also infrared- and collinear-safe.

In the body of the paper, we used $\alpha = 0$, $\beta=1$.  This choice corresponds to treating each subjet democratically, and using a $k_T$-like distance measure.  This distance measure makes $\tau_N$ similar to jet broadening \cite{Catani:1992jc},\footnote{By similar, we mean the distance measure has the same $\Delta R_{A,k} \rightarrow 0$ limit.  Because thrust-like observables are defined in a preferred rest frame and we are working with a longitudinally boost-invariant measure, the correspondence is inexact.} and we found that this was an effective choice for boosted object identification.  By varying $\beta$, we can change the angular weighting.   A thrust-like \cite{Farhi:1977sg} weighting corresponds to $\beta = 2$, while other angularities \cite{Berger:2003iw} with $-\infty < a < 2$ are given by $\beta = 2 - a$.  By varying $\alpha$, we can weight the distance measure by the hardness of the subjet directions.  Large positive (negative) $\alpha$ means that the minimum in \Eq{eq:tauNgen} is given by the distance to hardest (softest) candidate subjet.   Further studies of boosted object identification using different values of $\alpha$ and $\beta$ would be interesting, since studies of jet angularities have shown that additional information about jet substructure can be gleaned by combining different angular information \cite{Ellis:2010rw}.

\section{Additional Event Displays}
\label{app:additional_events}

\begin{figure}[tp]
  \begin{center}
    \subfigure[]{\label{fig:ed7}\includegraphics[trim = 3mm 0mm 3mm 0mm, clip, height=4.60cm]{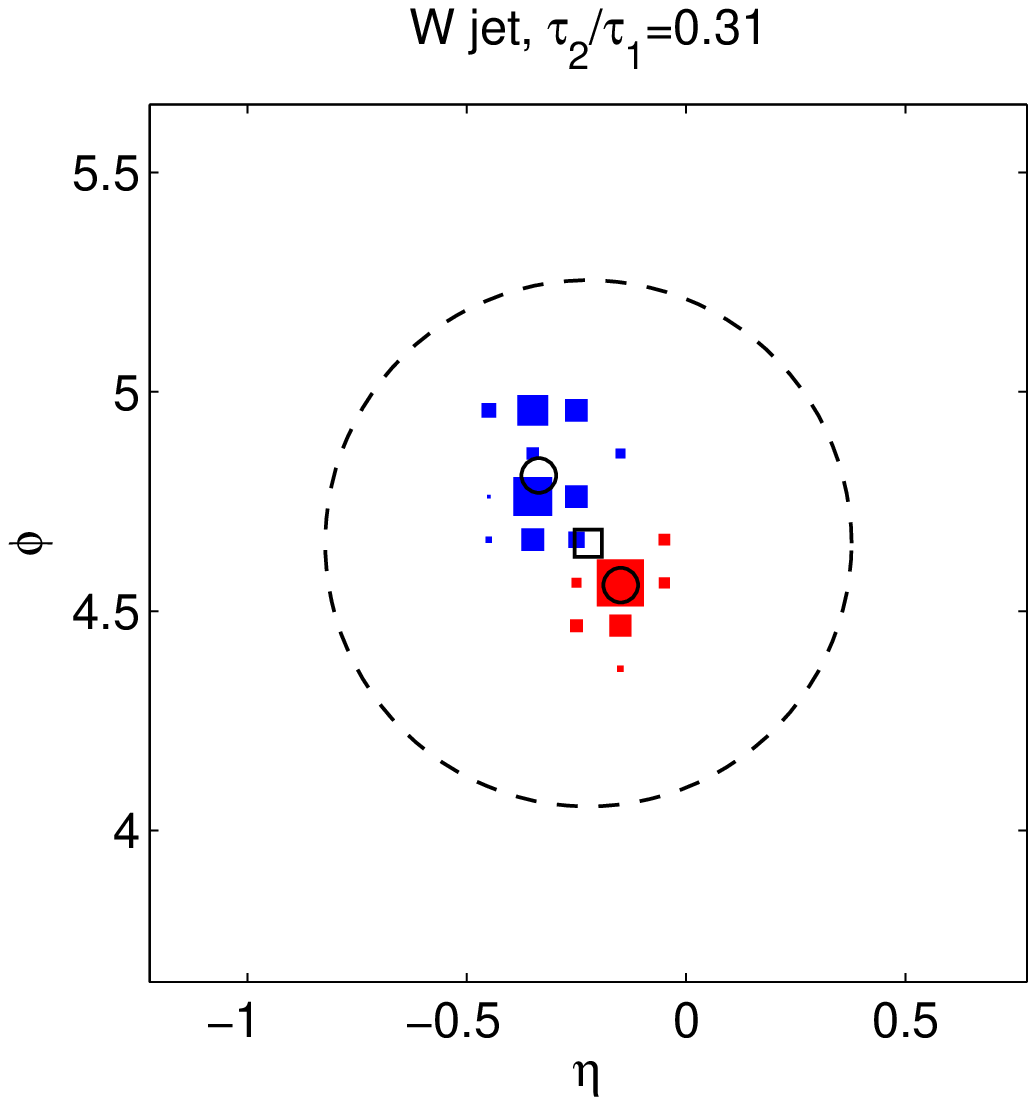}}
    \subfigure[]{\label{fig:ed8}\includegraphics[trim = 3mm 0mm 3mm 0mm, clip, height=4.60cm]{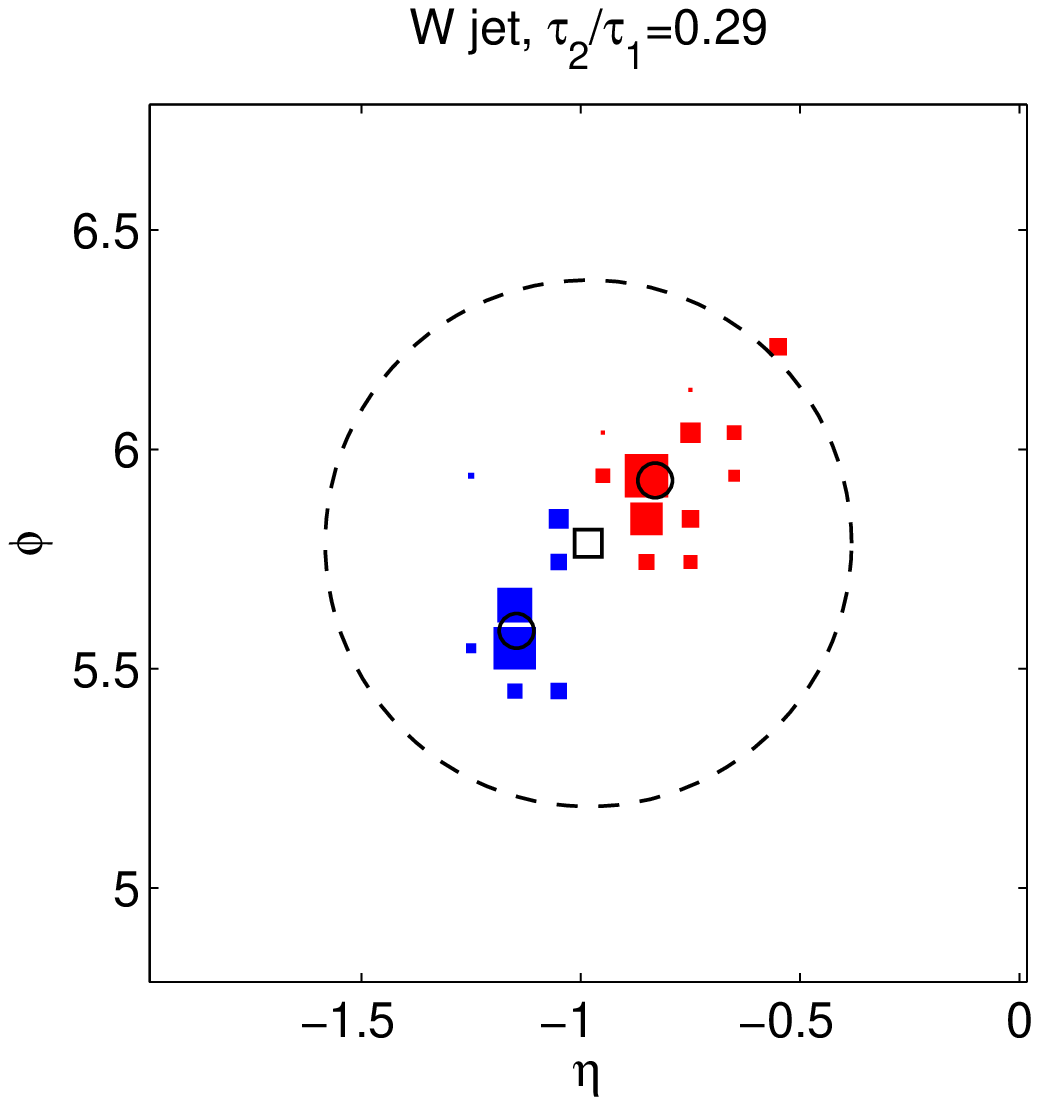}}
    \subfigure[]{\label{fig:ed9}\includegraphics[trim = 3mm 0mm 3mm 0mm, clip, height=4.60cm]{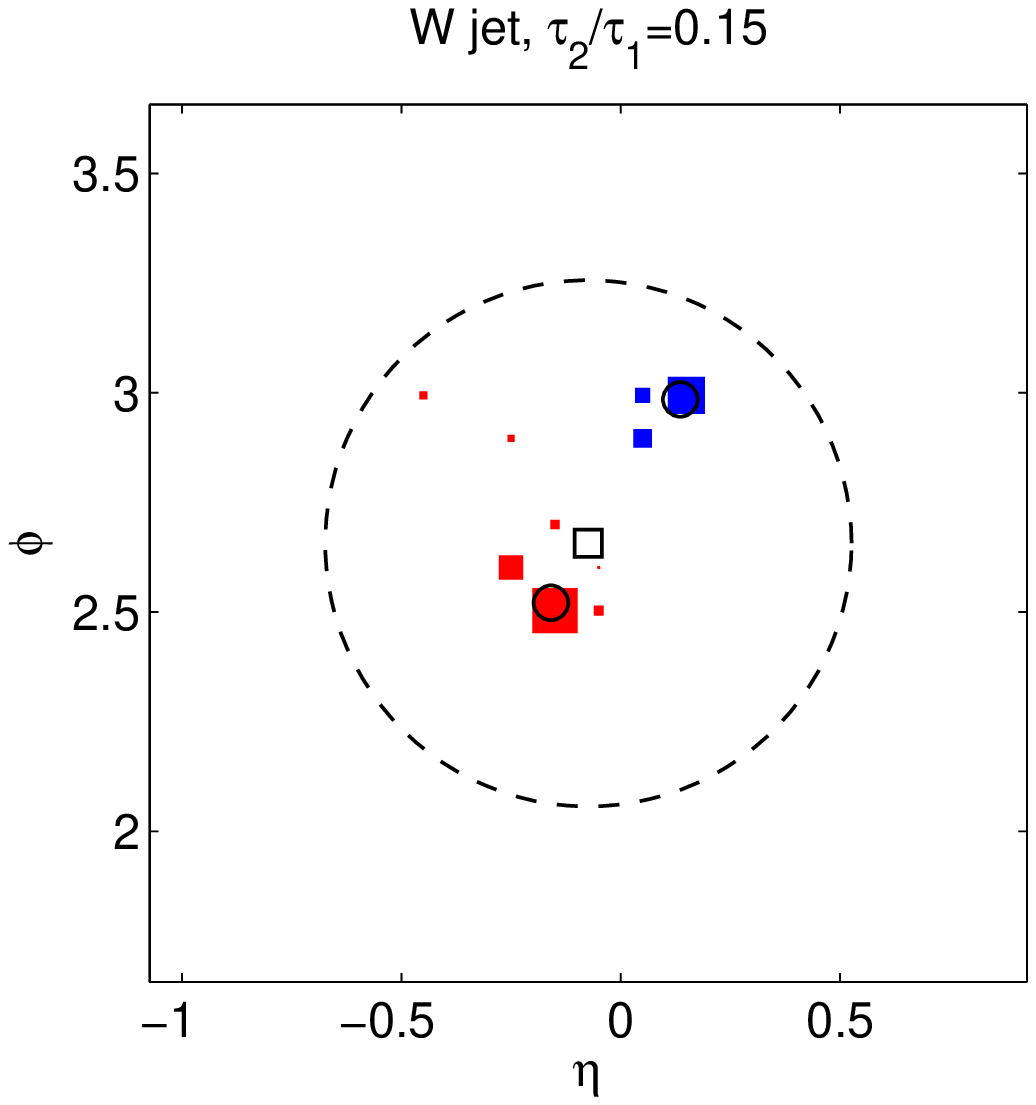}}
    \subfigure[]{\label{fig:ed10}\includegraphics[trim = 3mm 0mm 3mm 0mm, clip, height=4.60cm]{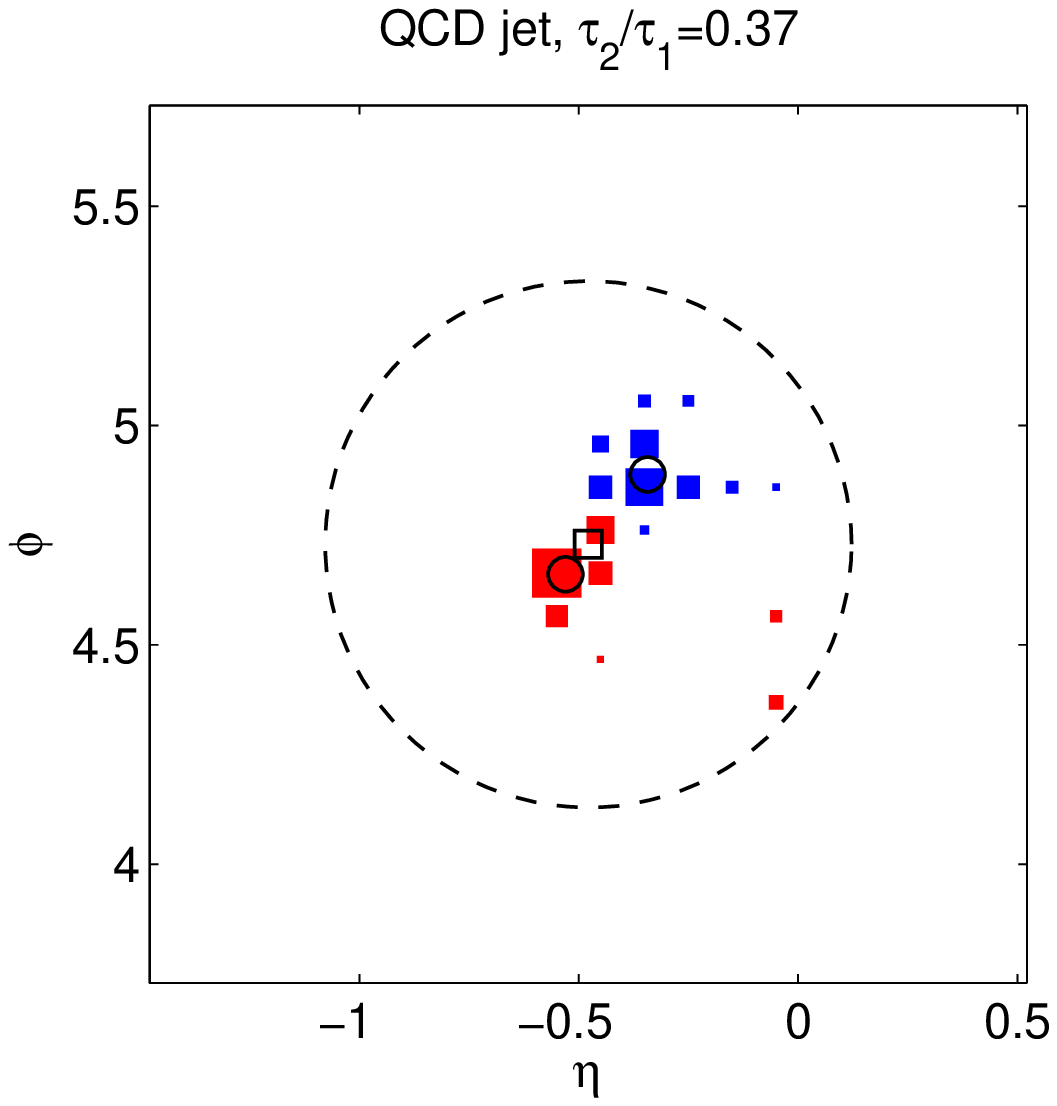}}
    \subfigure[]{\label{fig:ed11}\includegraphics[trim = 3mm 0mm 3mm 0mm, clip, height=4.60cm]{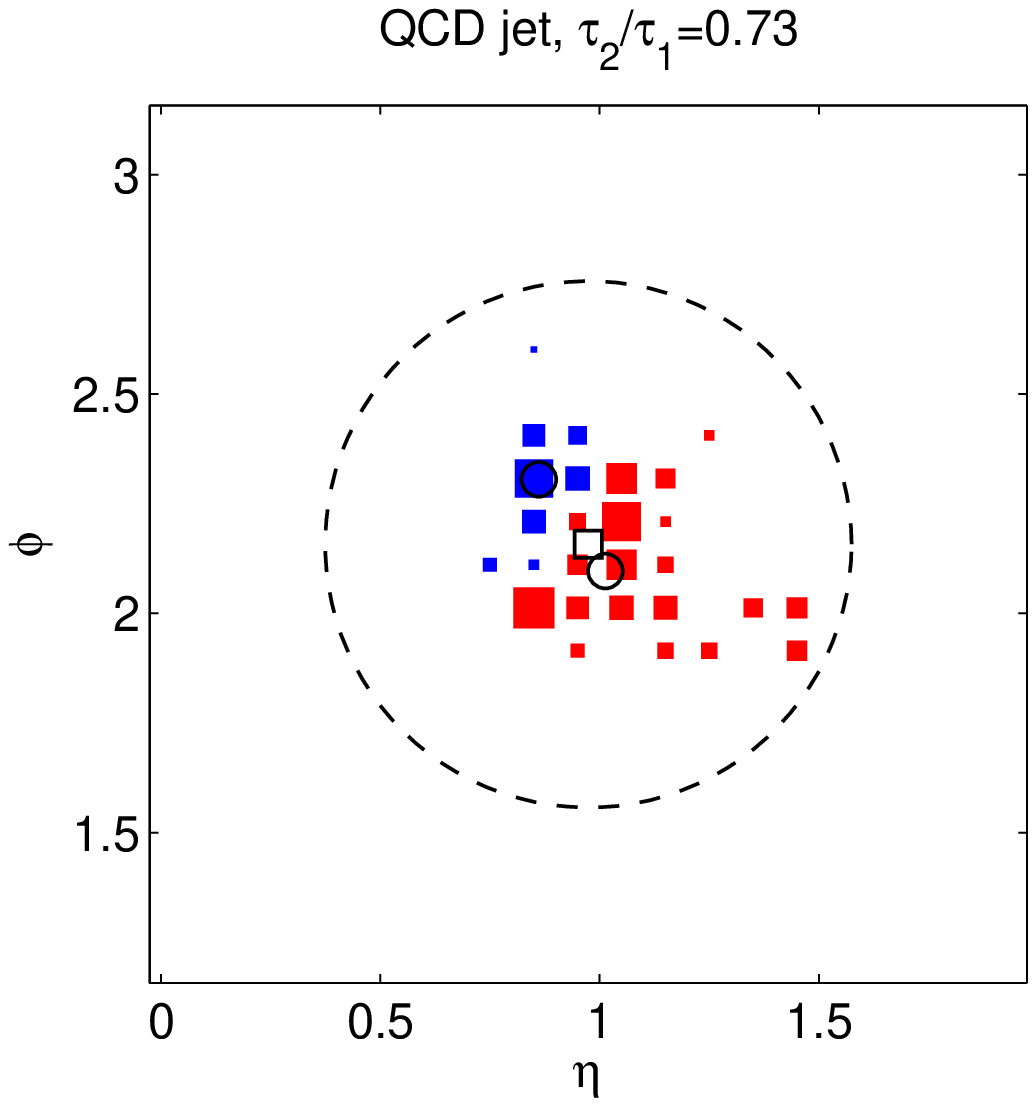}}
    \subfigure[]{\label{fig:ed12}\includegraphics[trim = 3mm 0mm 3mm 0mm, clip, height=4.60cm]{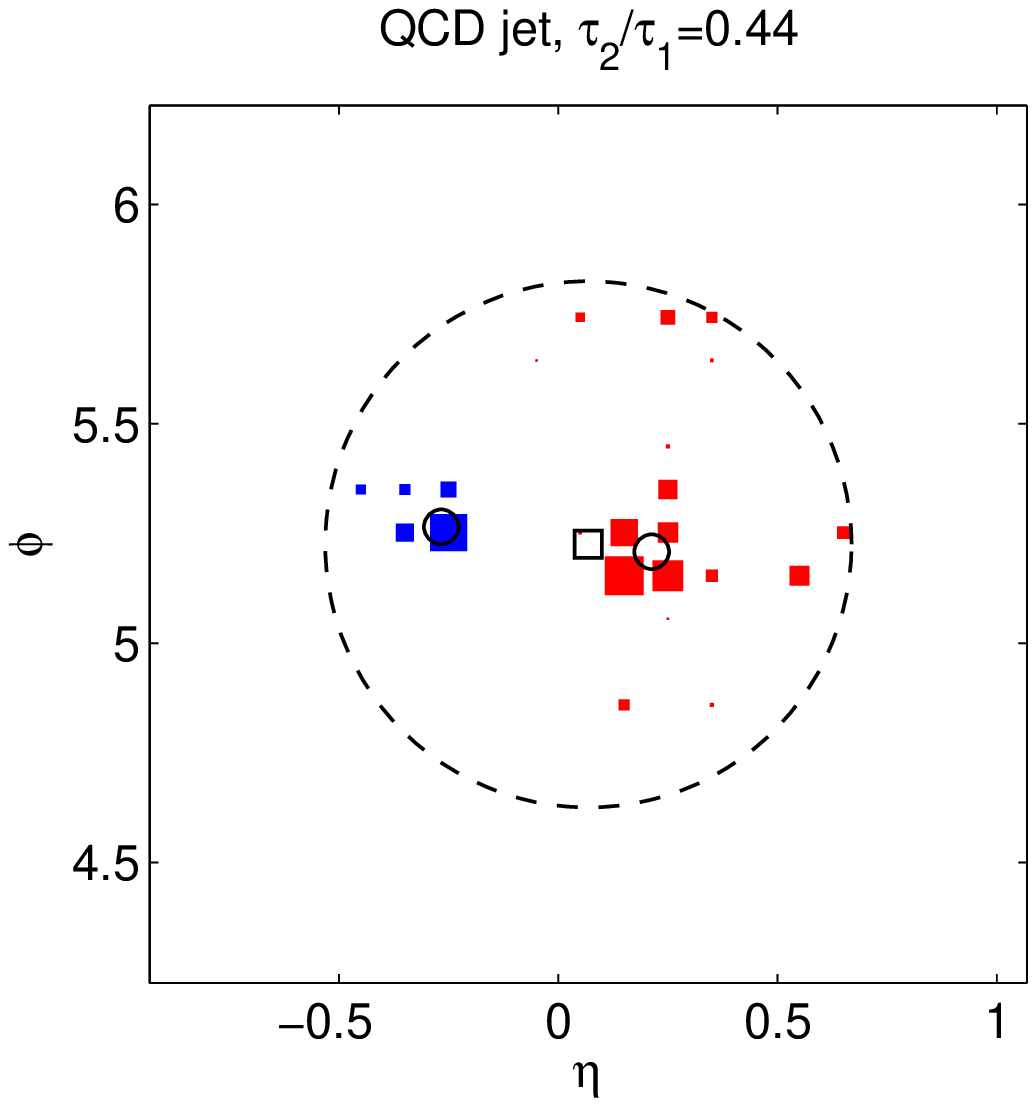}}
     \end{center}
     \vsh
    \caption{Top row:  $W$ jets.  Bottom row:  QCD jets with invariant mass close to $m_W$.  The coloring and labeling is the same as in \Fig{fig:eventDisplaysW}.  The title of the plot gives the calculated value of $\tau_2/\tau_1$, and generically $W$ jets have smaller $\tau_2/\tau_1$ values than QCD jets of comparable invariant mass.}
    \label{fig:6eventDisplaysW} 
\end{figure}

\begin{figure}[tp]
  \begin{center}
    \subfigure[]{\label{fig:ed1}\includegraphics[trim = 3mm 0mm 3mm 0mm, clip, height=4.60cm]{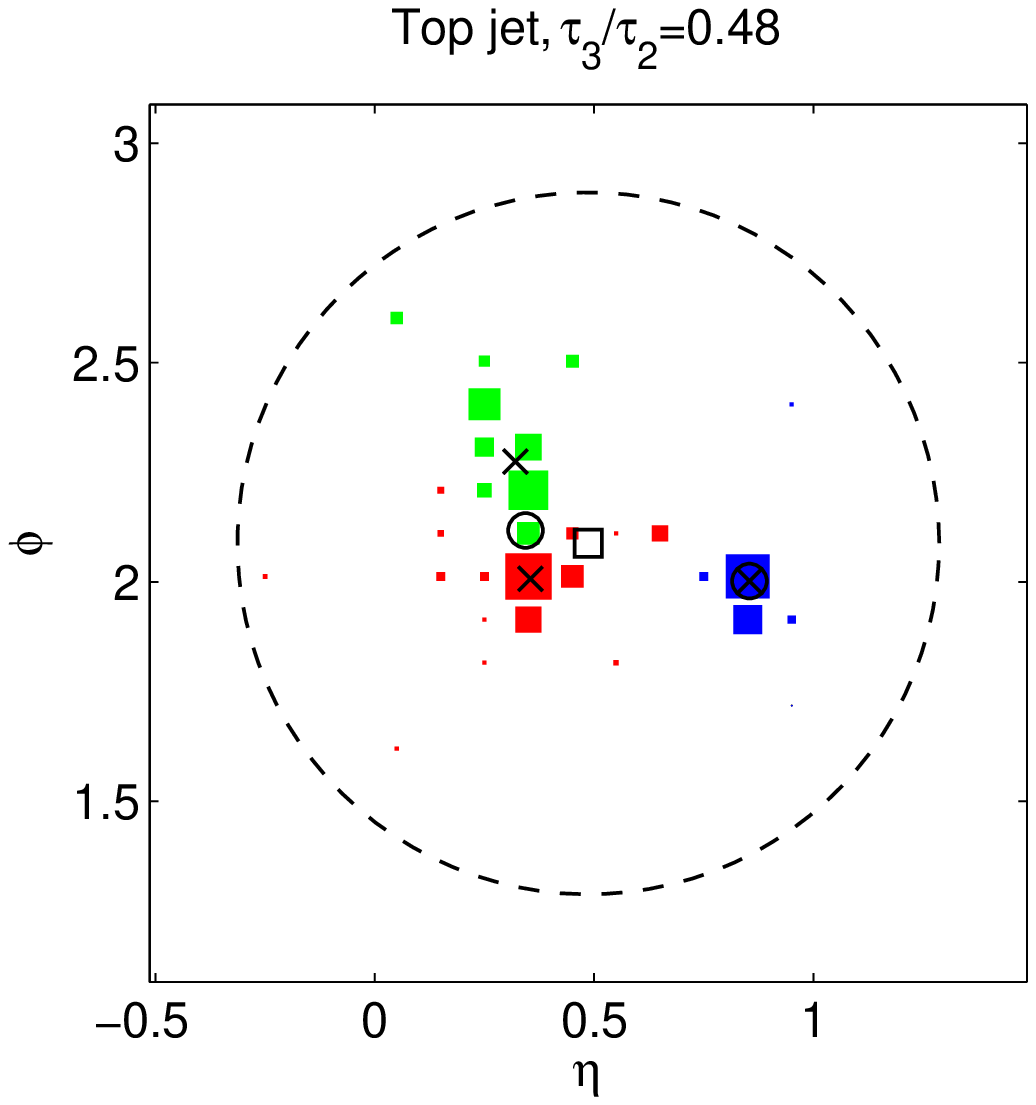}}
    \subfigure[]{\label{fig:ed2}\includegraphics[trim = 3mm 0mm 3mm 0mm, clip, height=4.60cm]{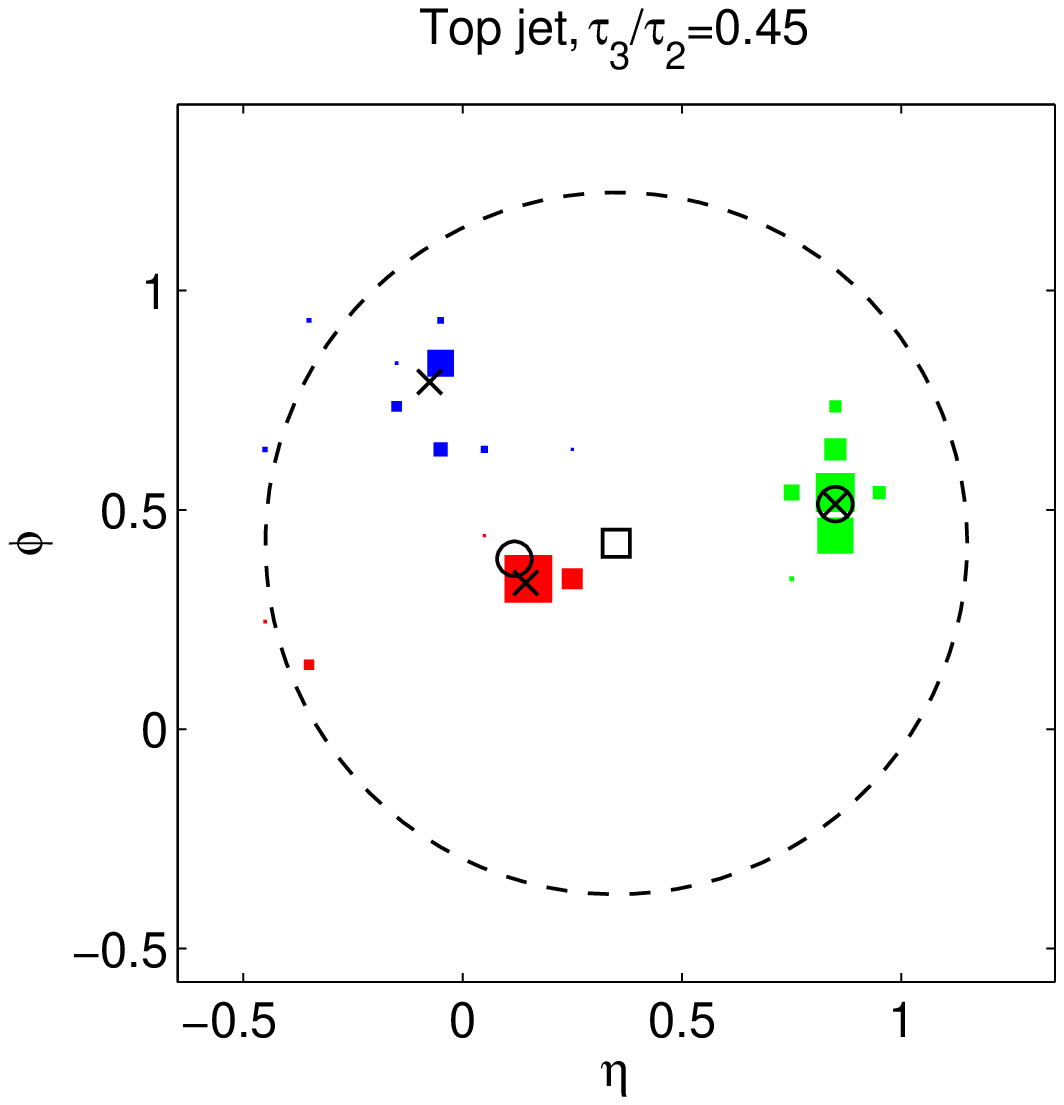}}
    \subfigure[]{\label{fig:ed3}\includegraphics[trim = 3mm 0mm 3mm 0mm, clip, height=4.60cm]{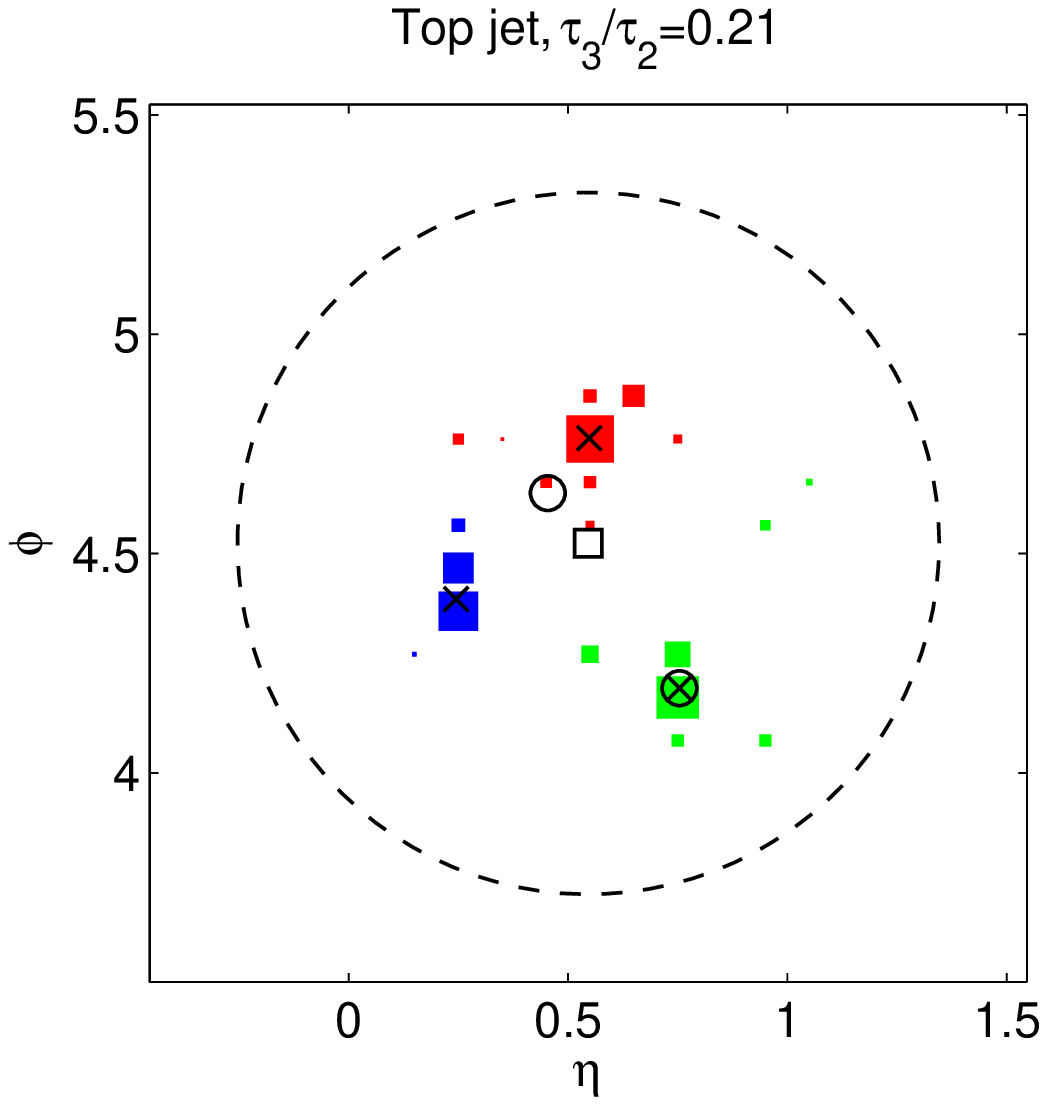}}
    \subfigure[]{\label{fig:ed4}\includegraphics[trim = 3mm 0mm 3mm 0mm, clip, height=4.60cm]{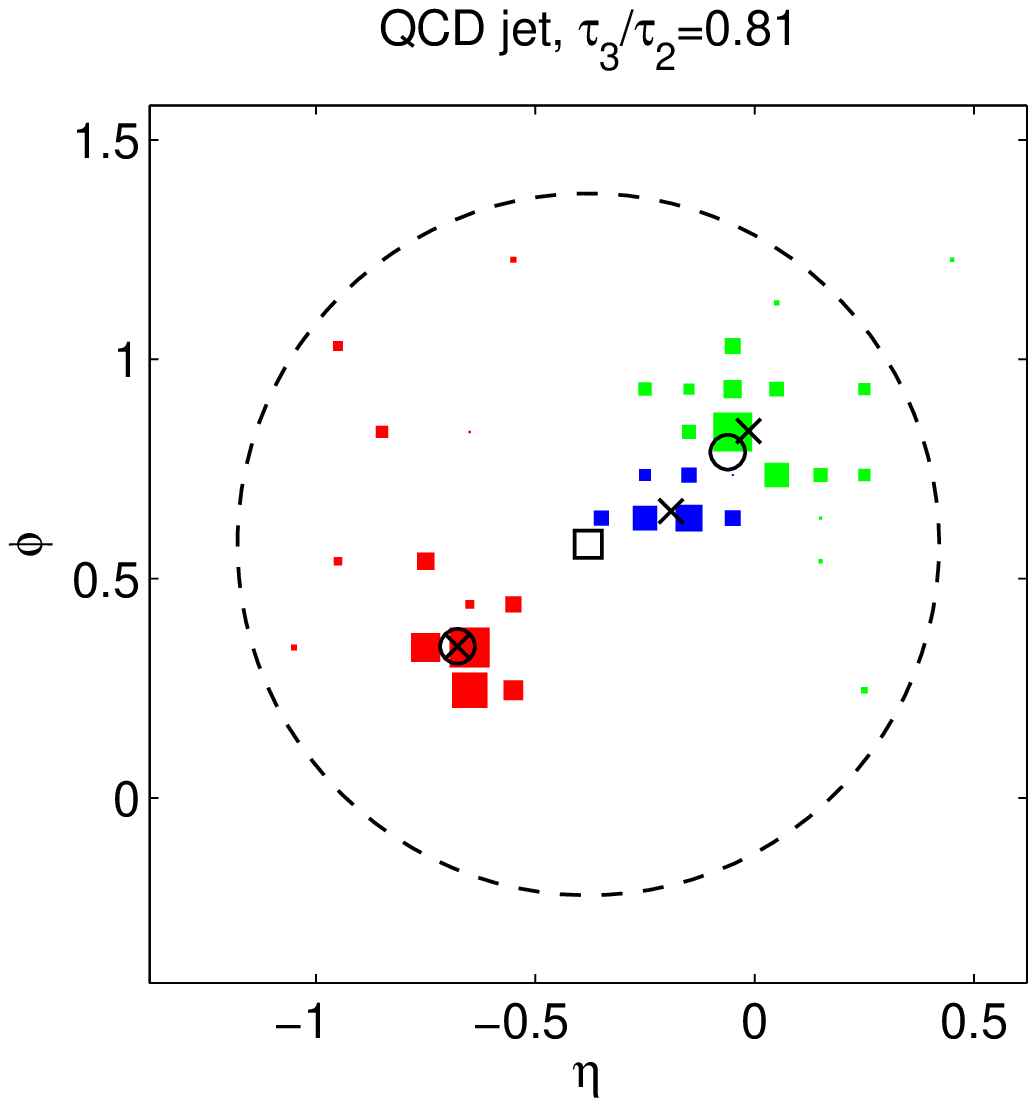}}
    \subfigure[]{\label{fig:ed5}\includegraphics[trim = 3mm 0mm 3mm 0mm, clip, height=4.60cm]{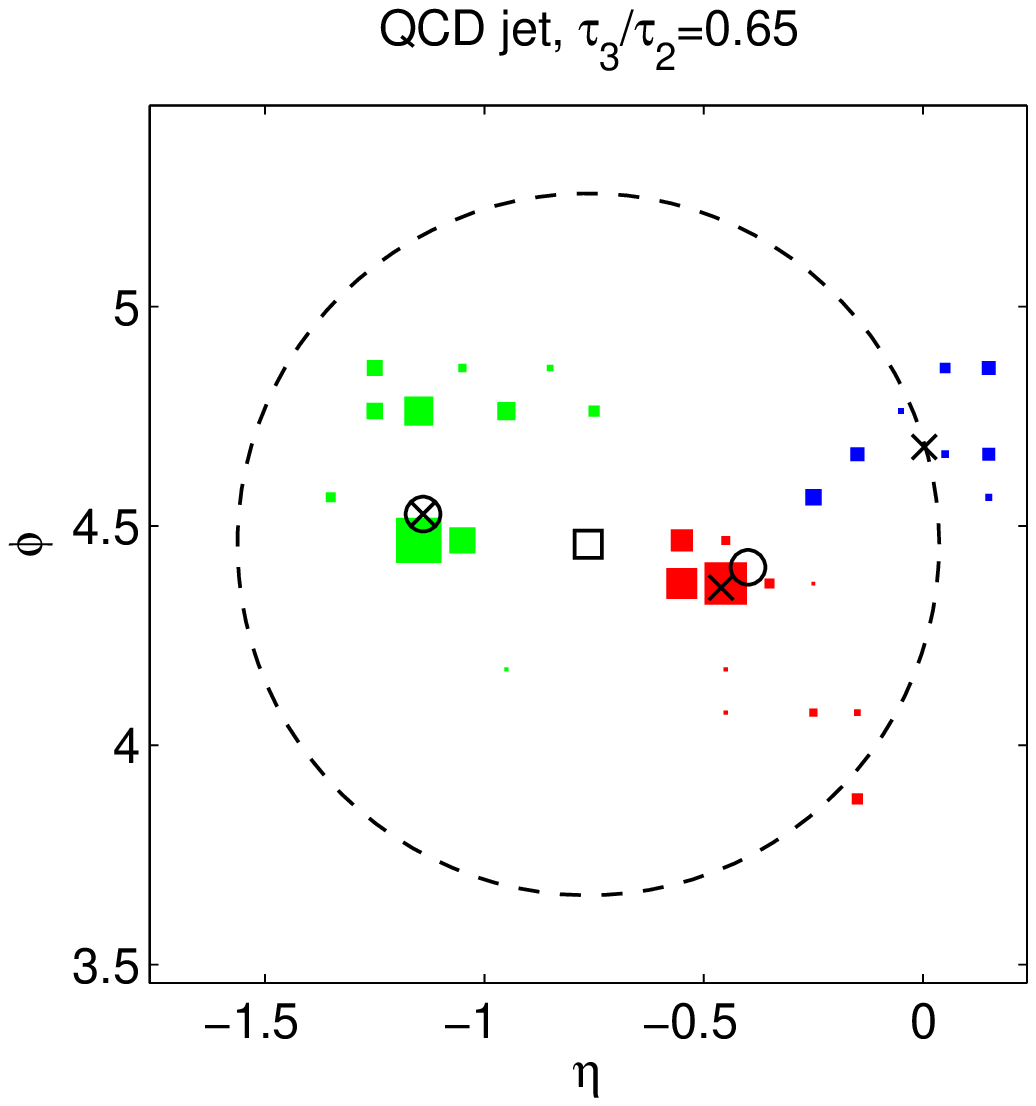}}
    \subfigure[]{\label{fig:ed6}\includegraphics[trim = 3mm 0mm 3mm 0mm, clip, height=4.60cm]{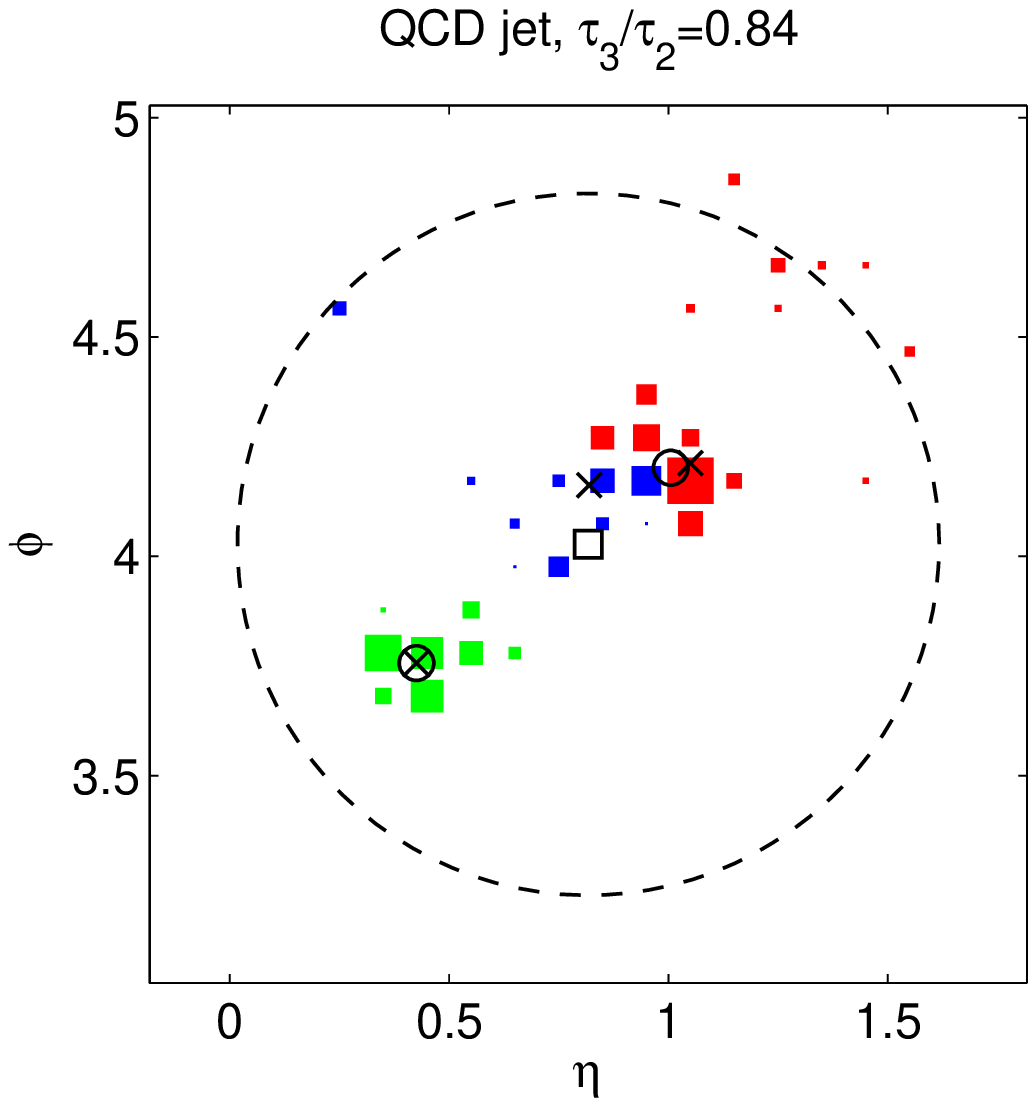}}
     \end{center}
     \vsh
    \caption{ Top row:  top jets.  Bottom row:  QCD jets with invariant mass close to $m_{\rm top}$.  The coloring and labeling is the same as in \Fig{fig:eventDisplaysTop}.  The title of the plot gives the calculated value of $\tau_3/\tau_2$, and generically top jets have smaller $\tau_3/\tau_2$ values than QCD jets of comparable invariant mass.}
    \label{fig:6eventDisplaysT}
\end{figure}

To further demonstrate our subjet finding method and the distinguishing power of $N$-subjettiness, we compare $W$ jets and QCD jets in \Fig{fig:6eventDisplaysW} and compare top jets and QCD jets in \Fig{fig:6eventDisplaysT}.  The discriminating variable $\tau_N/\tau_{N-1}$ measures the degree to which the jet energy is aligned along the $N$ candidate subjet directions compared to the $N-1$ candidate subjet directions.

\clearpage

\section{Comparison to YSplitter}
\label{app:ysplitter}

\begin{figure}[tp]
  \begin{center}
    \subfigure[][]{\label{fig:y1w}\includegraphics[trim = 00mm 0mm 00mm 0mm, clip,height=4.5cm]{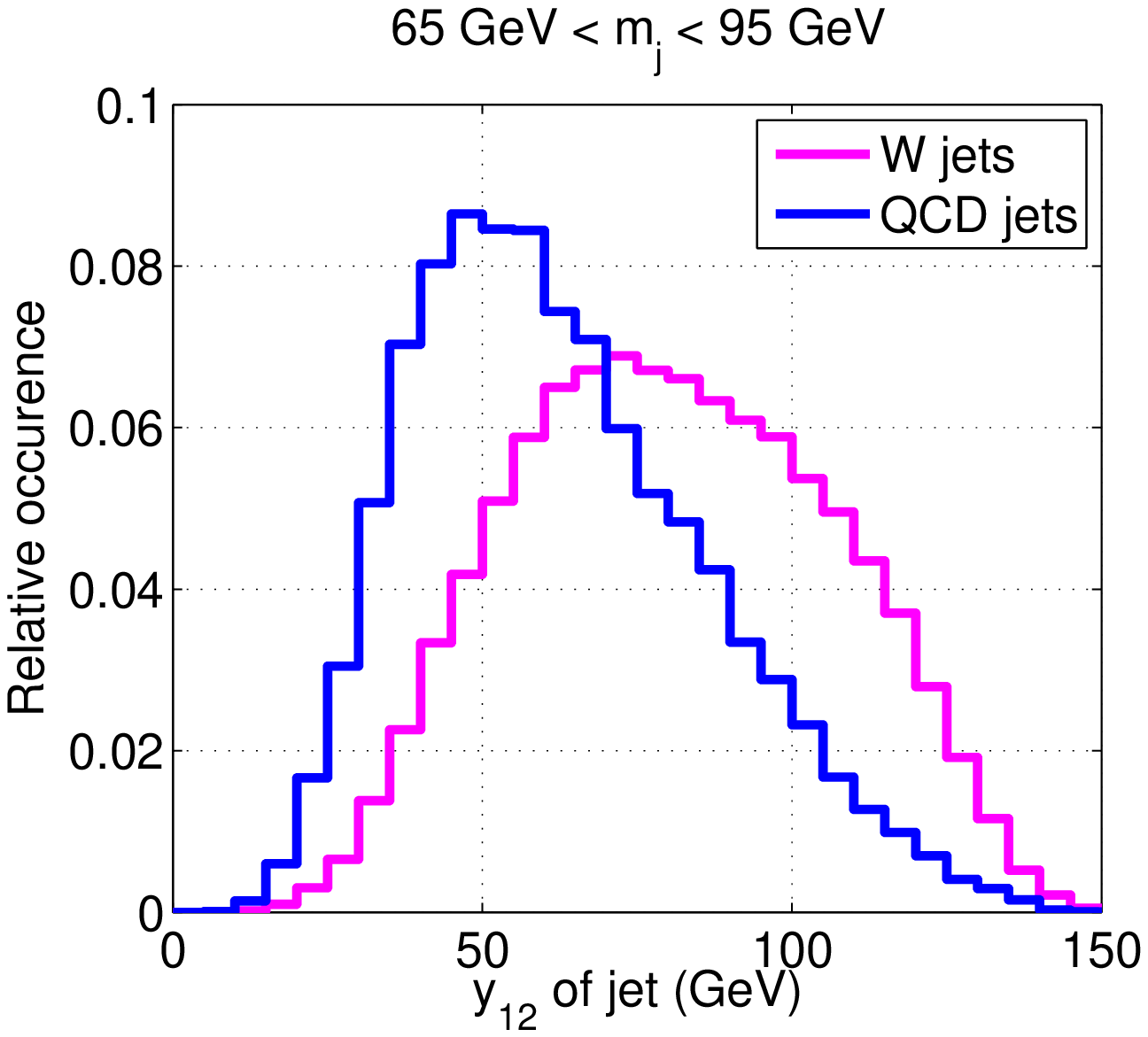}} 
    \subfigure[][]{\label{fig:y2w}\includegraphics[trim = 00mm 0mm 00mm 0mm, clip,height=4.5cm]{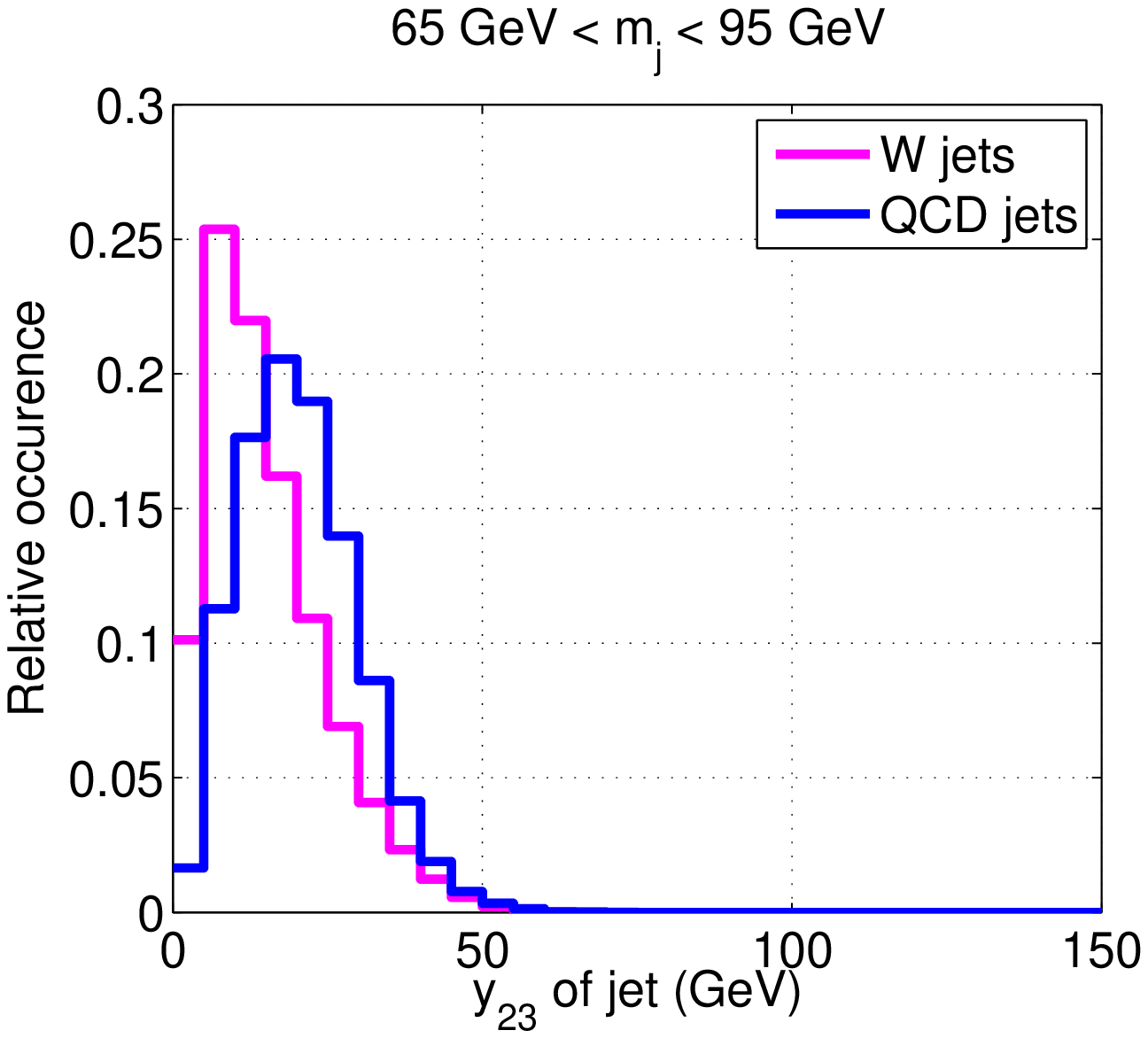}}
  \end{center}
  \vsh
\caption{Distributions of (a) $y_{12}$ and (b) $y_{23}$ for boosted $W$ and QCD jets.  For these plots, we impose an invariant mass window of $65 \text{ GeV} < m_{\text{jet}} < 95 \text{ GeV}$ on jets of $R = 0.6$, $p_T > 300$ GeV, and $|\eta| < 1.3$.}
  \label{fig:1Dy1y2y3w}
\end{figure}

\begin{figure}[tp]
  \begin{center}
      \subfigure[][]{\label{fig:y21w}\includegraphics[trim = 0mm 0mm 0mm 0mm, clip,height=4.5cm]{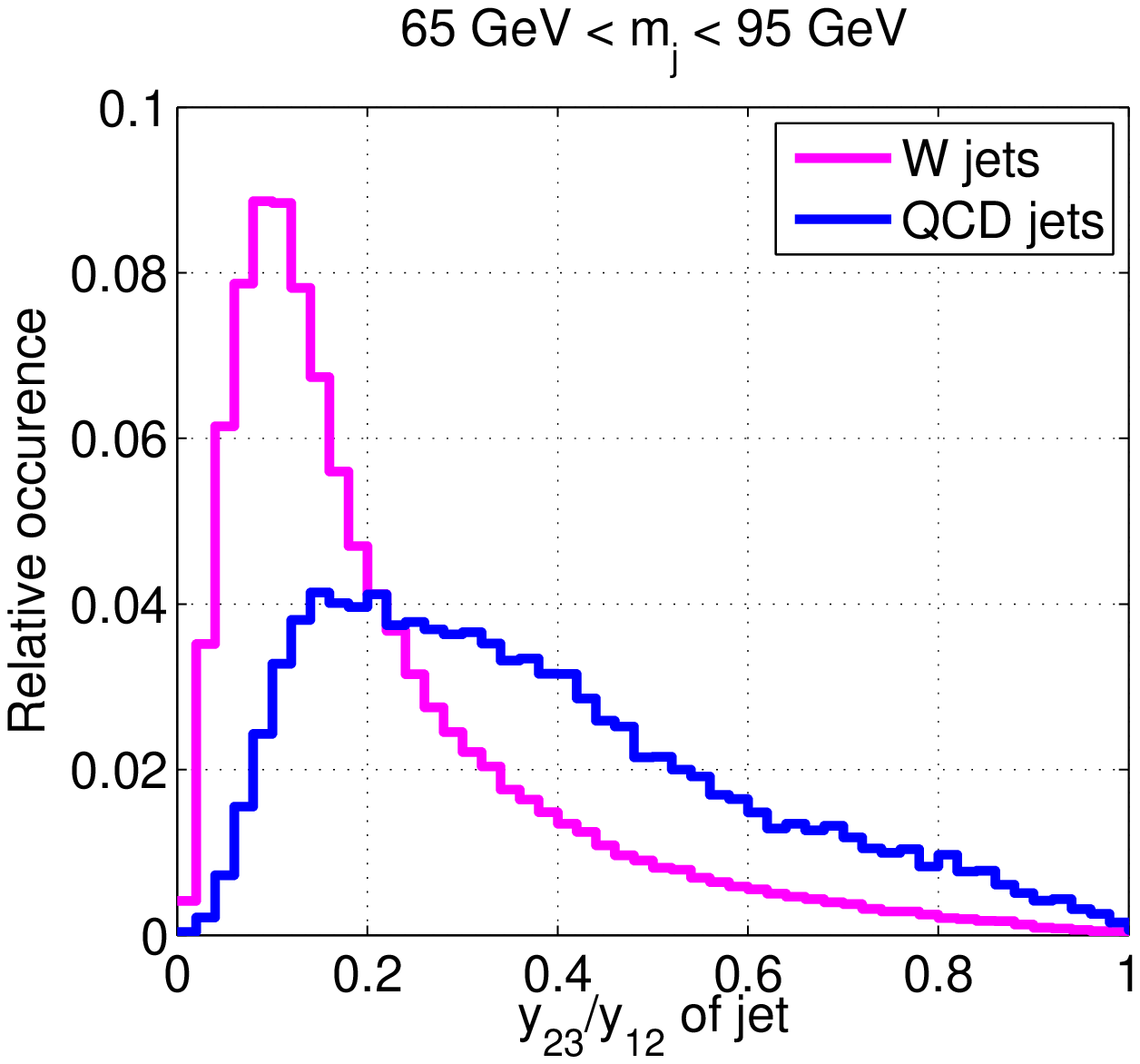}}
      \subfigure[][]{\label{fig:y21densityw}\includegraphics[trim = 0mm 0mm 0mm 0mm, clip, height=4.5cm]{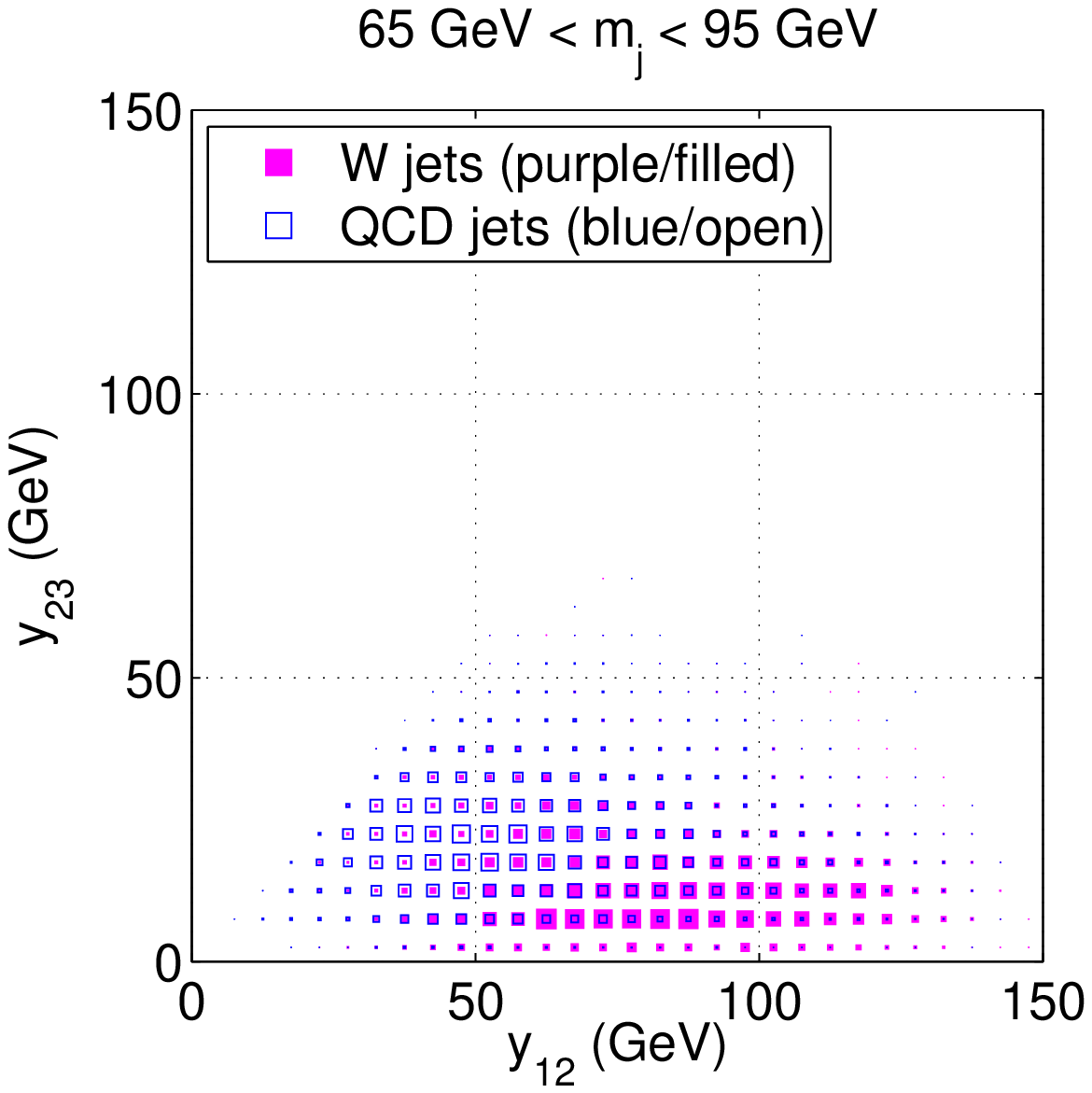}} \\
  \end{center}
  \vsh
  \caption{(a): Distribution of $y_{23}/y_{12}$ for boosted $W$ and QCD jets.  (b): Density plot in the $y_{12}$--$y_{23}$ plane.}
  \label{fig:W2Dy12}
\end{figure}

\begin{figure}[tp]
  \begin{center}
    \subfigure[][]{\label{fig:y1}\includegraphics[trim = 01mm 0mm 8mm 0mm, clip,height=4.5cm]{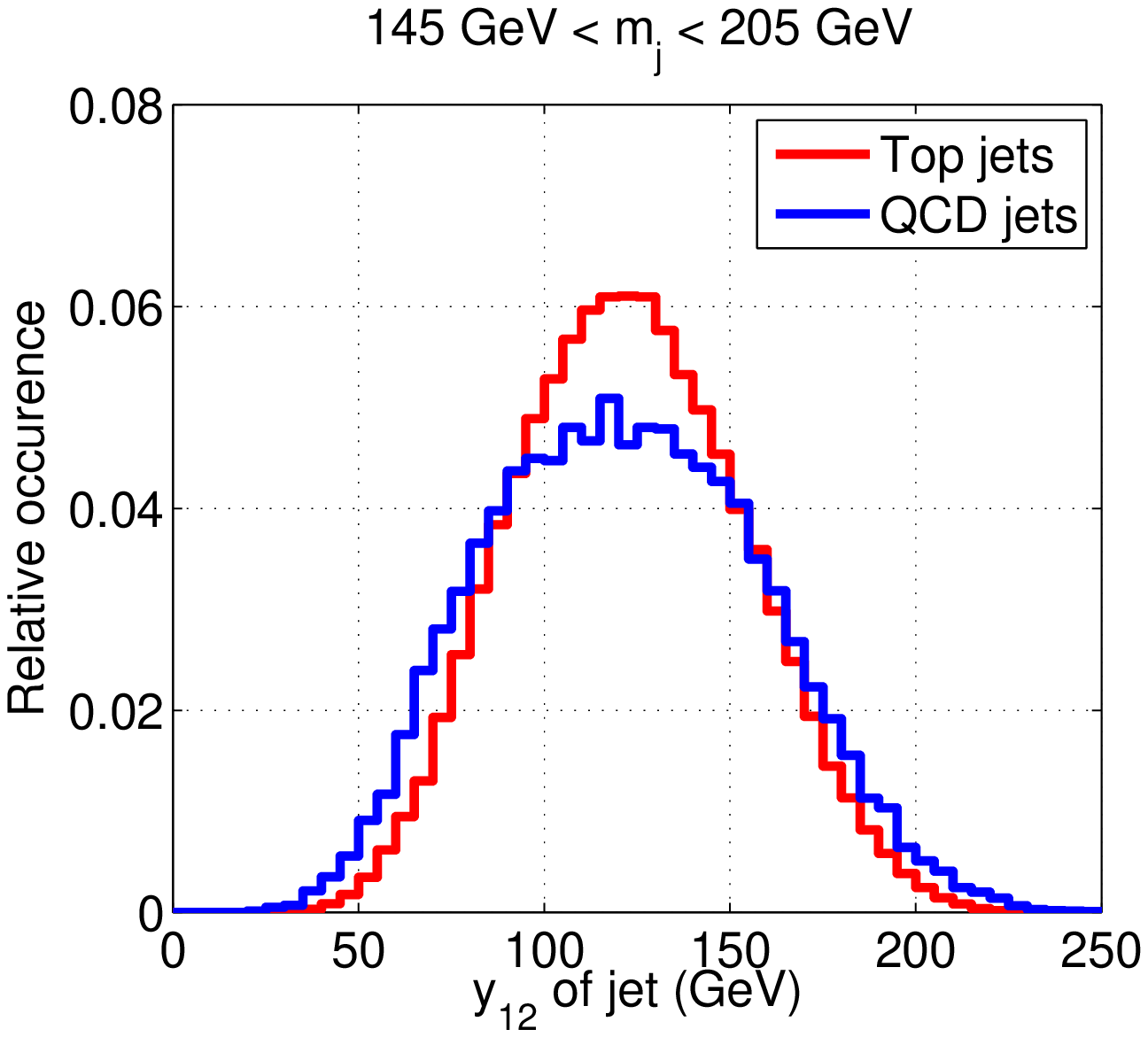}} 
    \subfigure[][]{\label{fig:y2}\includegraphics[trim = 01mm 0mm 8mm 0mm, clip,height=4.5cm]{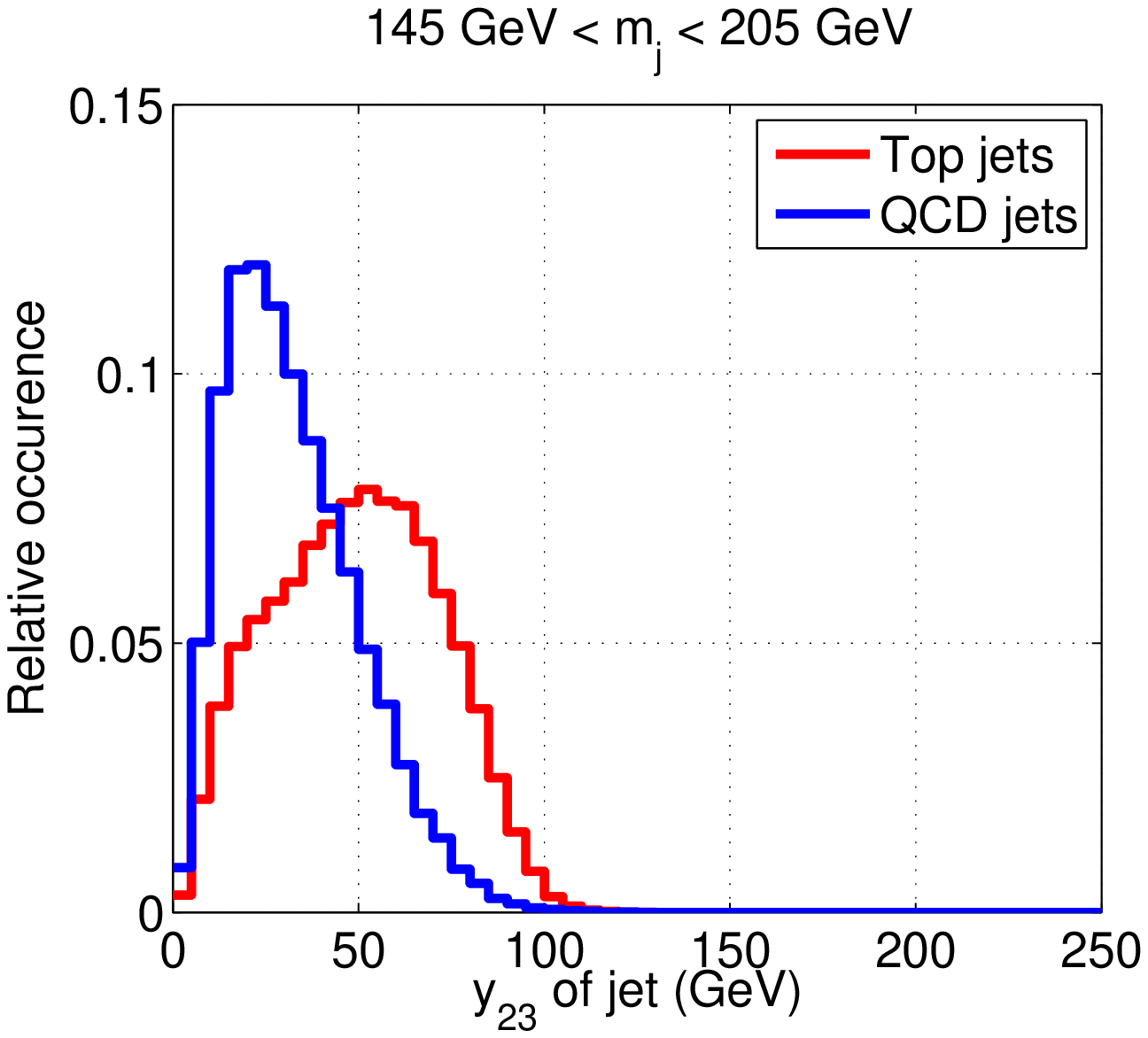}}
    \subfigure[][]{\label{fig:y3}\includegraphics[trim = 01mm 0mm 8mm 0mm, clip,height=4.5cm]{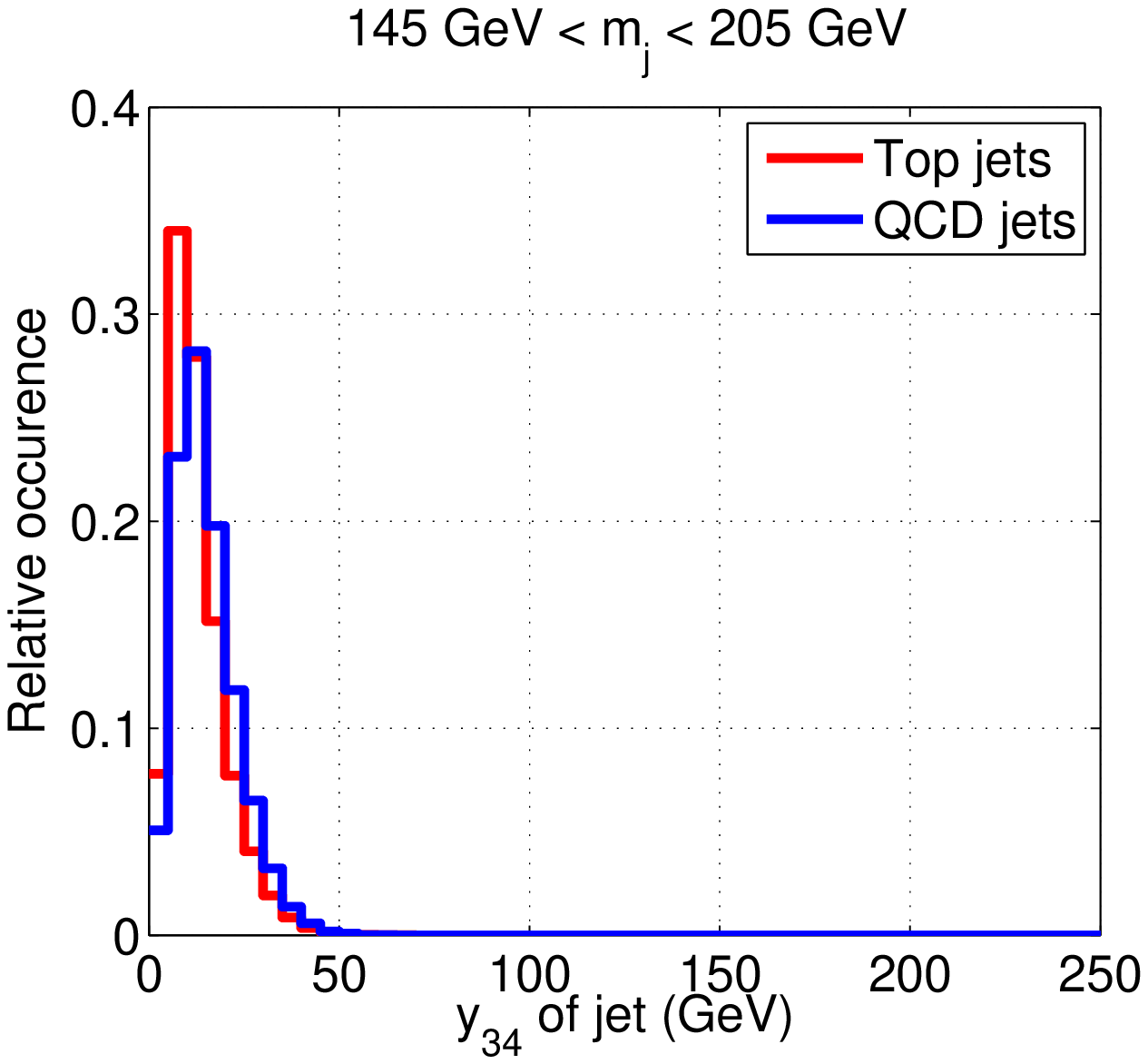}}
  \end{center}
  \vsh
\caption{Distributions of (a) $y_{12}$, (b) $y_{23}$, and (c) $y_{34}$ for boosted top and QCD jets.  For these plots, we impose an invariant mass window of $145 \text{ GeV} < m_{\rm jet} < 205 \text{ GeV} $ on jets with $R = 0.8$, $p_T > 300$ GeV, and $|\eta| < 1.3$.}
  \label{fig:1Dy1y2y3top}
\end{figure}

\begin{figure}[tp]
  \begin{center}
      \subfigure[][]{\label{fig:y21}\includegraphics[trim = 0mm 0mm 0mm 0mm, clip,height=4.5cm]{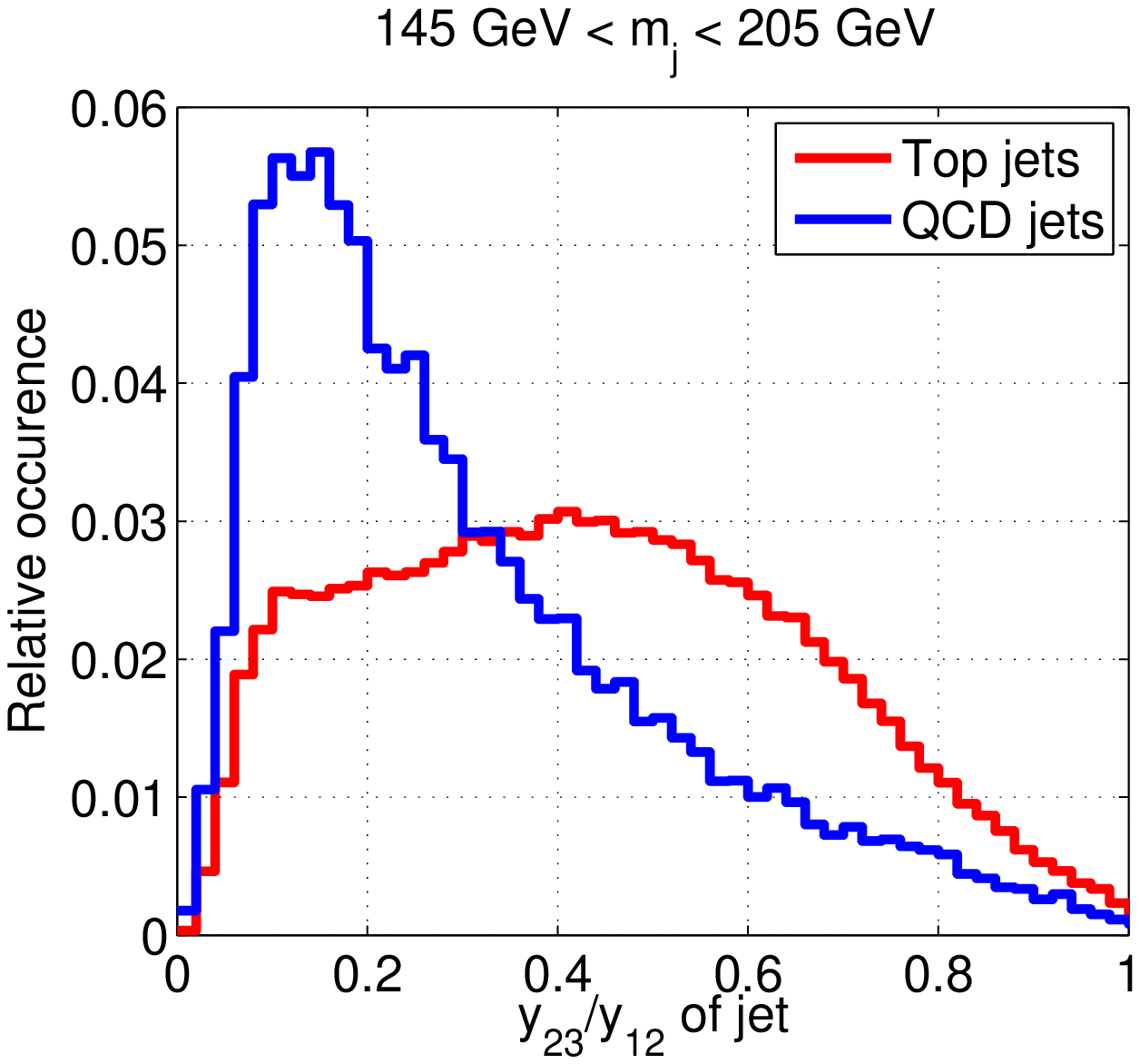}}
      \subfigure[][]{\label{fig:y32}\includegraphics[trim = 0mm 0mm 0mm 0mm, clip, height=4.5cm]{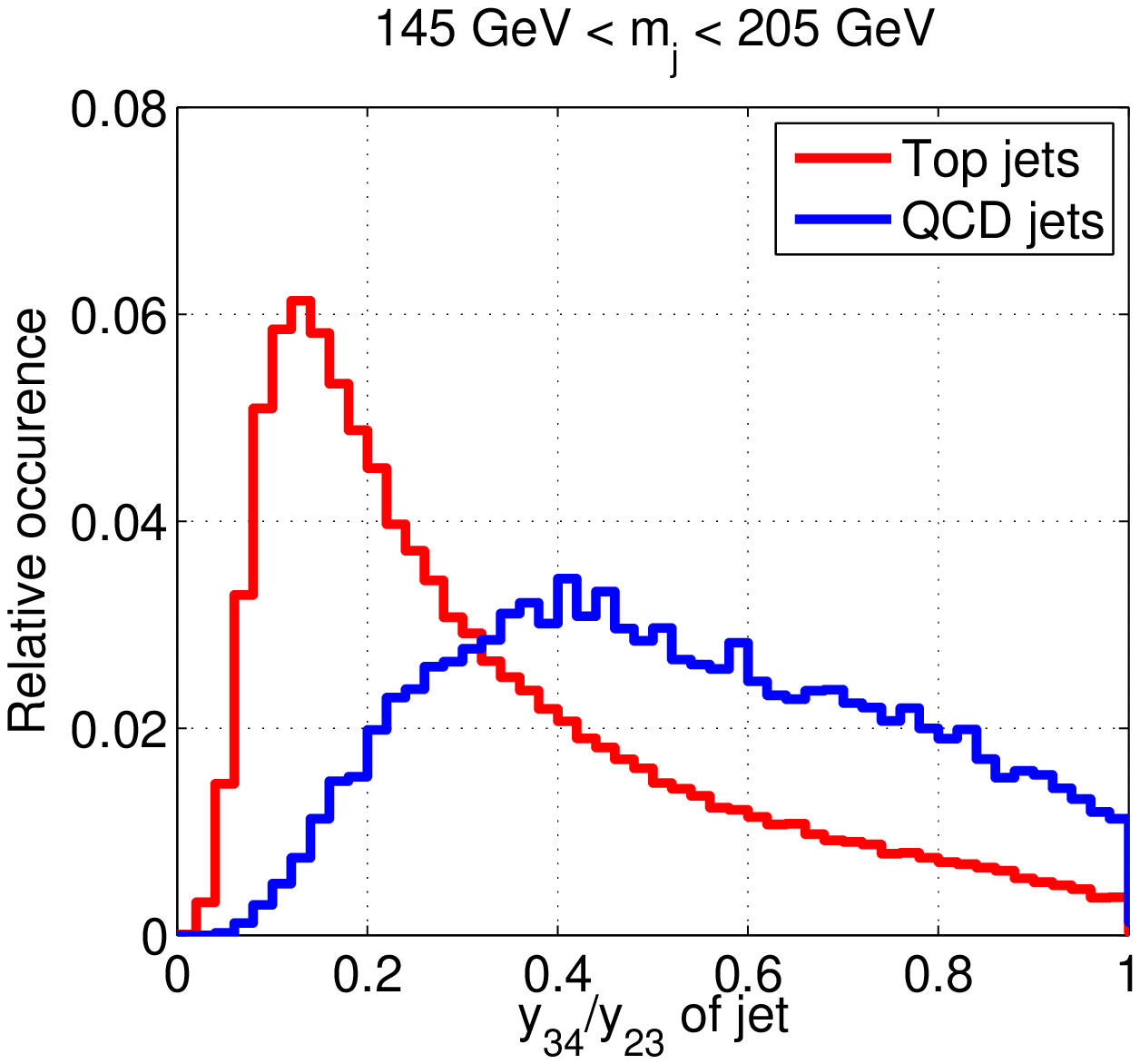}} \\
  \end{center}
  \vsh
  \caption{Distributions of  (a) $y_{23}/y_{12}$ and (b) $y_{34}/y_{23}$ for boosted top and QCD jets.  Analogous to \Fig{fig:1Dtau123ratios}, we do not use $y_{23}/y_{12}$ for top tagging in this paper, though it does contain additional information.}
  \label{fig:top1Dy12}
\end{figure}

\begin{figure}[tp]
  \begin{center}
      \subfigure[][]{\label{fig:y21density}\includegraphics[trim = 0mm 0mm 0mm 0mm, clip,height=5cm]{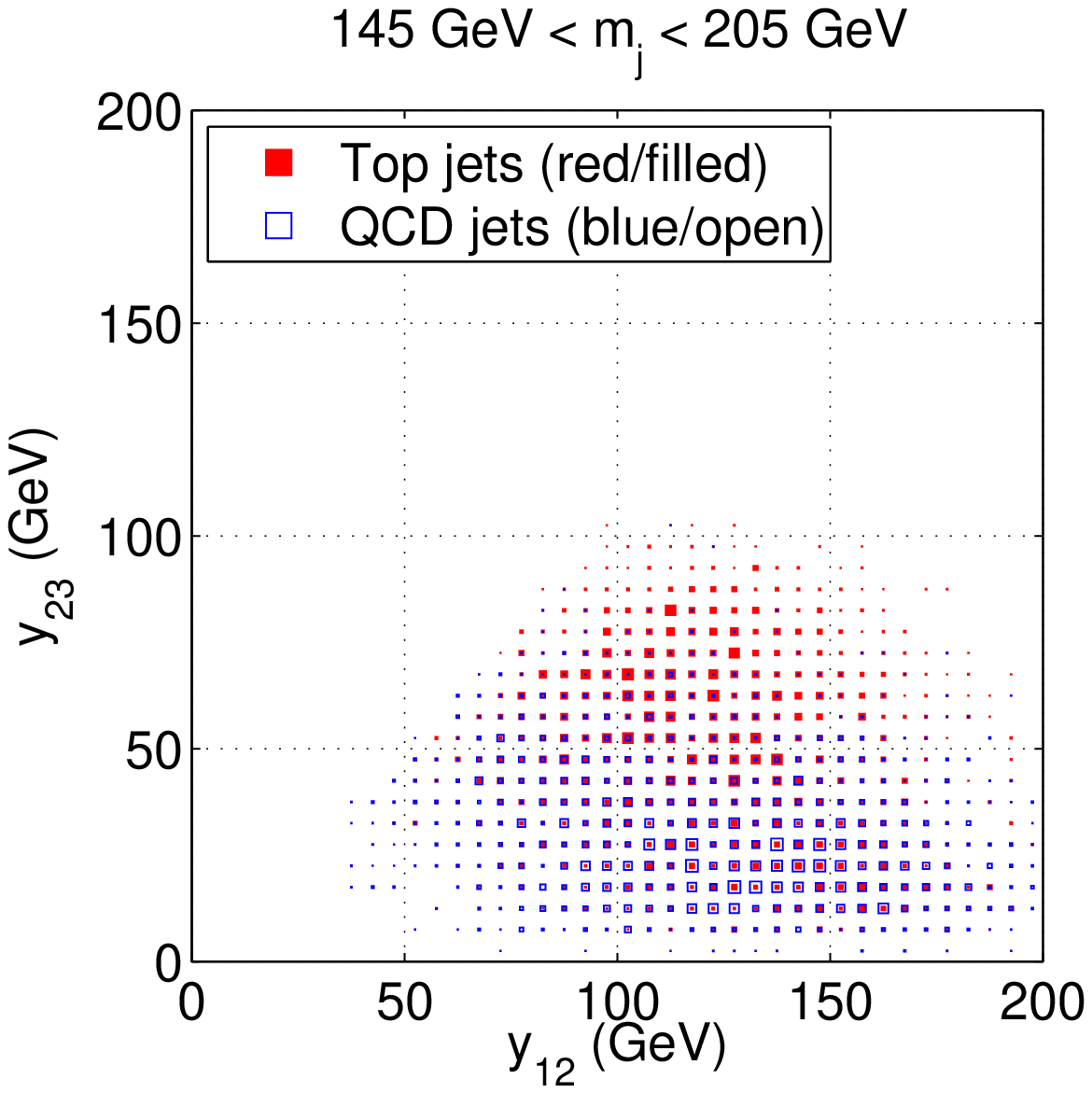}}
      \subfigure[][]{\label{fig:y32density}\includegraphics[trim = 0mm 0mm 0mm 0mm, clip, height=5cm]{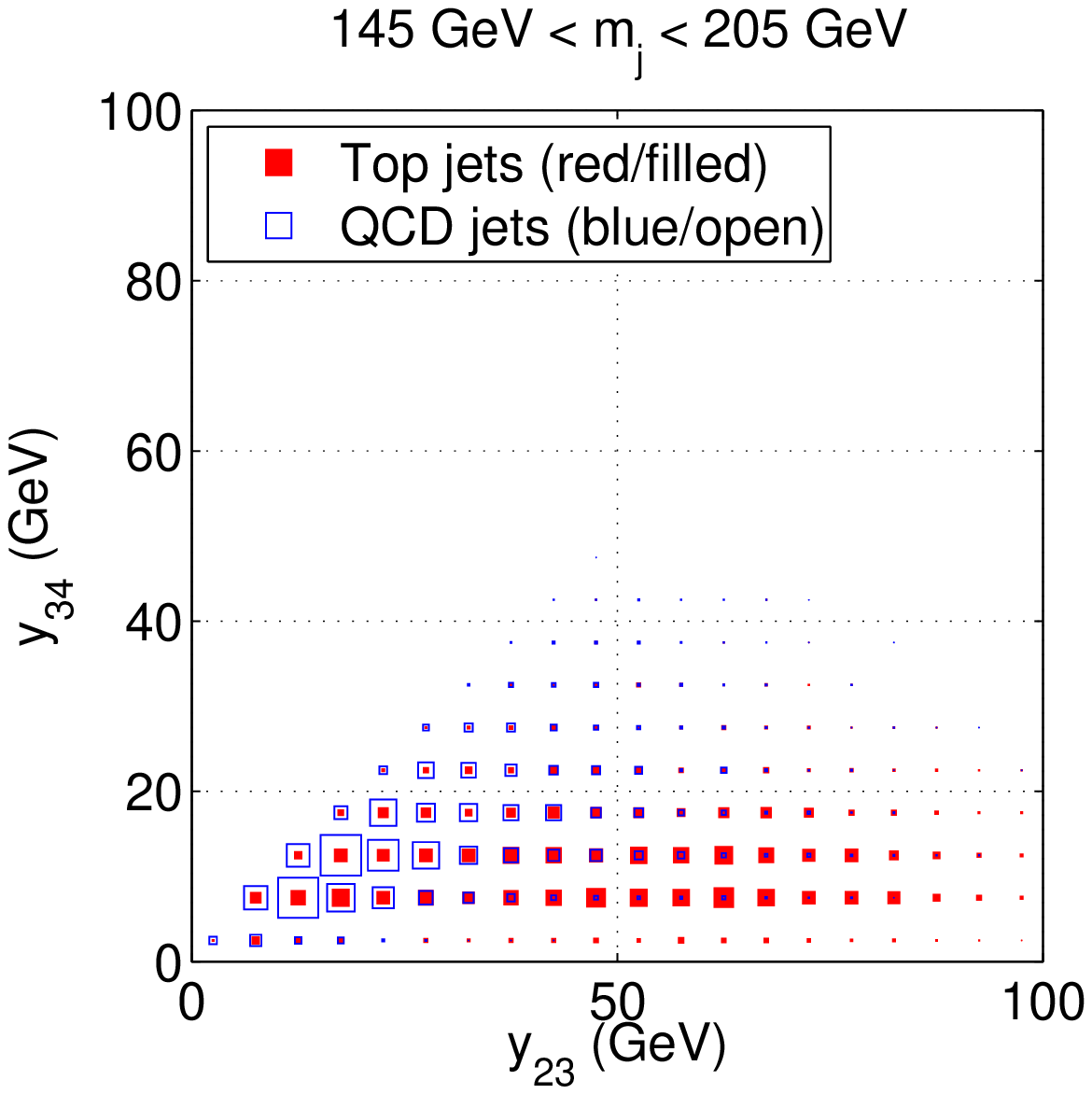}} \\
  \end{center}
  \vsh
  \caption{Density plots in the (a) $y_{12}$--$y_{23}$ plane and (b) $y_{23}$--$y_{34}$ plane for boosted top and QCD jets.}
  \label{fig:top2Dy12}
\end{figure}

In \Sec{sec:efficiency}, we compared $N$-subjettiness to a naive application of the YSplitter method \cite{Butterworth:2002tt,YSplitter,Brooijmans:2008}.  Here, we collect various plots of $y_{N,N+1}$ and their ratios so the reader can visually compare the discriminating power of  $N$-subjettiness and YSplitter. The results for $W$ jets are shown in Figs.~\ref{fig:1Dy1y2y3w} and \ref{fig:W2Dy12}, and top jets in Figs.~\ref{fig:1Dy1y2y3top}, \ref{fig:top1Dy12}, and \ref{fig:top2Dy12}.  As mentioned in the text, a full comparison of the two methods would require an optimization of all cuts, though it is encouraging that in \Sec{sec:optimization} we found that linear cuts in the $\tau_N$--$\tau_{N-1}$ plane are generically more effective than linear cuts in the $y_{N,N+1}$--$y_{N-1,N}$ plane.

\end{document}